\documentclass[a4paper, 12pt]{article}

\pdfoutput=1 

\usepackage{jheppub} 

\usepackage{
hyperxmp,		
mathtools,		
mleftright,		
orcidlink 		
}

\hypersetup{
    unicode=true,
	bookmarksdepth=3,
	bookmarksnumbered=true,
	pdftitle={On the open TS/ST correspondence},		
	pdfauthor={Matijn François, Alba Grassi},			
	pdfkeywords={topological string theory, supersymmetric gauge theory, spectral theory, quantum mechanics, quantum geometry, non-perturbative effects, difference equations, eigenfunctions},
	pdflang={en-GB},
	pdfcontactaddress={Section de Mathématiques, Université de Genève, Rue du Général Dufour 24},
	pdfcontactpostcode={1211},
	pdfcontactcity={Genève 4},
	pdfcontactregion={GE},
	pdfcontactcountry={CH},
	pdfcontactemail={matijn.francois@unige.ch, alba.grassi@cern.ch},
	pdfcontacturl={https://orcid.org/0009-0002-8099-3374, https://orcid.org/0000-0003-0654-1759},
	pdfversionid={4},									
	pdfdoi={10.48550/arXiv.2503.21762},					
	pdfurl={https://arxiv.org/abs/2503.21762},
	pdfpubtype={other},									
	pdfpubstatus={AO},									
	pdfapart=1,
	pdfaconformance=B,
}

\usepackage[type={CC}, modifier={by}, version={4.0}, imageposition=left]{doclicense}

\makeatletter 
\ifHy@unicode 
    \newcommand{\pdfhbar}{ℏ}
    \newcommand{\pdfin}{∈}
    \newcommand{\pdfpi}{π}
    \newcommand{\pdfrationals}{ℚ}
    \newcommand{\pdfxi}{ξ}
    \newcommand{\pdfsqrt}{√}
\else 
    \newcommand{\pdfhbar}{hbar}
    \newcommand{\pdfin}{in}
    \newcommand{\pdfpi}{pi}
    \newcommand{\pdfrationals}{Q}
    \newcommand{\pdfxi}{xi}
    \newcommand{\pdfsqrt}{sqrt}
\fi
\makeatother

\NewDocumentCommand{\NewMathFunction}{mo}{
	\expandafter\NewDocumentCommand\csname #1\endcsname{o}
	{\IfValueTF{#2}{#2}{\operatorname{#1}}\IfValueT{##1}{\mleft( ##1 \mright)}}
}

\NewDocumentCommand{\NewMathFunctionLim}{mo}{
	\expandafter\NewDocumentCommand\csname #1\endcsname{o}
	{\IfValueTF{#2}{#2}{\operatorname*{#1}}\IfValueT{##1}{\mleft( ##1 \mright)}}
}

\NewDocumentCommand{\RenewMathFunction}{mo}{
	\expandafter\RenewDocumentCommand\csname #1\endcsname{o}
	{\IfValueTF{#2}{#2}{\operatorname{#1}}\IfValueT{##1}{\mleft( ##1 \mright)}}
}

\NewDocumentCommand{\RenewMathFunctionLim}{mo}{
	\expandafter\RenewDocumentCommand\csname #1\endcsname{o}
	{\IfValueTF{#2}{#2}{\operatorname*{#1}}\IfValueT{##1}{\mleft( ##1 \mright)}}
}

\mleftright 

\RenewMathFunction{Re}
\RenewMathFunction{Im}

\RenewMathFunctionLim{min}
\RenewMathFunctionLim{max}
\RenewMathFunctionLim{inf}
\RenewMathFunctionLim{sup}
\RenewMathFunctionLim{lim}
\RenewMathFunctionLim{gcd}		
\RenewMathFunctionLim{det}		

\RenewMathFunction{arg}
\RenewMathFunction{deg}
\RenewMathFunction{dim}
\RenewMathFunction{ker}

\RenewMathFunction{exp}
\RenewMathFunction{ln}
\RenewMathFunction{log}
\RenewMathFunction{lg}

\RenewMathFunction{cos}
\RenewMathFunction{sin}
\RenewMathFunction{tan}
\RenewMathFunction{sec}
\RenewMathFunction{csc}
\RenewMathFunction{cot}

\RenewMathFunction{arccos}
\RenewMathFunction{arcsin}
\RenewMathFunction{arctan}

\RenewMathFunction{cosh}
\RenewMathFunction{sinh}
\RenewMathFunction{tanh}
\RenewMathFunction{coth}

\NewDocumentCommand{\rA}{}{\mathrm{A}}
\NewDocumentCommand{\rJ}{}{\mathrm{J}}

\NewDocumentCommand{\rd}{}{\mathrm{d}}
\NewDocumentCommand{\rp}{}{\mathrm{p}}

\NewDocumentCommand{\cO}{}{\mathcal{O}}
\NewDocumentCommand{\cT}{}{\mathcal{T}}

\DeclarePairedDelimiter{\pr}{(}{)}
\DeclarePairedDelimiter{\ps}{[}{]}
\DeclarePairedDelimiter{\pc}{\lbrace}{\rbrace}

\DeclarePairedDelimiter{\abs}{\lvert}{\rvert}
\DeclarePairedDelimiter{\norm}{\lVert}{\rVert}

\NewDocumentCommand{\br}{m}{\pr*{#1}}
\NewDocumentCommand{\squarebr}{m}{\ps*{#1}}
\NewDocumentCommand{\cbr}{m}{\pc*{#1}}

\NewDocumentCommand{\pints}{}{\mathbb{N}_{>0}}
\NewDocumentCommand{\nnints}{}{\mathbb{N}}
\NewDocumentCommand{\ints}{}{\mathbb{Z}}

\NewDocumentCommand{\prats}{}{\mathbb{Q}_{>0}}

\NewDocumentCommand{\rats}{}{\mathbb{Q}}

\NewDocumentCommand{\preals}{}{\mathbb{R}_{>0}}
\NewDocumentCommand{\nnreals}{}{\mathbb{R}_{\geqslant 0}}
\NewDocumentCommand{\reals}{}{\mathbb{R}}

\NewDocumentCommand{\comps}{}{\mathbb{C}}

\NewDocumentCommand{\pnaturals}{}{\pints}
\NewDocumentCommand{\naturals}{}{\nnints}
\NewDocumentCommand{\integers}{}{\ints}
\NewDocumentCommand{\prationals}{}{\prats}
\NewDocumentCommand{\rationals}{}{\rats}
\NewDocumentCommand{\complexes}{}{\comps}

\NewDocumentCommand{\IZ}{}{\mathbb{Z}}
\NewDocumentCommand{\IR}{}{\mathbb{R}}

\NewDocumentCommand{\IF}{}{\mathbb{F}}

\NewDocumentCommand{\re}{}{\mathrm{e}}
\NewDocumentCommand{\ri}{}{\mathrm{i}}

\NewMathFunction{bigo}[\mathcal{O}]
\NewMathFunction{sgn}

\NewMathFunction{sech}
\NewMathFunction{csch}

\NewMathFunction{arcsec}
\NewMathFunction{arccsc}
\NewMathFunction{arccot}

\NewMathFunction{arccosh}
\NewMathFunction{arcsinh}
\NewMathFunction{arctanh}
\NewMathFunction{arcsech}
\NewMathFunction{arccsch}
\NewMathFunction{arccoth}

\NewDocumentCommand{\real}{m}{\Re[#1]}
\NewDocumentCommand{\imaginary}{m}{\Im[#1]}
\NewDocumentCommand{\bigO}{m}{\bigo[#1]}

\NewDocumentCommand{\Li}{}{\operatorname{Li}}
\NewDocumentCommand{\dilog}{m}{\Li_2{\mleft( #1 \mright)}}

\NewDocumentCommand{\pFq}{mmmmm}{
  \, {}_{#1}F_{#2}\mleft[\begin{array}{c}#3\\[7pt]#4\end{array}\middle| \, #5\mright]
}

\NewDocumentCommand{\be}{}{\begin{equation}}
\NewDocumentCommand{\ee}{}{\end{equation}}
\NewDocumentCommand{\ba}{}{\begin{aligned}}
\NewDocumentCommand{\ea}{}{\end{aligned}}

\preprint{CERN-TH-2025-049}

\title{On the open TS/ST correspondence}
\author{Matijn Fran\c{c}ois\,\orcidlink{0009-0002-8099-3374}}
\author{and Alba Grassi\,\orcidlink{0000-0003-0654-1759}}
\affiliation{Section de Math\'ematiques, Universit\'e de Gen\`eve, 1211 Gen\`eve 4, Switzerland}
\affiliation{Theoretical Physics Department, CERN, 1211 Geneva 23, Switzerland}
\emailAdd{matijn.francois@unige.ch}
\emailAdd{alba.grassi@cern.ch}

\abstract{
The topological string/spectral theory correspondence establishes a precise, non-perturbative duality between topological strings on local Calabi-Yau threefolds and the spectral theory of quantized mirror curves.
While this duality has been rigorously formulated for the closed topological string sector, the open string sector remains less understood.
Building on the results of \cite{Marino:2016rsq, Marino:2017gyg, Francois:2023trm}, we make further progress in this direction by constructing entire, off-shell eigenfunctions for the quantized mirror curve from open topological string partition functions.
We focus on local $\mathbb{F}_0$, whose mirror curve corresponds to the Baxter equation of the two-particle, relativistic Toda lattice.
We then study the standard and dual four-dimensional limits, where the quantum mirror curve for local $\mathbb{F}_0$ degenerates into the modified Mathieu and McCoy-Tracy-Wu operators, respectively.
In these limits, our framework provides a way to construct entire, off-shell eigenfunctions for the difference equations associated with these operators.
Furthermore, we find a simple relation between the on-shell eigenfunctions of the modified Mathieu and McCoy-Tracy-Wu operators, leading to a functional relation between the operators themselves.}

\begin{document}

\phantomsection 
\pdfbookmark[1]{Abstract}{Abstract} 

\maketitle

\flushbottom

\clearpage

\section{Introduction}

The topological string/spectral theory (TS/ST) correspondence \cite{ghm, cgm2, Marino:2016rsq, Marino:2017gyg} establishes a precise non-perturbative relation between the partition functions of topological string theory on local Calabi-Yau (CY) threefolds and the spectral properties of certain quantum mechanical operators on the real line. These quantum operators are obtained through the quantization of mirror curves \cite{adkmv,acdkv} and correspond to Baxter equations for a class of relativistic integrable systems, such as cluster integrable systems \cite{Goncharov:2011hp, Fock:2014ifa} or elliptic Ruijsenaars-Schneider (RS) systems \cite{Ruijsenaars:1986vq}.

The TS/ST correspondence itself is structured in two parts: one relating closed strings to the spectrum, and the other relating open strings to the eigenfunctions.
A central feature of this duality is the relationship between the string coupling constant $g_s$ and the reduced Planck constant $\hbar$ given by
\be \label{eq:introsd} g_s = \frac{4\pi^2}{\hbar} \, . \ee
This implies that string perturbation theory naturally encodes non-perturbative effects on the spectral theory side. Conversely, the usual WKB expansion in spectral theory gives us the non-perturbative effects on the topological string side. Thus, the TS/ST correspondence bridges perturbative expansions in one theory with non-perturbative phenomena in its dual counterpart. This allows for the derivation of exact, closed-form expressions for many quantities on both sides of the correspondence.
Another important consequence of \eqref{eq:introsd} is the existence of the so-called self-dual, or maximally symmetric point \cite{cgm}, given by
\be
\hbar = 2 \pi = g_s \, .
\ee
At this special point, the TS/ST correspondence predicts remarkable simplifications, not only on the string theory side but also in operator theory, see e.g.~\cite{ghm, cgm2, Marino:2016rsq, Marino:2017gyg, Kashaev:2017zmv, Kashaev:2019gkn}.

In this work, we focus on the example where the underlying CY geometry is local $\IF_0$. In the closed string sector, one important statement of the TS/ST correspondence is  \cite{ghm}
\be\label{eq:detintro}
\det\left(1+\kappa \rho\right)=\sum_{k \in \IZ}\re^{{\rm J}(\mu + \ri 2 \pi k, \xi, \hbar)} \, ,
\qquad \qquad
\kappa=\re^{\mu} \, , \ee
where the operator $\rho: ~L^2(\IR)~\to~L^2(\IR)$ is the inverse of the quantum mirror curve to local $\IF_0$, that is $\rho={\rm O}^{-1}$ with
\be\label{eq:Ointro}
{\rm O} = \re^{\hat y} +\re^{-\hat y }+\re^{2\xi}\left(\re^{\hat x}+\re^{-\hat x}\right)
\, ,
\qquad \qquad
[\hat x, \hat y]=\ri \hbar
\, ,
\qquad \qquad
\xi \in \reals
\, , \,
\hbar \in \preals \, .
\ee
On the right-hand side of \eqref{eq:detintro}, \(\kappa\) parametrizes the closed string moduli space, while \(\xi\) is the so-called ``mass'' parameter, associated with the residue of the CY one-form at infinity.  The function
${\rm J}(\mu,\xi,\hbar)$  represents the topological string grand potential, incorporating both perturbative and non-perturbative contributions in $g_s= 4 \pi^2 / \hbar$,  see \eqref{eq:closed}.
From the perspective of the moduli space, ${\rm J}(\mu,\xi,\hbar)$ is defined around the large radius point, and its perturbative part in $g_s$ is given by the standard Gopakumar-Vafa (GV) free energy, see \eqref{eq:J_Closed_Instanton} and \eqref{eqF_Closed_GV_Instanton}.
The summation over $k$ in \eqref{eq:detintro} effectively smooths all the singularities across the closed string moduli space, parametrized by $\kappa$, allowing one to move away from the large radius point.
Indeed, we can prove that $\rho$ is a trace class operator on $L^2(\IR)$ \cite{Laptev:2015loa, Kashaev:2015kha}, and therefore its spectral determinant on the left-hand side of \eqref{eq:detintro} is an entire function of $\kappa$. We can expand it around $\kappa \to \infty$, the large radius point, or around $\kappa = 0$, the orbifold point \cite{ghm, Marino:2015ixa,cgm2}. Let us also note that the summation over $k$ in \eqref{eq:detintro} is also essential for ensuring good modular properties of the determinant \cite{cgm, Grassi:2016vkw} and for connecting it with $q$-isomonodromic $\tau$-functions \cite{bgt, Bershtein:2016aef, Bonelli:2017gdk, Gavrylenko:2023ewx, Bonelli:2022dse, Moriyama:2023mjx, Bonelli:2020dcp, Nosaka:2023xca, Nosaka:2020nuk, DelMonte:2022kxh, Bonelli:2024wha}.

The main motivation of this work is to extend the relation in \eqref{eq:detintro} to the open string sector. To this end, the first step is to define the open string grand potential ${\rm J}(x, \mu, \xi, \hbar)$, where $x$ is the open string modulus.
This was done in \cite{Marino:2016rsq, Marino:2017gyg} by focusing on the case where the brane is inserted in the outer leg of the toric diagram, the explicit form is given in \eqref{eq:Jopen}, \eqref{eq:Jfull}. Analogous to the closed string sector, ${\rm J}(x, \mu, \xi, \hbar )$ incorporates both perturbative and non-perturbative contributions in $g_s = 4 \pi^2 / \hbar$ and is defined in the large radius frame. However, one limitation of ${\rm J}(x, \mu, \xi, \hbar )$ is that it is not an entire function of the open string modulus $x$, which is a desirable property for a background-independent, non-perturbative formulation of open strings \cite{mmss}.
It was further suggested in \cite{Marino:2016rsq, Marino:2017gyg} that this requires a combination of the form
\be
\label{eq:sigmaint}
\psi(x, \kappa) = \sum_{k \in \IZ} \sum_{\sigma} \re^{{\rm J_\sigma}(x, \mu + \ri 2 \pi k, \xi, \hbar)}
\, ,
\qquad \qquad
\kappa=\re^{\mu}
\, ,
\ee
where the summation over $\sigma$ is expected to play a role analogous to the sum over $k$ in \eqref{eq:detintro}, but for the open string modulus $x$. In particular, just as the sum over  $k
$ smooths out all singularities in the closed string moduli, the sum over  $\sigma$  should similarly ensure that \eqref{eq:sigmaint} becomes an entire function of $x$.

In this paper, building on insights from \cite{Marino:2016rsq, Marino:2017gyg, Francois:2023trm}, we make this expectation precise by providing an explicit form for this summation over $\sigma$ in the case of local $\IF_0$, at generic values of $\hbar$ and the complex moduli $\mu$ and $\xi$. Specifically, we find that the precise combination to consider is
\be \label{eq:introspo}\psi(x, \kappa)= \sum_{k \in \IZ}\left(\re^{ {\rm J}(x, \mu + \ri 2 \pi k, \xi, \hbar)} + \re^{\frac{\ri}{\hbar} \frac{\pi^2}{2}+\frac{\pi x}{\hbar}+ {\rm J}(-x-\ri \pi, \mu + \ri \pi + \ri 2 \pi k, \xi, \hbar)} \right) \, .\ee
From the perspective of spectral theory, the combination \eqref{eq:introspo} is a solution of the functional difference equation corresponding to the quantized mirror curve \eqref{eq:Ointro},
\begin{equation}
    \label{eq:intro} \psi ( x + \ri \hbar , \kappa ) + \psi(x-\ri \hbar,\kappa) + 2 \re^{2\xi} \cosh(x) \psi(x,\kappa) +\kappa \psi(x,\kappa)=0
    \, .
\end{equation}
Difference equations of the type \eqref{eq:intro} admit many solutions.
The proposal \eqref{eq:introspo} stands out in three ways. Firstly, \eqref{eq:introspo} is always a well-defined function of $x, \kappa, \xi$ and $\hbar$ solving \eqref{eq:intro}, and not just a formal solution.\footnote{By ``formal solution'', we mean an object that solves \eqref{eq:intro} but that is not a well-defined function. For example, objects defined via divergent series expansions or those with a dense set of poles in their domain of definitions (e.g.~when $\hbar\in \preals$).} Secondly, \eqref{eq:introspo} becomes a proper eigenfunction of the operator \eqref{eq:Ointro} when $\kappa$ is a root of the spectral determinant \eqref{eq:detintro}. Thirdly, the eigenfunctions \eqref{eq:introspo} are entire in $x$ for all $\kappa \in \complexes$. We do not have a complete, mathematical proof for these statements, but we performed many analytic and numerical tests that support them.

One can further express such eigenfunctions in terms of $\mathrm{O}\br{2}$ matrix models by performing a canonical transformation \cite{Marino:2016rsq}. More specifically we have
 \begin{equation}
\label{eq:canonicalintro}
    \psi (x,\kappa) =  \int_{\mathbb{R}} \rd q \, U\left( x, q \right) \Xi(q,\kappa) \, ,
\end{equation}
where $U(x,q)$ is the kernel of a unitary transformation, see \eqref{eqused}, and we have
\begin{equation}\label{eq:kappaexin}
    \begin{split}
        \Xi(q, \kappa) & = \exp\left( \frac{\pi}{\sqrt{2}} \frac{q}{\hbar}\right) {\tt f}(q) \sum_{N = 0}^{\infty} \kappa^N \Psi_N(q) \, ,
    \end{split}
\end{equation}
where ${\tt f}(q)$ is given in \eqref{sqrtv} and $\Psi_N(q)$ is defined by the following unnormalized expectation value within the $\mathrm{O}\br{2}$ matrix model \eqref{eq:PsiNd}
\begin{equation}\label{eq:phiintro}
    \Psi_N(q) = \left\langle \prod_{k = 1}^N \tanh\left( \frac{\pi}{\sqrt{2}} \frac{q - q_k}{\hbar} \right) \right\rangle \, .
\end{equation}
The expression \eqref{eq:kappaexin} is particularly interesting because it allows one to easily connect with the conifold frame as we discuss later.

This paper is organized as follows.
In \autoref{sec:sec2}, we analyse the matrix model \eqref{eq:phiintro} in some detail and show how its canonical transformation \eqref{eq:canonicalintro} naturally leads to the symmetric structure of the two contributions in \eqref{eq:introspo}. In \autoref{sec:tsst}, we translate this into a conjecture for the eigenfunctions in terms of open topological strings, as given in \eqref{eq:introspo}, and we perform several detailed tests of the proposal.
In \autoref{sec:sec4}, we examine two specific limits of our construction: the standard four-dimensional limit \cite{kkv, Klemm:1996bj}, defined in \eqref{eq:4d}, and the dual four-dimensional limit \cite{bgt,bgt2}, defined in \eqref{eq:dual4d}.
In the standard four-dimensional limit, the difference equation \eqref{eq:intro} reduces to the Fourier-transformed Mathieu operator whose eigenvalue equation reads
\be \label{eq:matFintro}\sqrt{t} \left(\phi(x + \ri\epsilon, E)+\phi(x-\ri\epsilon, E)\right)+{x}^2\phi(x, E)- E \phi(x, E)=0 \, , \ee
where $t, E$ and $\epsilon$ correspond to the four-dimensional limits of $\xi$, $\kappa$ and $\hbar$ in \eqref{eq:intro}, respectively.
As with \eqref{eq:intro}, this equation has many formal solutions. However, our construction identifies a special class of eigenfunctions that are entire, even off-shell. These are given by
\be \label{eq:eigof}\phi (x, E)= \phi_{1}\left(\frac{x}{\epsilon},\frac{\sigma}{\epsilon},\frac{t}{ \epsilon^4}\right)+\phi_{2}\left(\frac{x}{\epsilon},\frac{\sigma}{ \epsilon},\frac{t}{\epsilon^4}\right) \ee
with
\be
\label{eq:s11int}
\ba
    \phi_{2}(x, \sigma, t)= &
    \phi_{1}(-x, \sigma, t)\left[  \frac{ \re^{- \frac{\ri}{2} \partial_\sigma F_\mathrm{NS}^{\rm 4d}\left( \sigma, t \right) }\left( \re^{2 \pi x} - \re^{2 \pi \sigma}  \right) - \re^{\frac{\ri}{2} \partial_\sigma F_\mathrm{NS}^{\rm 4d}\left( \sigma, t \right)} \left( \re^{2 \pi x} - \re^{- 2 \pi \sigma}  \right)}{\re^{2 \pi \sigma} - \re^{-2 \pi \sigma}}\right]
   \, ,
\ea\ee
where $\phi_{1}(x,\sigma,t)$ is defined in \eqref{eq:s12} with $\sigma$ and $E$ being related by \eqref{eq:quantumm}. The factor in square brackets in \eqref{eq:s11int} is crucial to ensuring that the off-shell function \eqref{eq:eigof} is entire.
When evaluated on-shell, this factor is $\pm 1$ depending on the parity of the eigenfunction, and our result reproduces the well-known expression for the on-shell eigenfunction in terms of 2d/4d surface defects in the Nekrasov-Shatashvili (NS) phase of the $\Omega$-background \cite{Kozlowski:2010tv, Alday:2010vg, Kanno:2011fw, Jeong:2021rll, Jeong:2018qpc, Jeong:2017pai, Alday:2009fs, Drukker:2009id, Sciarappa:2017hds}.
On the other hand, in the dual four-dimensional limit \eqref{eq:dual4d}, the operator \eqref{eq:Ointro} leads to the McCoy-Tracy-Wu operator \cite{bgt}
\be\label{eq:mtwope}  \re^{4 t^{1/4}\cosh(\hat x)} \cosh\left(\frac{\hat{y}}{2} \right) \re^{4 t^{1/4}\cosh(\hat x)}
\, ,
\ee
whose eigenfunctions are computed by 2d/4d surface defects in the GV (or self-dual) phase of the $\Omega$-background \cite{Francois:2023trm}. Even though the off-shell eigenfunctions of \eqref{eq:matFintro} and \eqref{eq:mtwope} are quite different, when evaluated on-shell, they are related in a remarkably simple way. This in turn provides a clear functional relation between the modified Mathieu \eqref{eq:matFintro} and McCoy-Tracy-Wu \eqref{eq:mtwope} operators, see \autoref{sec:relation} and equation \eqref{eq:functional}.
In \autoref{sec:conc}, we conclude and outline some open problems. We also have four appendices that provide technical details and definitions necessary for understanding the results in the main text.

\acknowledgments

We would like to thank Frederico Ambrosino, Giulio Bonelli, Shi Cheng, Fabrizio Del Monte, Pavlo Gavrylenko, Jie Gu, Saebyeok Jeong, Marcos Mari\~{n}o, Nicolas Orantin, Maximilian Schwick, Alessandro Tanzini, and Szabolcs Zakany for interesting and valuable discussions. We are also grateful to Fabrizio Del Monte and Marcos Mari\~{n}o, as well as two anonymous referees, for a careful reading of the draft.
This work is partially supported by the Swiss National Science Foundation Grant No.~185723 and the NCCR SwissMAP.

\paragraph{Version of Record.}

This version of the article has been accepted for publication, after peer review, but is not the Version of Record. The Version of Record of this article is published in Communications in Mathematical Physics, and is available online at \newline \url{https://doi.org/10.1007/s00220-026-05608-2}.

\paragraph{Data availability.}

No datasets were generated or analysed during the study, so data sharing is not applicable.

\paragraph{Conflict of interest.}

The authors have no conflict of interest to declare.

\section{Eigenfunctions and matrix models}\label{sec:sec2}

\subsection{The spectral problem}\label{sec:prep1}

The mirror curve to local $\IF_0$ reads \cite{kz,kkv}
\be  \label{eq:MC} \re^y + \re^{-y}+ \re^{2 \xi} \left(\re^x + \re^{-x} \right) + {\kappa} = 0 \, , \ee
where $\kappa$, $\xi$ are the complex structure moduli of local $\IF_0$. The eigenvalue equation corresponding to the quantization of the mirror curve \eqref{eq:MC} is \cite{acdkv,adkmv}
\begin{equation}
\label{eq:QuantizedMirrorCurve}
    \left(\re^{\hat y} +\re^{-\hat y } + \re^{2 \xi} \left(\re^{\hat x}+\re^{-\hat x}\right)\right)\psi(x, \kappa) + \kappa \, \psi(x, \kappa)  = 0
    \, ,
    \qquad \qquad
    [\hat x, \hat y]=\ri \hbar
    \, ,
\end{equation}
leading to the following difference equation
\begin{equation}
\label{eq:DifferenceEquation}
    \psi(x+\ri \hbar, \kappa) + \psi(x-\ri \hbar, \kappa) + 2 \re^{2 \xi} \cosh\left(x \right) \psi(x, \kappa) + \kappa \, \psi(x, \kappa) = 0 \, ,
\end{equation}
which also corresponds to the Baxter equation for the two-particle, relativistic, quantum Toda lattice \cite{reltoR}. In this paper, we always take
\begin{equation}
    \xi \in \reals
    \, ,
    \qquad \qquad
    \hbar\in \preals
    \, .
\end{equation}

Let us now look at the domain of the operators involved, acting as self-adjoint operators on the Hilbert space of square-integrable functions $L^2\br{\reals}$. The domain of the multiplication operator $\br{\re^{\hat{x}} + \re^{- \hat{x}}}$ contains then all functions $\psi \in L^2\br{\reals}$ for which $\re^{\pm x} \psi(x) \in L^2\br{\mathbb{R}}$, and the difference operator $\br{\re^{\hat{y}} + \re^{- \hat{y}}}$ acts similarly on the functions $\psi \in L^2\br{\reals}$ for which $\re^{\pm y} \widehat{\psi}(y) \in L^2\br{\mathbb{R}}$, with $\widehat{\psi}$ the Fourier transform of $\psi$. This condition on the Fourier transform $\widehat{\psi}$ is equivalent to the statement that $\psi$ admits an analytic continuation in the strip
\begin{equation} \label{qctopco}
    \cbr{x \in \mathbb{C} \mid \abs{\imaginary{x}} < \hbar}
    \, ,
\end{equation}
such that \( \psi(x) \) is square-integrable for all constant $\abs{\imaginary{x}} < \hbar$. In addition, one also requires that the limits
\begin{equation}
    \lim_{\epsilon\to 0^+} \psi(x - \ri\hbar + \ri \epsilon,\kappa)
    \, ,
    \qquad \qquad
    \lim_{\epsilon\to 0^+} \psi(x + \ri\hbar - \ri \epsilon,\kappa)
    \, ,
\end{equation}
exist in the sense of convergence in $L^2(\mathbb{R})$. The quantized mirror curve \eqref{eq:QuantizedMirrorCurve} is then a symmetric, strictly positive operator, defined on the intersection of the domains of the multiplication and difference operators. Hence, one can define its Friedrichs extension, see \cite{Laptev:2015loa}. It is this self-adjoint extension we will consider in everything that follows, and we refer to it as the quantum mirror curve.
This leads to a purely discrete spectrum $\{E_n\}_{n \in \naturals}$ for the quantum mirror curve, see \cite{ghm, Laptev:2015loa,Kashaev:2015kha}. Our conventions for the spectrum are
\be
\label{eq:spectralc}
\left(\re^{\hat y} +\re^{-\hat y } + \re^{2 \xi} \left(\re^{\hat x}+\re^{-\hat x}\right)\right)\psi(x, -\re^{E_n})=\re^{E_n}\psi(x, -\re^{E_n}) \, .
\ee
We refer to the variables $x,y$, and the corresponding operators $\hat x, \hat y $ in \eqref{eq:MC} as outer topological string coordinates for reasons that will become clear later.

Following \cite{Kashaev:2015kha, Kashaev:2015wia} we introduce the matrix model coordinates $q, p$,
\begin{equation}
\label{xytopq}
    \begin{split}
        x & = \frac{1}{\sqrt{2}} \left( q + p \right) + \xi \, ,
        \\
        y & =  \frac{1}{\sqrt{2}} \left( - q + p \right) + \xi \, .
    \end{split}
\end{equation}
After some creative algebra, one can show that the eigenvalue equation in the $q, p$ coordinates can be written as \cite[sec.~2.1]{Kashaev:2015wia}
\be
\label{mmc}
{\rm O} \, \Xi(q, \kappa)=-\kappa \, \Xi(q, \kappa) \, , \ee
\begin{equation}
    {\rm O} = \frac{\sqrt{2}}{\hbar}({{\tt f}^*(\hat{q})})^{-1} \, \cosh\left(\frac{\hat{p}}{\sqrt{2}}\right) \, ({\tt f}(\hat{q}))^{-1}
    \, ,
    \qquad \qquad
    \left[\hat{q},\hat{p}\right] = \ri \hbar
    \, ,
\end{equation}
where we used
\begin{equation}
\label{sqrtv}
    \begin{aligned}
        {\tt f}({q})
        & =
        \frac{2^{1/4}}{\sqrt{2 \pi b^2}} \exp\left( - \frac{\xi}{2} \right)
        \exp\left( \frac{q}{2 \sqrt{2}} \right) \frac{\Phi_{b}\left( \frac{q}{\sqrt{2} \pi b} - \frac{\xi}{\pi b} + \ri \frac{b}{4} \right)}{\Phi_{b}\left( \frac{q}{\sqrt{2} \pi b} + \frac{\xi}{\pi b} - \ri \frac{b}{4} \right)}
        \, ,
        \qquad \quad
        \hbar = \pi b^2
        \, ,
        \\
        {{\tt f}^*({q})}
        & =
        \frac{2^{1/4}}{\sqrt{2 \pi b^2}} \exp\left( - \frac{\xi}{2} \right)
        \exp\left( \frac{q}{2 \sqrt{2}} \right) \frac{\Phi_{b}\left( \frac{q}{\sqrt{2} \pi b} + \frac{\xi}{\pi b} + \ri \frac{b}{4} \right)}{\Phi_{b}\left( \frac{q}{\sqrt{2} \pi b} - \frac{\xi}{\pi b} - \ri \frac{b}{4} \right)}
        \, ,
        \qquad \quad
        \hbar = \pi b^2
        \, ,
    \end{aligned}
\end{equation}
with $\Phi_b$ Faddeev's non-compact quantum dilogarithm, see \autoref{sec:NCQuantDiLog}. Note that ${\tt f}^*(q) = \overline{{\tt f}(q)}$ only if $q, \xi, b \in \IR$.
It is convenient to introduce the inverse operator  $\rho=  {\rm O}^{-1} $ whose integral kernel is
\begin{equation}
\label{eq:rhok}
    \rho(q_1,q_2)= \frac{{{\tt f}(q_1)}{{\tt f}^*(q_2)}}{2 \cosh\br{\frac{q_1-q_2}{\sqrt{2} b^2}}}
    \, .
\end{equation}
One important property is that $\rho$ is of trace class and its Fredholm determinant admits an expansion in terms of an $\mathrm{O}\br{2}$ matrix model. More precisely
\be \label{eq:det}\det(1+{\kappa}\rho)=\prod_{n = 0}^{\infty}(1+{\kappa}\re^{- E_n})=\sum_{N = 0}^{\infty} \kappa^N Z(N, \hbar)\ee
where $E_n$ is determined as discussed around \eqref{eq:spectralc} and
\begin{equation}
    Z(N, \hbar)=
    \frac{1}{N!}\sum_{s\in S_N} (-1)^{\sgn[s]} \int_{\reals}  \rd^N x\prod_{k = 1}^N \rho \br{x_k, x_{s(k)}} \, ,
\end{equation}
where $S_N$ is the permutation group of $N$ elements.
By applying the Cauchy identity, one can further write $ Z(N,\hbar)$ as \cite{Kashaev:2015wia}
\begin{equation}
    \label{eq:ZNmm}
    Z(N,\hbar) =
    \frac{1}{2^N N!} \int_{\mathbb{R}^N} \rd^N q \prod_{k = 1}^N {\tt v}\left( q_k \right) \prod_{\ell = k+1}^N \tanh^2\left( \frac{q_k - q_\ell}{\sqrt{2} b^2} \right)
    \, ,
\end{equation}
\begin{equation}
\label{eq:MatrixModelPotential}
    {\tt v}\left( q \right) = {\tt f}(q){\tt f}^*(q) =
    \frac{\re^{-\xi}}{\sqrt{2} \pi b^2} \exp\left(\frac{q}{\sqrt{2}}\right)
    \frac{\Phi_b \left( \frac{q}{\sqrt{2} \pi b} - \frac{\xi}{\pi b} + \ri \frac{b}{4} \right) \Phi_b \left( \frac{q}{\sqrt{2} \pi b} + \frac{\xi}{\pi b} + \ri \frac{b}{4} \right)}
    {\Phi_b \left( \frac{q}{\sqrt{2} \pi b} + \frac{\xi}{\pi b} - \ri \frac{b}{4} \right) \Phi_b \left( \frac{q}{\sqrt{2} \pi b} - \frac{\xi}{\pi b} - \ri \frac{b}{4} \right)} \, ,
\end{equation}
where again $\hbar = \pi b^2$.
It is important to stress that \eqref{eq:det} is entire in $\kappa$, in particular, the sum on the right-hand side has an infinite radius of convergence. This is a standard result in Fredholm theory and follows from the trace class property of $\rho$, see \cite{ghm, Laptev:2015loa, Kashaev:2015kha}.

\subsection{Integrating quasi-periodic functions}
\label{sec:lemma}

In the following sections, we frequently encounter integrals involving Faddeev's quantum dilogarithm $\Phi_b$. To compute these integrals, we will make extensive use of Lemma 2.1 from \cite{Garoufalidis:2014ifa}, which we briefly review for future reference. Let $f : \mathcal{U} \to \mathbb{C}$ be an analytic function with $\mathcal{U} \subseteq \mathbb{C}$ open, $\mathcal{C} \subseteq \mathcal{U}$ an oriented path, and $a \in \mathbb{C} \setminus \left\{ 0 \right\}$ a constant such that
\begin{enumerate}
    \item $\mathcal{U} = a +  \mathcal{U}$ ,
    \item $f(z) \left( f(z + a) - f(z) \right) \neq 0$ for all $z \in \mathcal{C}$,
    \item $f(z+a) f(z-a) = f^2(z)$ for all $z \in \mathcal{U}$,
\end{enumerate}
then the following equality holds
\begin{equation}\label{eq:lemma:GaroufalidisKashaevQuasiPeriodicityLemma}
    \int_\mathcal{C} f(z) \rd z = \left( \int_\mathcal{C} - \int_{a+\mathcal{C}} \right) \frac{f(z)}{1 - f(z + a) / f(z)} \rd z \, .
\end{equation}
If we can close the contour on the right-hand side, then the integral reduces to a sum over residues. This will be the case for the integrals of interest to us.

\subsection{The eigenfunctions in matrix models coordinates}

\subsubsection{The general construction}

Off-shell eigenfunctions in the matrix model coordinates $q, p$ were found in
\cite[sec.~2]{Marino:2016rsq}, following \cite{Tracy:1995ax}. Let us define
\begin{equation}\label{eq:kappaex}
    \Xi^{\pm}(q; \kappa) = E^{\pm}(q) \sum_{N=0}^{\infty} \left( \pm \kappa \right)^N \Psi_N(q) \, ,
    \qquad \qquad
    E^{\pm}(q) = \exp\left( \pm \frac{q}{\sqrt{2} b^2}\right) {\tt f}(q)
    \, ,
\end{equation}
with ${\tt f}$ given in \eqref{sqrtv} and $\Psi_N(q)$ is defined by the following unnormalized expectation value
\begin{equation}
    \label{eq:PsiNd}
    \Psi_N(q) =
    \frac{1}{2^N N!} \int_{\mathbb{R}^N} \rd^N q \, \prod_{k = 1}^N \tanh\left( \frac{q - q_k}{\sqrt{2} b^2} \right)
    \,
    {\tt v}\left( q_k \right) \prod_{\ell = k+1}^N \tanh^2\left( \frac{q_k - q_\ell}{\sqrt{2} b^2} \right)
    \, ,
\end{equation}
where ${\tt v}$ is defined in \eqref{eq:MatrixModelPotential}.
Note that also \eqref{eq:kappaex} is entire in $\kappa$, in parallel with \eqref{eq:det}. This follows again from the trace class property of $\rho$.

Using \eqref{eq:kappaex}, we can write the eigenvalue equation \eqref{mmc} as
\begin{equation}
\label{eq2:Psidiff}
    \Omega^{\pm}\left(q + \ri \frac{\pi b^2}{\sqrt{2}}, \kappa \right) + \Omega^{\pm}\left(q - \ri \frac{\pi b^2}{\sqrt{2}}, \kappa \right)
    = - \sqrt{2} \pi b^2 \kappa \,  {\tt v}(q) \Omega^{\pm}(q, \kappa)
    \, ,
\end{equation}
where
\be\label{phimmd}\Omega^{\pm}(q, \kappa)= \exp\left(\pm \frac{q}{\sqrt{2} b^2}\right)\sum_{N=0}^{+\infty} \Psi_N(q) \left(\pm \kappa \right)^N \, . \ee
At the level of the $\Psi_N$, equation \eqref{eq2:Psidiff} reads
\be \label{eq:diffPsiN} \Psi_N\left(q + \ri \frac{\pi b^2}{\sqrt{2}}\right)-\Psi_N\left(q - \ri \frac{\pi b^2}{\sqrt{2}}\right)
=
\ri {\sqrt{2} \pi b^2} {\tt v}(q) \Psi_{N-1}(q) \, .
\ee
Therefore, the spectral problem in the matrix model coordinates $q,p$ can be formulated as follows. We look for solutions of \eqref{eq2:Psidiff} which are analytic in the strip $\abs{\imaginary{q}} < \pi b^2 / \sqrt{2}$ and which belong to $L^2(\IR)$. In the off-shell eigenfunctions \eqref{phimmd}, the first requirement is already implemented because of the specific form of $\Psi_N(q)$ given in \eqref{eq:PsiNd}, as we will verify in the case $N=1$ below. As for the $L^2(\IR)$ requirement, we have
\begin{equation}
    \Omega^{\pm}(q, \kappa)  \simeq
    \begin{dcases}
        \det(1\pm\kappa\rho) \exp\br{\pm \frac{q}{\sqrt{2}b^2}} \quad & q \to + \infty
        \\
        \det(1\mp\kappa\rho) \exp\br{\pm \frac{q}{\sqrt{2}b^2}} \quad & q \to - \infty
    \end{dcases}
\end{equation}
leading to the quantization condition $ \det(1+\kappa\rho)=0$ as expected and in agreement with the discussion in \autoref{sec:prep1}.
Hence, we can regard $\Xi_{\pm}$ as an analogue of the Jost functions in one-dimensional scattering theory \cite{Tracy:1995ax,Marino:2016rsq}: the functions $\Xi_{\pm}(q;\kappa)$ become genuine, square-integrable eigenfunctions of \eqref{mmc} when $\kappa = -\exp(E_n) < 0$ is on-shell,
\begin{equation}
    \Xi_{+}\br{q; -\re^{E_n}} = (-1)^n \, \Xi_{-}\br{q;-\re^{E_n}} \in L^2\br{\reals} \, .
\end{equation}
Their canonical transformation then yields the eigenfunctions $\psi(x; -\re^{E_n})$ of \eqref{eq:spectralc} in the topological string $(x,y)$-coordinates, as discussed in \autoref{sec:eigts}.

\subsubsection{\texorpdfstring{The case $N = 1$ and $\hbar \in \pi \prationals$}{The case N = 1 and \pdfhbar~\pdfin~\pdfpi \pdfrationals}}

In this section, we make use of \autoref{sec:lemma} to compute $\Psi_1(q)$ explicitly, and to test its analytic properties.
We have from \eqref{eq:PsiNd}
\begin{equation}
\label{eq:Psi1Integral}
    \Psi_1(q) = \frac{1}{2} \int_{\mathbb{R}} \rd p \, \tanh\left( \frac{q - p}{\sqrt{2} b^2} \right)  {\tt v}(p)
    \, ,
    \qquad \qquad
    \hbar = \pi b^2
    \, ,
\end{equation}
where ${\tt v}$ is given in \eqref{eq:MatrixModelPotential}.
The function ${\tt v}$ inherits some quasi-periodicity from the quantum dilogarithms \eqref{eq:NCQDiLogQuasiPer}, namely
\begin{multline}
    \frac{ { \tt v}\left(p + \ri \sqrt{2} \pi b^2 k \right)}{ { \tt v}(p) }
    = \re^{\ri \pi b^2 k}
    \\
    \prod_{\ell = 0}^{\left| k \right| - 1}
    \left\{
    \frac{
    \left( 1 + \re^{s_k \ri \pi b^2 \left( 2 \ell + \frac{1}{2} \right)} \re^{\sqrt{2} p - 2 \xi }\right)
    \left( 1 + \re^{s_k \ri \pi b^2 \left( 2 \ell + \frac{1}{2} \right)} \re^{\sqrt{2} p + 2 \xi }\right)}
    {\left( 1 + \re^{s_k \ri \pi b^2} \re^{s_k\ri \pi b^2 \left( 2 \ell + \frac{1}{2} \right)} \re^{\sqrt{2} p - 2 \xi }\right)
    \left(1 + \re^{s_k\ri \pi b^2} \re^{s_k \ri \pi b^2 \left( 2 \ell + \frac{1}{2} \right)} \re^{\sqrt{2} p + 2 \xi }\right)}
    \right\}
\end{multline}
for $k \in \mathbb{Z}$ and $s_k = \sgn[k]$.
One can then see that the integrand of \eqref{eq:Psi1Integral} is quasi-periodic in the sense of the lemma in \autoref{sec:lemma} when $b^2 \in \mathbb{Q}_{>0}$: we have quasi-periodicity under shifts by $\ri \sqrt{2} \pi b^2 m = \ri \sqrt{2} \pi n$ when $b^2 = n / m$ with $n, m$ positive coprime integers.
However, the quasi-periodic shift is trivial when $n \in 2 \mathbb{N}_{>0}$, so we can only use the lemma for $b^2 = ( 2 n + 1) / m$. In that case, one gets
\begin{equation}
\label{eq:Psi1DeformedIntegrand}
    \Psi_1(q) = \frac{1}{2} \left( \int_{\mathbb{R} + \ri 0} - \int_{\mathbb{R} + \ri \sqrt{2} \pi \left( 2 n + 1 \right) + \ri 0} \right) \rd p
    \,
    \tanh\left( \frac{q - p}{\sqrt{2} b^2} \right)
    \frac{{\tt v}(p)}{1 - \frac{{\tt v}\left(p + \ri \sqrt{2} \pi \left( 2 n + 1 \right) \right)}{{\tt v}(p)}}
    \, ,
\end{equation}
\begin{equation}
    \frac{1}{1 - \frac{{\tt v}\left(p + \ri \sqrt{2} \pi \left( 2 n + 1 \right) \right)} {{\tt v}(p)}}
    =
    - \ri \frac{
    \left( 1 - \ri \br{-1}^{n+m} \re^{- m \left( \sqrt{2} p - 2 \xi \right)} \right)
    \left( 1 + \ri \br{-1}^{n+m} \re^{m \left( \sqrt{2} p + 2 \xi \right)} \right)}
    {\br{-1}^{n+m} 4 \re^{2 m \xi} \sinh \left( \sqrt{2} m p \right)}
    \, .
\end{equation}
The integrand has an essential singularity at complex infinity, but it doesn't contribute to the integral since $ {\tt v}(p)  \propto \exp\left( \mp p / \sqrt{2}\right)$ when $\real{p} \to \pm \infty$ with $\imaginary{p}$ constant. Hence, the integration over $p$ reduces to a sum over the residues of the integrand in \eqref{eq:Psi1DeformedIntegrand} with $\Im(p) \in \left] 0, \sqrt{2} \pi \left( 2 n + 1 \right) \right]$.

Note that ${\tt v}(p)$ has poles at
\begin{equation}
    {\tt v}(p):
    \qquad
    \frac{p}{\sqrt{2}} = s \xi \pm \ri \pi \left[ b^2 \left( k + \frac{1}{4} \right) + \left( \ell + \frac{1}{2} \right) \right]
    \qquad
    \text{poles}
    \qquad
    k , \ell \in \mathbb{N}
    \, ,
\end{equation}
where $s \in \left\{ \pm 1 \right\}$, and the upper sign is coming from the numerator and the lower sign from the denominator of ${\tt v}$. Note that all the poles inside the integration contour are simple, due to the observation made around equation \eqref{eq:FaddeevQuantumDilogOrderOfRationalPoles}. Likewise, one finds
\begin{equation}
\label{eq:Psi1:Zeros&PolesDenominator}
    \frac{1}{1 -  \frac{{\tt v}\left(p + \ri \sqrt{2} \pi \left( 2 n + 1 \right) \right)}{  {\tt v}(p)}} :
    \qquad
    \begin{cases}
        \frac{p}{\sqrt{2}} = \pm \xi + \ri \frac{\pi}{m} \left[ k + \br{-1}^{n+m} \frac{1}{4} \right]
        & \text{roots}
        \\
        \frac{p}{\sqrt{2}} = \ri \frac{\pi}{2} \frac{k}{m}
        & \text{poles}
    \end{cases}
    \qquad
    k \in \mathbb{Z}
    \, .
\end{equation}
One can check that all the poles of ${\tt v}$ with positive imaginary part coincide with roots of the denominator, and hence are not realized as poles of the integrand when they are inside the integration contour, where they are simple.
However, we do have $m$ poles from the hyperbolic tangent and $2 \left( 2 n + 1 \right) m$ poles from the denominator at
\begin{align}
    p
    & = q + \ri \frac{\pi}{\sqrt{2}} b^2 \left( 2 k + 1 \right) \, ,
    & k \in \left\{ 0 , \cdots , m - 1 \right\} \, ,
    \\
    p
    & = \ri \frac{\pi}{\sqrt{2}} \frac{\ell}{m} \, ,
    & \ell \in \left\{ 1 , \cdots , 2 \left( 2 n + 1 \right) m \right\} \, ,
\end{align}
respectively, which have residues
\begin{equation}
    - \sqrt{2} b^2 \, ,
    \qquad \qquad
    \text{and}
    \qquad \qquad
    \br{-1}^{n + m + \ell + 1} \, \ri \, \frac{\cosh\left( 2 m \xi \right)}{2 \sqrt{2} m} \, .
\end{equation}
This gives finally the following expression for $ \Psi_1$
\be\label{eq:phi1f}
\boxed{
    \Psi_1(q) =\Psi_1^{(1)}(q)+  \Psi_1^{(2)}(q)
}
\ee
\be
\label{eq:psi2r}
\boxed{\ba
    \Psi_1^{(1)}(q) & =-\ri \frac{\pi}{\sqrt{2}} b^2
    \left[ 1 + \ri \br{-1}^{n + m} \cosh \left(2 m \xi \right) \csch \left(\sqrt{2} m q\right)\right]
    \\
    &
    \qquad \qquad \qquad \qquad \qquad
    \qquad \qquad \qquad
    \sum_{k = 0}^{m - 1} {\tt v}\left(q + \ri \frac{\pi}{\sqrt{2}} b^2 \left(2 k + 1 \right) \right)
    \, .
    \\
    \Psi_1^{(2)} \br{q} & =  \br{-1}^{n + m + 1} \frac{\sqrt{2} \pi}{4 m} \cosh \left(2 m \xi \right)
    \sum_{\ell =-2 n}^{2 n + 1} \br{-1}^\ell \coth \left(\frac{q}{\sqrt{2} b^2} - \ri \frac{\pi}{2}  \frac{\ell}{2 n + 1} \right)
    \\
    &
    \qquad \qquad \qquad \qquad \qquad
    \qquad \qquad \qquad
    \sum_{k = 0}^{m-1}
    {\tt v}\left(\ri \frac{\pi}{\sqrt{2}} \left(\frac{\ell}{m} + b^2 (2 k + 1) \right) \right)
    \, .
\ea}\ee
and we remind the reader that we took $\hbar / \pi = b^2 = (2 n + 1) / m$ with $2n+1$ and $m$ coprime. Furthermore, it is noteworthy that these functions are real along the real line.

Let us look at the analytic properties of $\Psi_1$.
We can then make the following considerations.
\begin{enumerate}

    \item Let us consider $\Psi_1^{(1)}(q)$. The simple poles of ${\tt v}\left( q + \ri \pi b^2 \left(2 k + 1 \right) / \sqrt{2} \right)$ with $\abs{\imaginary{q}} \leqslant \pi b^2 / \sqrt{2}$ coincide with the simple roots of the factor in square brackets, and are hence not realized.

    \item There are true simple poles for $\Psi_1^{(1)}(q)$ originating from $\csch[\sqrt{2} m q]$ at
    \begin{equation}
    \label{eq:polesc}
        q = \ri \frac{\pi}{\sqrt{2}} \frac{r}{m}
        \qquad \qquad
        r \in \left\{ - \left( 2 n + 1 \right) , \cdots  , + \left( 2 n + 1 \right)  \right\} \, ,
    \end{equation}
    where the upper and lower bound on $r$ come from the requirement of being inside the strip $\abs{\imaginary{q}} \leqslant \pi b^2 /  \sqrt{2}$. The residue of $\Psi_1^{(1)}(q)$ at these poles is
    \begin{equation}
    \label{eq:polespsi(1)1}
        \br{-1}^{r+n+m} \frac{\pi}{2} \frac{b^2}{m} \cosh(2 m \xi) \sum_{k = 0}^{m - 1} {\tt v}\left( \ri \frac{\pi}{\sqrt{2}} \left( \frac{r}{m} + b^2 \left( 2 k + 1 \right) \right) \right) \, .
    \end{equation}

    \item One can see that also $\Psi_1^{(2)} (q)$ has simple poles at \eqref{eq:polesc}, originating from the term $\ell=r$ in \eqref{eq:psi2r}. Moreover, $\Psi_1^{(2)}  (q)$ has the same residue \eqref{eq:polespsi(1)1} with the opposite overall sign.

\end{enumerate}
Hence, we can conclude that $\Psi_1 (q )$ is analytic on the strip $- \pi b^2 / \sqrt{2} \leqslant \Im(q) \leqslant \pi b^2 / \sqrt{2}$ as expected. Note however that outside the strip there are higher order poles in $\Psi_1^{(1)}$, coming from ${\tt v}$, which do not get cancelled by the simple roots of the factor in square brackets or by any poles coming from the periodic part. Hence, $\Psi_1$ is analytic on the strip, but not entire.
One can also check that our solution \eqref{eq:phi1f} satisfies
\begin{equation}
    \Psi_1 \left( q + \ri \frac{\pi b^2}{\sqrt{2}}\right) - \Psi_1 \left( q - \ri \frac{\pi b^2}{\sqrt{2}} \right) = \ri \sqrt{2} \pi b^2 { \tt v}\left( q  \right)
    \, ,
\end{equation}
which is the difference equation we expect for $\Psi_1$ from \eqref{eq:diffPsiN}.

Since $\Psi_1(q)$ reduces to the first spectral trace $Z(1,\hbar)$ in the $q \to + \infty$ limit, one gets from \eqref{eq:phi1f}
\begin{equation}\label{eq:z1}
    \boxed{
    Z\left(1, \br{\frac{2 n + 1}{m}} \pi\right) = \br{-1}^{n+m} \frac{\sqrt{2} \pi}{4 m} \cosh \left( 2 m \xi \right) \sum_{\ell = 1}^{2 (2 n + 1) m} \br{-1}^\ell {\tt v}\left( \ri \frac{\pi}{\sqrt{2}} \frac{\ell}{m} \right)
    \, ,
    }
\end{equation}
where ${\tt v}$ is defined in \eqref{eq:MatrixModelPotential} and $2n+1$, $m$ are coprime. This is also the result one gets when applying the residue technique above directly to the integral defining $Z(1,\hbar)$ in \eqref{eq:ZNmm}.
The expression \eqref{eq:z1} is, in some sense, complementary to that of \cite[eq.~(3.55)]{Gu:2021ize}, which holds for ${\rm Im} ( b^2 ) > 0$.

\subsection{The eigenfunctions in outer topological string coordinates}\label{sec:eigts}

\subsubsection{The general construction and symmetric structures}

One motivation for considering the outer topological string $(x, y)$-coordinates is that they establish a direct connection with the open topological string in the presence of a D-brane on the external leg of the toric diagram, as we will discuss in \autoref{sec:tsst}. Consequently, in these coordinates, we obtain a particularly explicit framework for computing these eigenfunctions using topological string partition functions.\footnote{The matrix model coordinates seem more natural to describe a brane on the internal leg of the toric diagram \cite{Aganagic:2000gs,akv}.}

The eigenfunctions $\psi(x,\kappa)$ in the topological string $(x,y)$-coordinates and the eigenfunctions $\Xi(q,\kappa)$ in the matrix model $(q,p)$-coordinates are then related by a canonical transformation. More precisely,
\begin{equation}
\label{eq:canonical}
    \psi^{\pm}(x,\kappa) =  \int_{\mathbb{R}} \rd q \, U\left( x, q \right) \Xi^{\pm}(q,\kappa) \, ,
\end{equation}
where from \cite{Marino:2016rsq}
\begin{equation}\label{eqused}
    U\left( x, q \right) =
    \frac{2^{1/4}}{\sqrt{2 \pi \hbar}}
    \exp \left( \frac{\ri}{\hbar} \left( \frac{x^2}{2} - \sqrt{2} \br{x - \xi} q + \frac{q^2}{2} \right) \right) \, .
\end{equation}
Hence, if we take the eigenfunctions \eqref{eq:kappaex}, the corresponding eigenfunctions in topological string coordinates are
\begin{equation}
\label{eq:def:CanTransEigenFunQPtoXY}
    \psi^\pm(x, \kappa)=\sum_{N\geqslant0}(\pm \kappa)^N \psi_{N}^\pm(x) \, ,
    \qquad \qquad
    \psi_{N}^\pm(x) =  \int_{\mathbb{R}} \rd q \, U\left( x, q \right) E^{\pm}(q) \Psi_N(q) \, .
\end{equation}
Note that this integral is only well-defined when $E^{\pm}(q) \Psi_N(q)$ is integrable. However, we have
\begin{equation}
\label{eq:AsymptoticsInQP}
    \begin{split}
        E^{+}(q) \Psi_N(q) & \simeq
        \begin{dcases}
            \exp\left[ \left( \frac{b^2 - 2}{2 \sqrt{2} b^2} \right) (- q) \right] \exp\left( - \ri \frac{2 \sqrt{2}}{\pi} \frac{\xi q}{b^2} \right) \quad & q \to + \infty
            \\
             \exp\left[ \left( \frac{b^2 + 2}{2 \sqrt{2} b^2} \right) q \right] \quad & q \to - \infty
        \end{dcases}
        \quad ,
        \\
        E^{-}(q) \Psi_N(q) & \simeq
        \begin{dcases}
            \exp\left[ \left( \frac{b^2 + 2}{2 \sqrt{2} b^2} \right) (-q) \right] \exp\left( - \ri \frac{2 \sqrt{2}}{\pi} \frac{\xi q}{b^2} \right) \quad & q \to + \infty
            \\
            \exp\left[ \left( \frac{b^2 - 2}{2 \sqrt{2} b^2} \right) q \right] \quad & q \to - \infty
        \end{dcases}
        \quad .
    \end{split}
\end{equation}
Hence, the unitary transformation \eqref{eq:canonical} is well-defined only when $b^2>2$, as also noted in \cite{Marino:2016rsq}.
This raises the question of how to make sense of the canonical transformation \eqref{eq:def:CanTransEigenFunQPtoXY} when $\hbar = b^2 \pi \leqslant 2 \pi$.\footnote{Curiously, if it were not for the $q^2$ term in the canonical transformation, the $N = 0$ version of this transform would be very close to the ``integral analogue of the $_1\psi_1$-summation formula of Ramanujan'' as given in \cite[eq.~(51)]{EllegaardAndersen:2011vps}, which can be explicitly computed in closed form using the method of residues. } Our strategy to address this issue is the following.
In this region, we will momentarily set aside convergence issues and directly apply the Lemma from \autoref{sec:lemma}.
This approach gives a finite result for any value of $b^2$, even if the starting point was problematic for $ b^2  \leqslant 2 $.
The case $\xi=0$, $b^2=2$ was analysed in \cite[p.~15]{Marino:2016rsq} using the same strategy.

Let us first make some simple observations regarding the relation between $\psi^+$ and $\psi^-$.
Using the parity properties of the quantum dilogarithm \eqref{eq:NCQuantDiLogParity} and the expression \eqref{eq:PsiNd}, we have
\begin{equation}\label{eq:shiftprp}
    \begin{split}
        E^{\pm}(-q) \Psi_N(-q) & = \exp \left( \mp \frac{\sqrt{2} \pi}{\hbar} q \right) \exp \left( \frac{\ri}{\hbar} 2 \sqrt{2} \xi q \right) E^{\pm}(q)\br{-1}^N \Psi_N(q)
        \\
        & = \exp \left( \frac{\ri}{\hbar} 2 \sqrt{2} \xi q \right) E^{\mp}(q) \br{-1}^N \Psi_N(q) \, ,
        \\
        \psi_{N}^\pm(-x) & = \exp\left(\frac{\ri}{\hbar} \frac{\pi^2}{2}\right) \exp\left( \mp \frac{\pi x}{\hbar} \right) \br{-1}^N  \psi_{N}^\pm(x \mp \ri \pi ) \, ,
        \\
        & = \br{-1}^N \psi_{N}^\mp(x) \, .
    \end{split}
\end{equation}
The eigenfunctions behave then as
\begin{equation}\label{eq:parity}
    \begin{split}
        \Xi^{\pm}(-q, \kappa) & = \exp \left( \mp \frac{\sqrt{2} \pi q}{\hbar} \right) \exp \left( \ri \frac{2 \sqrt{2}}{\hbar} \xi q \right) \Xi^{\pm}(q, - \kappa)
        \\
        & = \exp \left( \ri \frac{2 \sqrt{2}}{\hbar} \xi q \right) \Xi^{\mp}(q, \kappa)
        \\
        \psi^{\pm}(-x, \kappa) & = \exp\left(\frac{\ri}{\hbar} \frac{\pi^2}{2}\right) \exp\left( \mp \frac{\pi x}{\hbar} \right) \psi^{\pm}(x \mp \ri \pi, - \kappa)
        \\
        & = \psi^{\mp}(x, \kappa) \, ,
    \end{split}
\end{equation}
and we see in particular that the on-shell eigenfunctions in the $(x,y)$-coordinates have a well-defined parity while there is a local phase for generic $\xi$ in the $(q,p)$-coordinates. Because of the simple relation between the two eigenfunctions, we will often work with
\begin{equation}
\label{eq:conventionsub}
    \psi_{N}(x)\equiv  \psi_{N}^+(x) \, ,
    \qquad \qquad
    \psi(x) \equiv \psi^+(x) \, ,
\end{equation}
and $\psi_{N}^-(x)$ and $\psi^-(x)$ can then simply be found from \eqref{eq:shiftprp} and \eqref{eq:parity} respectively.

We are now going to study \eqref{eq:canonical} and \eqref{eq:def:CanTransEigenFunQPtoXY} in detail.
First, it was conjectured in \cite{Marino:2016rsq, Marino:2017gyg} that this integral can be written as the sum of two contributions, which are in a one-to-one correspondence with the two saddles of the integrand on the right-hand side of \eqref{eq:def:CanTransEigenFunQPtoXY}.
Second, in \cite{Francois:2023trm}, a special scaling limit of \eqref{eq:def:CanTransEigenFunQPtoXY} was analysed in detail, and it was found that the two saddles are related in a simple way. By combining these two observations, we can propose the following ansatz for the expression of the eigenfunctions in the topological string coordinates:
\begin{equation}
    \label{eq:intermo}
    \psi^{ \pm}(x, \kappa)
    =
    \omega^{ \pm }(x, \kappa)
    +
    \exp\left(\frac{\ri}{\hbar} \frac{\pi^2}{2}\pm\frac{\pi x}{\hbar}\right) \omega^{ \pm }( - x \mp \ri \pi , - \kappa)
    \, ,
\end{equation}
for some functions $\omega^{\pm}(x,\kappa)$.
Our result reads then
\begin{equation}
\label{eq:ProposedStructureEigenFun}
\boxed{
    \psi(x, \kappa)
    =
    \omega(x, \kappa)
    +
    \exp\left(\frac{\ri}{\hbar} \frac{\pi^2}{2}+\frac{\pi x}{\hbar}\right) \omega( - x  -  \ri \pi , - \kappa)
        \, ,
    }
\end{equation}
for some function $\omega(x,\kappa)$. It is important to note that the parity relation for $\psi^{\pm}(x,\kappa)$ in \eqref{eq:parity} is the same as the one relating the two terms in \eqref{eq:ProposedStructureEigenFun}, ensuring the self-consistency of our proposal \eqref{eq:ProposedStructureEigenFun}.
At the level of  components in the $\kappa$ expansion \eqref{eq:kappaex}, equation \eqref{eq:intermo} becomes
\be \label{eq:intermocomp} \psi_{N}^\pm(x)=  \omega_{N}^\pm(x) + \br{-1}^N \exp\left(\frac{\ri}{\hbar} \frac{\pi^2}{2}\pm\frac{\pi x}{\hbar}\right) \omega_{N}^\pm(-x \mp\ri \pi)\, , \ee
for some functions $\omega_{N}^{\pm}(x)$ where $\omega_{N}^{-}(x) = (-1)^N \omega_{N}^{+}(-x)$. Equation \eqref{eq:intermocomp} reads equivalently
\be\label{eq:ProposedStructureEigenFun2}  \boxed{  \psi_{N}(x)=  \omega_{N}(x)+ \br{-1}^N \exp\left(\frac{\ri}{\hbar} \frac{\pi^2}{2}+\frac{\pi x}{\hbar}\right) \omega_{N}(-x -\ri \pi)} \, , \ee
for some function $\omega_{N}(x)$.

We will test this proposal in several ways, and we also give explicit expressions for $\omega$ by using topological string theory, see \autoref{sec:opents}. As we discuss in \autoref{app:mass1}, the case $\xi=0$, $\hbar=2\pi$ analysed in \cite{Marino:2016rsq, Marino:2017gyg} is special and the above structure is hidden.

\subsubsection{\texorpdfstring{The case $N = 0$ and $\hbar \in \pi \prationals$}{The case N = 0 and \pdfhbar~\pdfin~\pdfpi \pdfrationals}}

In this subsection, we compute $\psi_{N} = \psi_{N}^+$ for $N = 0$ and $\hbar \in \pi \prationals$,\footnote{The result for $\psi_{N}^-$ follows immediately from \eqref{eq:shiftprp}. Hence, we adopt the notation \eqref{eq:conventionsub}.} which serves two goals. Firstly, it will provide some evidence that we can expect the off-shell eigenfunctions of the quantum mirror curve to be entire in $x$,\footnote{The observation that the eigenfunctions are entire was also made in \cite{Marino:2016rsq}, based on a computation of $\psi_N$ for several $N \in \naturals$ for $\xi = 0$ and $\hbar = 2 \pi$ \cite[eqs.~(2.95),~(2.96)]{Marino:2016rsq}. See also \cite[eqs.~(4.20),~(4.21)]{Marino:2017gyg} for the case $\xi = 0$, $\hbar = 4 \pi$ and $\hbar = 2 \pi / 3$.} and secondly, we will use it as an analytical check of our conjecture \eqref{eq:ProposedStructureEigenFun2}.
We are interested in
\begin{equation}
\label{eq:def:psi0x}
    \psi_0\br{x}
    =
    \int_\reals \rd q \, U\br{x, q} E{\br{q}}
    \, ,
    \qquad \qquad
    E{\br{q}} = E^+{\br{q}}
    \, .
\end{equation}
It is important to note that the integrand in \eqref{eq:def:psi0x} is a meromorphic function of $q$, which inherits some quasi-periodicity from the quantum dilogarithms
\eqref{eq:NCQDiLogQuasiPerRepeated},
\begin{multline}
    \frac{U{\br{x, q + \ri \sqrt{2} \pi k b^2}} E{\br{q  + \ri \sqrt{2} \pi k b^2}}}{U{\br{x, q}} E{\br{q}}}
    =
    \re^{- \ri \pi  k^2 b^2} \re^{-2 k \xi} \re^{2 k x} \re^{-\sqrt{2} k q}
    \re^{ \ri \pi  k} \re^{\ri \frac{\pi}{2}  k b^2}
    \\
    \prod_{\ell = 0}^{\abs{k} - 1}
    \br{\frac{1 + \re^{\ri \frac{\pi}{2} \sgn[k] b^2 \br{4 \ell + 1}} \re^{\sgn[k] 2 \xi} \re^{\sqrt{2} q}}
    {1 + \re^{\ri \frac{\pi}{2} \sgn[k] b^2 \br{4 \ell + 3}} \re^{- \sgn[k] 2 \xi} \re^{\sqrt{2} q}}}
    \, ,
\end{multline}
where $\hbar = \pi b^2$ and $k \in \integers$.
When we take
\begin{equation}
    b^2 = \frac{n}{m} \in \prationals \, ,
    \qquad \qquad
    n, m \in \pnaturals \text{ and coprime} \, ,
    \qquad \qquad
    \text{and } k = \pm m \, ,
\end{equation}
then this simplifies further to
\begin{multline}
    \frac{U{\br{x, q \pm \ri \sqrt{2} \pi n}} E{\br{q \pm \ri \sqrt{2} \pi n}}}{U{\br{x,q}} E{\br{q}}}
    =
    \\
    \br{-1}^{\br{n+1} m} \re^{\pm \ri \frac{\pi}{2} n} \re^{\mp 2 m \xi} \re^{\pm 2 m x} \re^{\mp \sqrt{2} m q}
    \br{\frac{1 - \br{-1}^m \re^{\ri \frac{\pi}{2} n} \re^{2 m \xi} \re^{\sqrt{2} m q}}{1 - \br{-1}^m \re^{- \ri \frac{\pi}{2} n} \re^{- 2 m \xi} \re^{\sqrt{2} m q}}}^\pm
    \, .
\end{multline}
Hence we find that the integrand defining $\psi_0$ in \eqref{eq:def:psi0x} is quasi-periodic in the sense of the quasi-periodic integrand lemma of \autoref{sec:lemma}. Following the lemma, we can rewrite \eqref{eq:def:psi0x} as
\begin{multline}
\label{eq:psi0x_DeformedIntegral}
    \psi_0{\br{x}}
    =
    \br{\int_{\reals + \ri 0} - \int_{\reals + \ri \sqrt{2} \pi n + \ri 0}}
    \rd q
    \\
    \frac{U{\br{x,q}} E{\br{q}}}
    {1 - \br{-1}^{\br{n+1} m} \re^{\ri \frac{\pi}{2} n} \re^{-2 m \xi} \re^{2 m x} \re^{- \sqrt{2} m q}
    \br{\frac{1 - \br{-1}^m \re^{\ri \frac{\pi}{2} n} \re^{2 m \xi} \re^{\sqrt{2} m q}}{1 -  \br{-1}^m \re^{- \ri \frac{\pi}{2} n} \re^{- 2 m \xi} \re^{\sqrt{2} m q}}}}
    \, .
\end{multline}
The integrand above is a meromorphic function of $q$ with a finite number of poles inside the integration contour and an essential singularity at infinity. Note furthermore from \eqref{eq:AsymptoticsInQP} that the integrand decays exponentially inside the whole contour for $\real{q} \to \pm \infty$. Hence, we can close the contour at complex infinity and the integral reduces to a sum over the residues of the poles.

\begin{figure}
    \centering
    \includegraphics[width=0.49\linewidth]{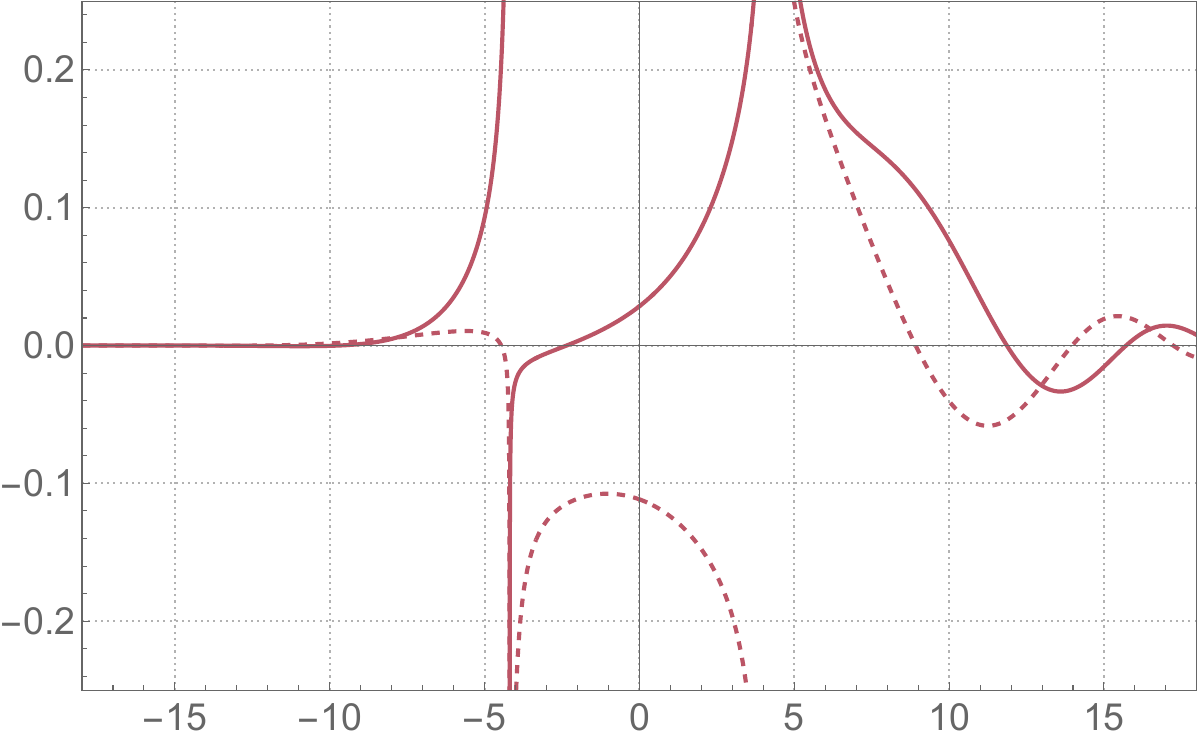}
    \includegraphics[width=0.49\linewidth]{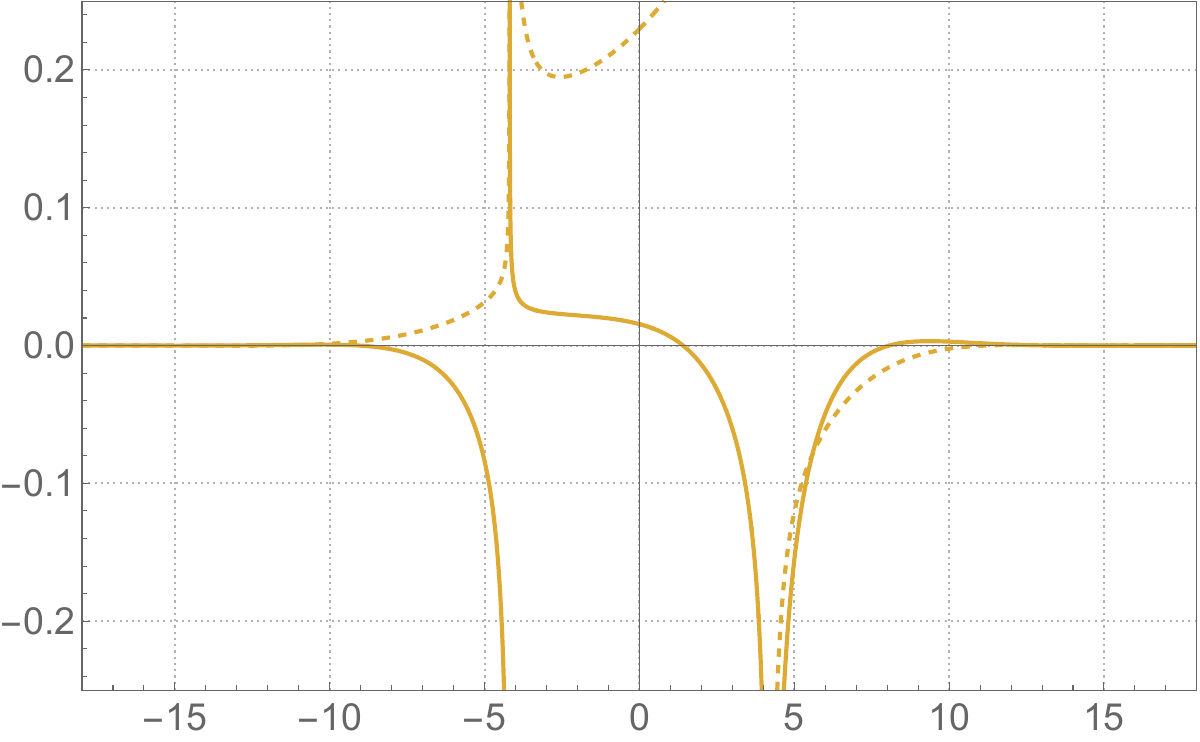}
    \includegraphics[width=0.49\linewidth]{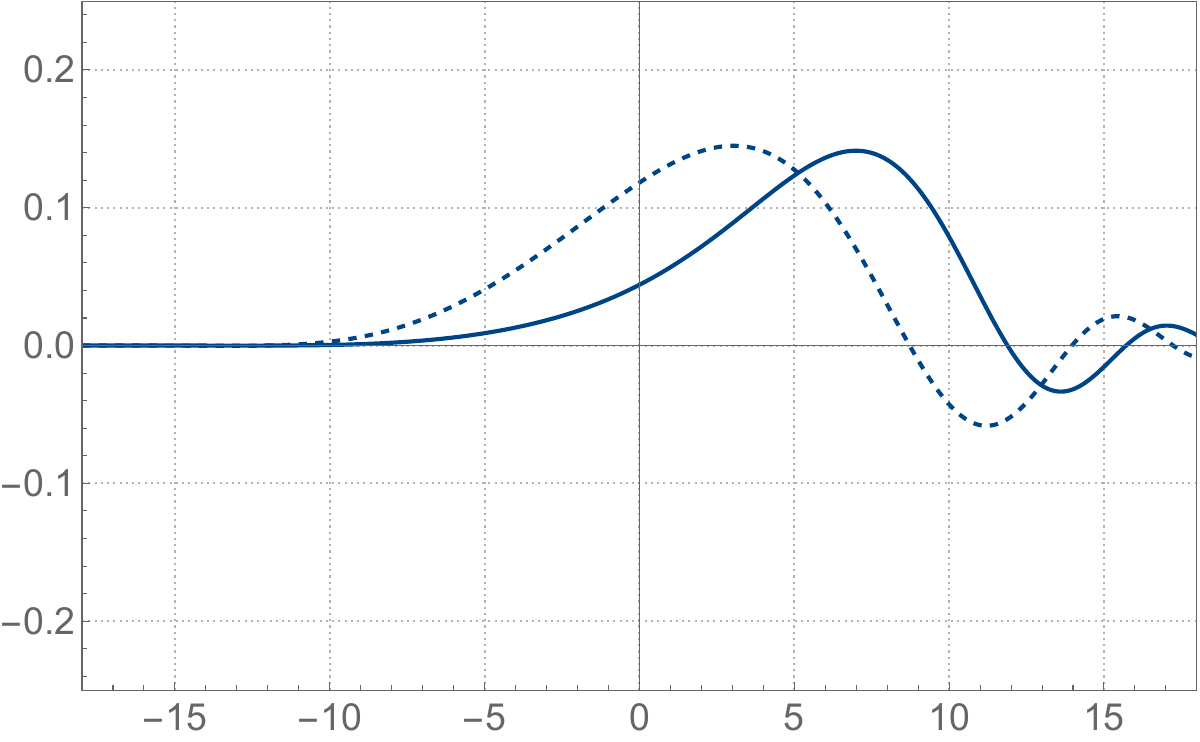}
    \caption{From top left to bottom: the first term $\omega_{0,+}$ on the right-hand side of \eqref{eq:Psi0x_RationalB_Result}, the second term $\omega_{0,-}$, and their sum $\psi_{0}$ for $\xi = - 7 / 4$ and $\hbar = 4 \pi$. The solid and dashed lines correspond to the real and imaginary parts, respectively.}
    \label{fig:b2eq4}
\end{figure}

To simplify the discussion, we will assume that either $\xi \neq 0$ or $n \in \br{2 \naturals + 1}$.\footnote{\label{footnote:TrivialMass&EvenN}The derivation for $\xi = 0$ and $n \in 2 \pnaturals$ is done analogously, yielding the same result as obtained by taking the $\xi \to 0$ limit for $n \in 2 \pnaturals$ on the solution \eqref{eq:Psi0x_RationalB_Result}. This case agrees with the results of \cite{Marino:2016rsq, Marino:2017gyg}, see also \autoref{app:mass1}.}
There are potential poles of the integrand coming from $E$ at \eqref{eq:poles&rootsNCQDiLog}
\begin{equation}
\label{eq:PolesEsq}
    q = \pm \sqrt{2} \xi \pm \ri \sqrt{2} \pi \br{\br{k + \frac{1}{2}} + \frac{n}{m} \br{\ell + \frac{1}{4}}} \, ,
    \qquad \qquad
    k , \ell \in \naturals \, ,
\end{equation}
and all the poles inside the integration contour in \eqref{eq:psi0x_DeformedIntegral} are simple as a direct consequence of the observation made around \eqref{eq:FaddeevQuantumDilogOrderOfRationalPoles}. However, for each such pole of $E$ there is a coinciding simple pole for the denominator in \eqref{eq:psi0x_DeformedIntegral}, and hence the integrand in \eqref{eq:psi0x_DeformedIntegral} is analytic around these points. The only poles inside the integration contour are hence coming from the roots of the denominator and are located at
\begin{multline}
\label{eq:DeformedIntegralPsi0x_Poles}
    \sqrt{2} m q_{\pm, k}\br{x}
    =
    \ln \Biggl[\br{-1}^m \re^{\ri \frac{\pi}{2} n} \br{\frac{\re^{2 m \xi}}{2}} \re^{m x}
    \Biggl( \br{-1}^{n\br{m+1}} \re^{m x} + \re^{- m x}
    \\
    \pm \sgn[\arg[x + \ri \frac{\pi}{2}]] \sqrt{\br{\br{-1}^{n\br{m+1}} \re^{m x} + \re^{- m x} }^2 - \br{-1}^{n m} 4 \re^{- 4 m \xi}} \Biggr) \Biggr]
    + \ri 2 \pi k
    \, ,
\end{multline}
where $k \in \integers$ should be such that $0 < \imaginary{q_{\pm, k}\br{x}} \leqslant \sqrt{2} \pi n$ and we use the convention $\sgn[0] = -1$. It is important for later to note that these points are by construction a solution to
\begin{equation}
\label{eq:QuasiPeriodicityQs}
    \begin{split}
        U\br{x, q_{\pm, k}\br{x}} E\br{q_{\pm, k}\br{x} }
        & = U\br{x,  q_{\pm, k}\br{x} + \ri \sqrt{2} \pi n} E\br{ q_{\pm, k}\br{x} + \ri \sqrt{2} \pi n}
        \\
        & = U\br{x,  q_{\pm, k + n m}\br{x}} E\br{ q_{\pm, k + n m}\br{x}}
        \, .
    \end{split}
\end{equation}
The residues corresponding with \eqref{eq:DeformedIntegralPsi0x_Poles} and coming from the denominator are given by
\begin{multline}
\label{eq:DeformedIntegralPsi0x_Residues}
    \mathrm{Res}_\pm{\br{x}}
    =
    \frac{1}{2 \sqrt{2} m}
    \\
    \pm \frac{\sgn[\arg[x + \ri \frac{\pi}{2}]]}{2 \sqrt{2} m} \br{\frac{\br{-1}^{n\br{m+1}} \re^{m x} - \re^{- m x} }{\sqrt{\br{\br{-1}^{n\br{m+1}} \re^{m x} + \re^{- m x} }^2 - \br{-1}^{n m} 4 \re^{- 4 m \xi}}}} \, .
\end{multline}
It should be noted that the particular choice of branches in \eqref{eq:DeformedIntegralPsi0x_Poles} and \eqref{eq:DeformedIntegralPsi0x_Residues} is to some extent purely conventional. However, we will relate $q_+$ and $q_-$ in \eqref{eq:ReltionBetweenTheTerms_Qpm&Respm}, and for this purpose, it is important to use the particular choice made in \eqref{eq:DeformedIntegralPsi0x_Poles} and \eqref{eq:DeformedIntegralPsi0x_Residues}, or something equivalent.
In the end, we find that \eqref{eq:def:psi0x} is given by\footnote{The subscript $\pm$ in $\omega_{0, \pm}$ is logically independent of the superscript $\pm$ in $\omega^\pm$ that appeared in \eqref{eq:intermo}.}
\begin{equation}
\label{eq:Psi0x_RationalB_Result}
    \psi_0\br{x} = \omega_{0,+}\br{x} +  \omega_{0,-}\br{x} \, ,
\end{equation}
\begin{equation}
    \qquad
    \omega_{0,\pm}\br{x} =  \ri 2 \pi \mathrm{Res}_\pm{\br{x}} \sum_{k = 0}^{n m - 1} U{\br{x, q_{\pm, k}{\br{x}}}} E{\br{q_{\pm, k}{\br{x}}}}
    \, ,
\end{equation}
where $\hbar = \pi b^2 = \pi n/m$ with $n,m \in \pnaturals$ and coprime. It is noteworthy that $\omega_{0,\pm}$ and hence $\psi_0$ can be expressed entirely in terms of elementary functions and the classical dilogarithm $\Li_2$ by using \eqref{eq:FQDiLogSD}. See \autoref{fig:b2eq4} for some plots of the functions defined above.

Let us look at the analytic properties of $\omega_{0,\pm}$ and $\psi_0$.
Note first that $\omega_{0,\pm}$ and $\psi_0$ are analytic along the logarithmic branch cut of $q_{\pm, k}$, since crossing the branch simply amounts to shifting the range of $k$ in the sum in \eqref{eq:Psi0x_RationalB_Result}, which doesn't affect $\omega_{0,\pm}$ or $\psi_0$ by \eqref{eq:QuasiPeriodicityQs}. Note furthermore that $\psi_0$ is analytic along the square root branches of $q_{\pm, k}$ and $\mathrm{Res}_\pm$ as well, since crossing the branches simply interchanges the $\pm$-signs.
The only remaining potential singularities are the branch points of the logarithm in $q_{\pm, k}$ and the square roots, and the poles of $E\br{q_{\pm, k}\br{x}}$. The logarithmic branch point of $q_{\pm, k}$ is never realized, and the square root branch points in $\mathrm{Res}_\pm\br{x}$ cancel between the $\omega_{0,\pm}$. Furthermore, the poles of $E$ are never reached by $q_{\pm, k}$ when either $\xi \neq 0$ or $n \in \br{2 \naturals + 1}$ as we assumed.\footnote{Though the $\xi \to 0$ limit for even $n$ behaves well, see \autoref{footnote:TrivialMass&EvenN}.}
Hence, we conclude that $\psi_0\br{x}$ is an entire function of $x$, even though  neither $\omega_{0,+}$ nor $\omega_{0,-}$ is entire, see \autoref{fig:b2eq4}.
The same structure will reappear when we express the eigenfunctions in terms of the grand potential of topological strings on local $\mathbb{F}_0$ in \eqref{tsstopenf}.

One can furthermore check that $\psi_0$ solves the difference equation associated with the quantized mirror curve at $\kappa = 0$, that is
\begin{equation}
\label{eq:DiffEq_Psi0x}
    \psi_0\br{x + \ri \hbar} + \psi_0\br{x - \ri \hbar} + 2 \re^{2 \xi} \cosh\br{x} \psi_0\br{x} = 0 \, .
\end{equation}
However, $\psi_0$ is not an eigenfunction of the quantized mirror curve, since it is not square integrable in the strip \eqref{qctopco}.

Let us end with the observation that \eqref{eq:Psi0x_RationalB_Result} is a well-defined, entire function of $x$ that solves \eqref{eq:DiffEq_Psi0x} for all $b^2 = n / m \in \prationals$, also $b^2 \leqslant 2$, even though the original integral transform as given in \eqref{eq:def:psi0x} is only well-defined for $b^2 > 2$.

\subsubsection{\texorpdfstring{Relating the two terms for $N = 0$ and $\hbar \in \pi \prationals$}{Relating the two terms for N = 0 and \pdfhbar~\pdfin~\pdfpi \pdfrationals}}

We now want to show that \eqref{eq:Psi0x_RationalB_Result} can be written as in \eqref{eq:ProposedStructureEigenFun2} for $N=0$.
For all $x \in \complexes$, $\xi \in \reals$, coprime $n, m \in \naturals_{>0}$, and $k \in \integers$,\footnote{The branches in \eqref{eq:DeformedIntegralPsi0x_Poles} have to be chosen differently when $x = - \ri \pi / 2$.} one finds the following relation for \eqref{eq:DeformedIntegralPsi0x_Poles} and \eqref{eq:DeformedIntegralPsi0x_Residues}
\begin{equation}
\label{eq:ReltionBetweenTheTerms_Qpm&Respm}
    \begin{split}
        q_{\pm,k}{\br{x}} = - q_{\mp, - k - \ell}{\br{- x - \ri \pi}} \, ,
        \qquad \qquad
        \mathrm{Res}_\pm\br{x} = \mathrm{Res}_\mp\br{- x -  \ri \pi} \, ,
    \end{split}
\end{equation}
where $\ell \in \cbr{-1, 0, +1}$ should be chosen appropriately according to the branches. It should be noted that this symmetry is only visible upon the specific choice of the branch structure we made in \eqref{eq:DeformedIntegralPsi0x_Poles} and \eqref{eq:DeformedIntegralPsi0x_Residues}. One can use the parity structure for $q$ in \eqref{eq:shiftprp} to rewrite
\begin{multline}
    U\br{x, q_{\pm, k}\br{x}} E\br{q_{\pm, k}\br{x}}
    =
    U\br{x, - q_{\mp, - k - \ell}\br{ - x - \ri \pi}} E\br{- q_{\mp, - k - \ell}\br{ - x - \ri \pi}}
    \\
    = \exp\br{\frac{\ri}{\hbar}\frac{\pi^2}{2} + \frac{\pi x}{\hbar}}
    U\br{- x - \ri \pi, q_{\mp, - k - \ell }\br{ - x - \ri \pi}} E\br{q_{\mp, - k - \ell}\br{ - x - \ri \pi}} \, .
\end{multline}
Note that at the level of $\omega_{0,\mp}$ we can replace $q_{\mp, -k - \ell}$ by $q_{\mp, k}$ because of \eqref{eq:QuasiPeriodicityQs}.
Hence, we find that \eqref{eq:Psi0x_RationalB_Result} can be written as
\begin{equation}
    \begin{split}
        \psi_0\br{x}
        & =
        \omega_{0,+}\br{x} +
        \exp\br{\frac{\ri}{\hbar}\frac{\pi^2}{2} + \frac{\pi x}{\hbar}}  \omega_{0,+}\br{- x - \ri \pi}
        \\
        & =
        \exp\br{\frac{\ri}{\hbar}\frac{\pi^2}{2} + \frac{\pi x}{\hbar}} \omega_{0,-}\br{-x - \ri \pi}
        + \omega_{0,-}\br{x}
    \end{split}
\end{equation}
which is precisely the conjectured structure in \eqref{eq:ProposedStructureEigenFun2}, here for $N = 0$ and $\hbar \in \pi \prationals$.

\subsection{The 't Hooft expansion}

An important way to test the conjectured structure in \eqref{eq:ProposedStructureEigenFun2} is by analysing ${\psi_{N}^\pm\br{x}}$ \eqref{eq:def:CanTransEigenFunQPtoXY} in a 't Hooft limit. This limit was studied in detail in \cite[sec.~3]{Marino:2016rsq} for the case $\xi = 0$, see also \cite{Zakany:2018dio}. We will closely follow their approach and use it to argue for the structure in \eqref{eq:ProposedStructureEigenFun2}. The 't Hooft limit is defined by taking
\begin{equation}
\label{eq:tHooftLimit}
    \hbar, N, \xi, \abs{q}, \abs{x} \to + \infty
    \, ,
\end{equation}
while keeping the following ratios constant,
\begin{equation}
\label{eq:tHooftLimitScalings}
    \lambda = \frac{N}{\hbar}
    \, ,
    \qquad \quad
    \xi_D = \frac{2 \pi}{\hbar} \xi
    \, ,
    \qquad \quad
    q_D = \frac{2 \pi}{\hbar} q
    \, ,
    \qquad \quad
    x_D = \frac{2 \pi}{\hbar} x
    \, .
\end{equation}

\subsubsection{Preparation}

Later on, we will need the solutions of the classical mirror curve \eqref{eq:MC} in the matrix model coordinates $q, p$ \eqref{xytopq} and the topological string coordinates $x, y$, which read
\begin{equation}
\label{eq:MCsolutions}
    \begin{split}
       \re^{p_\sigma\br{q, \xi, \kappa}/\sqrt{2}} & = \, {P_\sigma\br{\re^{q/\sqrt{2}}, \xi, \kappa}}
        \, ,
        \\
        \re^{y_\sigma\br{x, \xi, \kappa}} & = { Y_\sigma\br{\re^{x}, \xi, \kappa}}
        \, ,
    \end{split}
\end{equation}
where $\sigma=\pm 1$ and
\begin{equation}
    \begin{split}
        P_\pm\br{Q, \xi, \kappa} & = \frac{- \re^{-\xi} \kappa \pm \sqrt{ \br{\re^{-\xi} \kappa}^2 - 4 \br{\re^\xi Q + \re^{-\xi} Q^{-1}} \br{\re^{- \xi} Q + \re^\xi Q^{-1}}}}{2 \re^\xi \br{\re^\xi Q + \re^{-\xi} Q^{-1}}} \, ,
        \\
        Y_\pm(X, \xi, \kappa) &
        =
        - \br{\frac{\re^{2 \xi}}{2} \br{X + \frac{1}{X}} + \frac{\kappa}{2}} \pm \sqrt{\br{\frac{\re^{2 \xi}}{2} \br{X + \frac{1}{X}} + \frac{\kappa}{2}}^2 - 1} \, .
    \end{split}
\end{equation}
Let us also introduce the convenient shorthand notation
\begin{equation}
    \label{eq:MCsolutionsDual}
    {p_\pm\br{q_D}} = {p_\pm\br{q_D, \xi_D, \kappa_D}}
    \, ,
    \qquad \qquad
    {y_\pm\br{x_D}} = {y_\pm\br{x_D, \xi_D, \kappa_D}}
    \, .
\end{equation}
As we discuss below \eqref{eq:T0}, there is an implicit definition of $\kappa_D$ in terms of $\lambda$ and $\xi_D$.

\subsubsection{In matrix model coordinates}

Let us now consider the 't Hooft limit as defined in \eqref{eq:tHooftLimit} and \eqref{eq:tHooftLimitScalings} on the matrix model \be
    {E\br{q}} {\Psi_N\br{q}} / {Z\br{N,\hbar}}
    \, ,
\ee
where the relevant functions are defined in equations \eqref{sqrtv}, \eqref{eq:ZNmm}, \eqref{eq:kappaex}, and below. One finds the asymptotic expansion \cite{Kashaev:2015wia}
\begin{equation}
    \begin{split}
        E(q) & \simeq {\exp\br{\frac{\mathrm{i}}{g_s} \mathcal{E}_0 (q_D) + \mathcal{E}_1 (q_D) + \bigO{g_s}}} \, ,
        \\
        \frac{E(q) \Psi_N(q)}{Z(N)} & \simeq \exp\br{\frac{\mathrm{i}}{g_s} \mathcal{T}_0 (q_D) + \mathcal{T}_1(q_D) + \bigO{g_s}} \, .
    \end{split}
\end{equation}
The precise form of the functions $\mathcal{T}_1$ and $\mathcal{E}_1$ is not essential for our discussion. Thus, we focus on the leading-order terms.
Using the quasi-classical expansion of the quantum dilogarithm \eqref{eq:NonCompQuantDiLog_QuasiClassicalExpansion} gives
\begin{equation}
    \mathcal{E}_0(q_D) = \ri \pi \xi_D - \ri \pi \frac{q_D}{\sqrt{2}} - 2 \Li_2\left( - \ri \re^{q_D/\sqrt{2}} \re^{-\xi_D} \right) + 2 \Li_2\left( \ri \re^{q_D/\sqrt{2}} \re^{\xi_D} \right)
    \, .
\end{equation}
The computation of $\mathcal{T}_0$ is more involved and requires various matrix model techniques developed in \cite[sec.~3.2]{Marino:2016rsq} and \cite{Zakany:2018dio, Kashaev:2015wia}. Following \cite[sec.~3.2]{Marino:2016rsq} we get
\begin{equation}
\label{eq:T0}
    \mathcal{T}_0(q_D) =  \int^{q_D} p_\sigma ( q_D' ) \mathrm{d} q_D' + \ri \pi \xi_D
\end{equation}
where $p_{\sigma}$ is defined through \eqref{eq:MCsolutions} and \eqref{eq:MCsolutionsDual}.
The correct sign $\sigma = {\sigma\br{q_D}} \in \cbr{ \pm }$ depends on the region of the complex $q_D$-plane. However, we will not need it for what follows.
In \eqref{eq:T0}, we implicitly use the matrix model relation between the ’t Hooft coupling  $\lambda$  and $\kappa_D$. This relation is obtained as follows:
\begin{itemize}
\item[-] There is an explicit relation between the ’t Hooft coupling $\lambda$ and the endpoints of the eigenvalue density, denoted by $a^{\pm}$. For our matrix model, this relation is given in \cite[eqs.~(2.76)–(2.80)]{Kashaev:2015wia}.
\item[-] The endpoints of the cuts, $a^{\pm}$, are related to the mirror curve parameters $\xi_D$ and $\kappa_D$ as
\begin{equation}
\label{eq:BP}
    a^{\pm 2} = {a^{\pm 2}\br{\xi_D, \kappa_D}} = \frac{\re^{-2\xi_D}\kappa_D^2}{8} - \cosh(2\xi_D) \pm \sqrt{\left( \frac{\re^{-2\xi_D}\kappa_D^2}{8} - \cosh(2\xi_D) \right)^2 - 1} \, .
\end{equation}
These correspond to the branch points of the mirror curve in the $(p,q)$ coordinates.
\end{itemize}
By combining the above points, we obtain an explicit relation between $\kappa_D$, $\xi_D$, and $\lambda$, leading to \eqref{eq:T0}. We refer to \cite{Kashaev:2015wia} for more details.

\subsubsection{In outer topological string coordinates}

We are interested in the 't Hooft limit of \eqref{eq:def:CanTransEigenFunQPtoXY}. By recalling the conventions \eqref{eq:conventionsub} we have
\begin{equation}
\label{eq:def:CanTransEigenFunQPtoXYRepeated}
    \frac{\psi_{N}\br{x}}{Z\br{N}} =
    \int_{\reals} \rd q \, {U\br{x, q}} \frac{{E\br{q}} {\Psi_N\br{q}}}{Z\br{N}}
    =
    \int_{\reals} \rd q_D \, \frac{U^D\br{x_D, q_D}}{\sqrt{2 \pi g_s}} \frac{{E^D\br{q_D}} {\Psi_N^D\br{q_D}}}{Z\br{N}} \, ,
\end{equation}
where we defined for future convenience
\begin{equation}
    E^D\br{q_D}
    = E\br{\frac{2 \pi}{g_s} q_D}
    \, ,
    \qquad \qquad
    \Psi_N^D\br{q_D}
    = \Psi_N\br{\frac{2 \pi}{g_s} q_D}
    \, ,
\end{equation}
\begin{equation}
    {U^D\br{x_D, q_D}} = \sqrt{\frac{\br{2 \pi}^3}{g_s}} {U\br{\frac{2 \pi}{g_s} x_D, \frac{2 \pi}{g_s} q_D}}
    \, .
\end{equation}
The 't Hooft limit of \eqref{eq:def:CanTransEigenFunQPtoXYRepeated} becomes then a simple application of the stationary phase method. The essential ingredient is the saddle point equation,
\begin{equation}
\label{eq:SPEq}
    \partial_{q_D} F\br{x_D, q_D} + \cT_0 ' \br{q_D}
    =
    - \sqrt{2} ( x_D - \xi_D) + q_D + p_{\sigma}(q_D) = 0
    \, ,
\end{equation}
where we defined from \eqref{eqused}
\begin{equation}
    F\br{x_D, q_D} = \frac{x_D^2}{2} - \sqrt{2} \br{x_D - \xi_D} q_D + \frac{q_D^2}{2}
    \, .
\end{equation}
This should then be solved to find $q^D(x_D)$. Note that the saddle point equation is precisely the transformation of $x$ in the canonical transformation \eqref{xytopq}, which is a direct consequence of the construction of $U$ \eqref{eqused} \cite[sec.~2.5]{Marino:2016rsq} and the form of $\cT_0$ in \eqref{eq:T0}. Combining this with the transformation of $y$ in \eqref{xytopq} and using the fact that $p_\sigma$ and $y_\sigma$ in \eqref{eq:MCsolutions} are both solutions of the classical mirror curve \eqref{eq:MC} yields,\footnote{Up to an integer multiple of $\ri \sqrt{2} \pi$ if we are outside the strip $\abs{\imaginary{x_D}} < 2 \pi $.}
\begin{equation}
    {q_\pm^D\br{x_D}} = \frac{\sqrt{2}}{2} \br{x_D - {y_\pm\br{x_D}}}
    \, ,
    \qquad \qquad
    \abs{\imaginary{x_D}} < 2 \pi \, , \,
    \xi_D \in \mathbb{R} \, ,
\end{equation}
where $y_\pm$ is given in \eqref{eq:MCsolutions} and \eqref{eq:MCsolutionsDual}. Note that the restriction on $x_D$ is the usual domain restriction of the eigenfunctions \eqref{qctopco} in the dual variables. Observe that $q_\pm^D$ has a simple parity symmetry,
\begin{equation}
\label{eq:THooftSaddlesParity}
    q_\pm^D(x_D) = - q_\mp^D(-x_D) \, ,
\end{equation}
which will be important soon.

As a result of the stationary phase method one finds \cite[ch.~5]{miller2006applied}
\begin{equation}
    \frac{\psi_{N}\br{x}}{Z\br{N}} \simeq
    \sum_{\sigma \in \cbr{\pm}} \left(
    \frac{U^D\left(x_D,q_\sigma^D(x_D)\right)}{\sqrt{- \ri \left(\cT_0 + F \right)''\left( q_\sigma^D(x_D) \right)}}
    \frac{E^D (q_\sigma^D(x_D)) \Psi_N^D(q_\sigma^D(x_D))}{{Z\br{N}}}
    + \cO\left(g_s\right) \right) \, .
\end{equation}
It should be noted that we didn't make the 't Hooft limit expansion in $g_s$ explicit, and the functions involved are still complicated functions of $g_s$. However, the form given above is the most convenient one to understand the relation between the saddles.\footnote{See \cite[sec.~3.3,~app.~A]{Marino:2016rsq} for an explicit construction of the subleading order in the 't Hooft limit.}
We found in \eqref{eq:THooftSaddlesParity} that $q_\pm^D$ has a simple parity symmetry
and one can see that the denominator has similarly
\begin{equation}
    \left(\cT_0 + F \right)''(q_\pm^D(x_D))
    = {p_\pm^D}'(q_\pm^D(x_D)) + 1
    = \frac{\sqrt{2}}{{q_\pm^D} '(x_D)}
    = \left(\cT_0 + F \right)''(q_\mp^D(-x_D)) \, ,
\end{equation}
where the second equality follows from the $x_D$ derivative of the saddle point equation \eqref{eq:SPEq}.
We can then use the parity symmetries of the functions involved to get rid of the minus sign in front of the $q_\mp^D\br{-x_D}$ in \eqref{eq:THooftSaddlesParity} to find that the saddles are related by
\begin{multline}
    \frac{U^D\left(x_D,q_\pm^D(x_D)\right)}{\sqrt{- \ri \left(\cT_0 + F \right)''\left( q_\pm^D(x_D) \right)}}
    \frac{E^D (q_\pm^D(x_D)) \Psi_N^D(q_\pm^D(x_D))}{{Z\br{N}}}
    =
    \\
    \br{-1}^N \re^{\ri \frac{g_s}{8}} \re^{ \frac{x_D}{2}}
    \frac{U^D\left( - x_D -  \ri \frac{g_s}{2}  , q_\mp^D( - x_D )\right)}{\sqrt{- \ri \left(\cT_0 + F \right)''\left( q_\mp^D( - x_D) \right)}}
    \frac{E^D (q_\mp^D( - x_D)) \Psi_N^D(q_\mp^D( - x_D))}{Z\br{N}}
    \, .
\end{multline}
It is then again a consequence of the saddle point equation \eqref{eq:SPEq} that we can shift the argument of $q_\mp(-x_D)$ as well to get
\begin{multline}
    \frac{U^D\left(x_D,q_\pm^D(x_D)\right)}{\sqrt{- \ri \left(\cT_0 + F \right)''\left( q_\pm^D(x_D) \right)}}
    \frac{E^D (q_\pm^D(x_D)) \Psi_N^D(q_\pm^D(x_D))}{{Z\br{N}}}
    =
    \br{-1}^N \re^{\ri \frac{g_s}{8}} \re^{ \frac{x_D}{2}}
    \\
    \frac{ U^D\left( - x_D -  \ri \frac{g_s}{2} , q_\mp^D( - x_D -  \ri \frac{g_s}{2} )\right)}{\sqrt{- \ri \left(\cT_0 + F \right)''\left( q_\mp^D( - x_D -  \ri \frac{g_s}{2} ) \right)}}
    \frac{E^D (q_\mp^D( - x_D -  \ri \frac{g_s}{2}  )) \Psi_N^D(q_\mp^D( - x_D -  \ri \frac{g_s}{2} ))}{{Z\br{N}}}
    + \bigO{g_s} \, ,
\end{multline}
which is precisely the proposed structure \eqref{eq:ProposedStructureEigenFun2} in the dual variables, up to potential corrections of $\bigO{g_s}$. Note that factors like ${\exp\br{\ri g_s / 8}}$ or the shifts of the argument of $q_\pm^D$ are not visible at this level, but are part of the $\bigO{g_s}$ corrections.

\section{\texorpdfstring{The TS/ST correspondence for local $\mathbb{F}_0$}{The TS/ST correspondence for local F0}}
\label{sec:tsst}

In this section, we first review some aspects of the TS/ST correspondence for the closed string sector and then discuss the generalization to the open string sector. We focus on the particular case where the toric CY threefold is local $\mathbb{F}_0$.

\subsection{The quantum mirror map and Wilson loop}

Mirror symmetry plays an important role in the TS/ST correspondence, with one of its key components being the quantum mirror map. Originally introduced from a geometrical perspective, this map was defined as the quantization of the A-period in the mirror curve \cite{mirmor,acdkv}. Subsequently, it was understood that, from a geometric engineering perspective, this map could be identified with a Wilson loop in a corresponding five-dimensional gauge theory, see \cite{Gaiotto:2014ina, Bullimore:2014awa, Fucito:2015ofa, Sciarappa:2017hds}.

The quantum mirror map for local $\IF_0$ is given in \cite[sec.~7.2]{acdkv} or \cite[eq.~(3.58)]{Hatsuda:2013oxa}. The first few orders in a large $\mu$ expansion are
\begin{equation}
\label{eq:QuantumMirrorMap}
    t_B(\hbar)= 2 \mu
    - 2 \br{\re^{4 \xi} + 1} \re^{-2 \mu}
    - \left({3}\re^{8 \xi} + 2 \br{\re^{\ri \hbar} + 4 + \re^{-\ri \hbar}} \re^{4 \xi} + {3} \right) \re^{-4 \mu}
    + \bigO{\re^{-6 \mu}}
\end{equation}
 where $\kappa = \exp\br{\mu}$ and $\xi$ are the complex moduli of the mirror curve \eqref{eq:MC}.
 We will also use
 \begin{equation}
 \label{eq:QuantumMirrorMapForTheFibre}
      t_F\br{\hbar} = t_B\br{\hbar} - 4 \xi
      \, .
 \end{equation}
If we send $\hbar \to 0$ we recover the classical mirror map relating the K\"ahler parameters of local $\IF_0$,  to the complex moduli $\mu$ and $\xi$. For example, if $\xi = 0$ the classical mirror map is simply given by
\begin{equation}
    t_B \br{0}
    =
    2 \mu
    - 4 \, \re^{- 2 \mu} \,
    \pFq{4}{3}{ 1 , \, 1 , \, \frac{3}{2} , \, \frac{3}{2} }{ 2 , \, 2 , \, 2}{16 \, \re^{- 2 \mu}}
    \, ,
\end{equation}
where $\, _4F_3$ is the generalized hypergeometric function.

From the perspective of geometric engineering, topological string theory on local $\mathbb{F}_0$ engineers a five-dimensional, $\mathcal{N}=1$, $\mathrm{SU}\br{2}$ SYM theory in the $\Omega$-background \cite{kkv, Klemm:1996bj}. In this gauge theory context, the quantum mirror map corresponds to the inverse of the Wilson loop in the fundamental representation. The latter can be computed via supersymmetric localization \cite{Gaiotto:2014ina, Bullimore:2014awa} and the first few terms read
\begin{multline}
\label{WL}
    W\br{t_F, t_B, \hbar} = \re^{t_F/2} + \re^{- t_F / 2} + \squarebr{\frac{\br{\re^{t_F/2} + \re^{- t_F / 2}}}{\br{1 - \re^{\ri \hbar} \re^{- t_F}}\br{1 - \re^{-\ri \hbar} \re^{- t_F}}}} \re^{-t_B} +
    \\
    \Biggl[ \frac{
    \re^{- 2 t_F} \br{\re^{t_F/2} + \re^{- t_F / 2}}}{ \br{1 - \re^{\ri 2 \hbar} \re^{- t_F}} \br{1 - \re^{\ri \hbar} \re^{- t_F}}^3\br{1 - \re^{-\ri \hbar} \re^{- t_F}}^3 \br{1 - \re^{-\ri 2 \hbar} \re^{- t_F}}}
    \biggl( - \br{3 \re^{\ri \hbar} + 4 + 3 \re^{- \ri \hbar}}
    \\
    + \br{ \re^{\ri 2 \hbar} + \re^{\ri \hbar} + 1 + \re^{- \ri \hbar} +  \re^{- \ri 2 \hbar}} \br{\re^{t_F} + \re^{- t_F}} \biggr)
    \Biggr] \re^{- 2 t_B} + \bigO{\re^{- 3 t_B}}
\end{multline}
where $t_{B,F}$ are the Kähler parameters.
We refer to \cite[eq.~(3.22)]{gmcomp} for the full definition in this specific example. It is easily verified that setting
\begin{equation}
    t_B = t_B(\hbar) \, ,
    \qquad \qquad
    t_F = t_F(\hbar) \, ,
\end{equation}
in equation \eqref{WL} gives
\begin{equation} \label{mmdic}
    W\left( t_F(\hbar), t_B(\hbar), \hbar \right) = \re^{-2 \xi} \kappa \, .
\end{equation}
One can easily check, at least numerically, that both expansions in \eqref{eq:QuantumMirrorMap} and in \eqref{WL} are convergent, see e.g.~\cite{Hatsuda:2013oxa,gmcomp}.

\subsection{The closed string sector and the spectral determinant}
\label{sec:closts}

Let us first review some important elements of the TS/ST correspondence for the closed string sector \cite{ghm,cgm2}, see \cite{mmrev,mmrev2} for a review. Note that we focus on the specific case where the CY threefold is local $\IF_0$.

One feature of the TS/ST correspondence is that, on the topological string side, the relevant quantities involve a special combination of refined topological string partition functions in the GV ($- \epsilon_1 = \epsilon_2 = g_s$) and NS limit ($\epsilon_1 \to 0, \epsilon_2 = \hbar$) respectively with the relation
\be \label{eq:gsh}g_s = \frac{4 \pi^2}{\hbar} \, . \ee
The arguments of these two sets of special functions are typically rescaled with respect to each other.
Hence, it is convenient to define
\be \label{eq:dsnot}\alpha^D=\br{\frac{2\pi}{\hbar}} \alpha\,.\ee
The self-dual, or maximally supersymmetric point, is defined at $\hbar=g_s=2\pi$ \cite{cgm,ghm}.

The main quantity is the closed topological string grand potential ${\rm J}(\mu, \xi, \hbar)$. This quantity encapsulates both perturbative and non-perturbative contributions to the closed topological string free energy near the large radius point \cite{Hatsuda:2013oxa, ghm, Gu:2015pda}.
More precisely, we have
\begin{equation}
\label{eq:closed}
    {\rm J}(\mu, \xi, \hbar)
    =
    \rA\br{\xi, \hbar}
    +{\rm J}_{\rm p}\left(\mu, \xi, \hbar \right)
    + \rJ_\text{1-loop}\br{\mu, \xi , \hbar}
    + {\rm J}_{\rm inst}(\mu, \xi, \hbar)
    \, .
\end{equation}
Let us define these functions:
\begin{itemize}
    \item We denote with $\rA\br{\xi, \hbar}$ the constant map contribution whose closed-form reads \cite{yhum, Kashaev:2015wia}
    \begin{equation}
        \rA\br{\xi, \hbar} = \frac{4 \xi^3}{3 \pi \hbar} + \frac{\hbar \xi}{12 \pi} + A_c \br{\frac{\hbar}{\pi}} - F_\mathrm{CS}\br{ \xi , \hbar}
        \, ,
    \end{equation}
    \begin{equation}
        A_c\br{k} = \frac{2 \zeta\br{3}}{\pi^2 k} \br{1 - \frac{k^3}{16}} + \br{\frac{k}{\pi}}^2 \int_0^{+ \infty} \rd x \ \frac{x}{\re^{k x} - 1} \ln\br{1 - \re^{-2 x}}
        \, ,
    \end{equation}
    \begin{multline}
        F_\mathrm{CS}\br{\xi , \hbar}
        =
        \frac{\hbar^2}{8 \pi^4} \squarebr{\Li_3\br{- \re^{2 \br{\frac{2 \pi}{\hbar}} \xi}} + \Li_3\br{- \re^{- 2  \br{\frac{2 \pi}{\hbar}} \xi}} - 2 \zeta\br{3}}
        \\
        + \int_0^{+ \infty} \rd x \ \frac{x}{\re^{2 \pi x} - 1} \ln \squarebr{\frac{\sinh^2\br{\frac{\pi^2 x}{\hbar}}}{\sinh^2\br{\frac{\pi^2 x}{\hbar}} + \cosh^2\br{\br{\frac{2 \pi}{\hbar}} \xi}}}
        \, ,
    \end{multline}
    where $\Li_3$ is the polylogarithm of order 3 and $\zeta$ is the Riemann zeta function.

    \item
    The polynomial part $\rJ_\rp$ is given by
    \begin{equation}
        \label{eq:J5DClosedPolynomial}
        \rJ_\rp\br{\mu, \xi, \hbar}
        =
        \frac{t_B^3\br{\hbar}}{12 \pi \hbar} - \frac{\xi t_B^2\br{\hbar}}{2 \pi \hbar} + \br{\frac{\pi}{6 \hbar} - \frac{\hbar}{24 \pi}} t_B\br{\hbar} - \frac{\pi \xi}{3 \hbar}
        \, ,
    \end{equation}
    where the quantum mirror map $t_B\br{\hbar}$ is given in \eqref{eq:QuantumMirrorMap}.

    \item The one-loop part consists of two contributions: one coming from the one-loop part of the free energy in the NS phase and one being the one-loop part of the free energy in the GV phase of the $\Omega$-background. The sum of these two contributions reads then
    \begin{multline}
    \label{eq:J_Closed_1Loop}
        \rJ_\text{1-loop} \br{\mu , \xi, \hbar}
        =
        \sum_{k = 1}^{+ \infty} \squarebr{\frac{1}{2 \pi k^2} \cot{\br{\hbar \frac{k}{2}}} \br{1 + k t_F\br{\hbar}} + \frac{\hbar}{4 \pi k} \csc^2\br{\hbar \frac{k}{2}}} \re^{- k t_F\br{\hbar}}
        \\
        - \sum_{k = 1}^{+ \infty} \frac{1}{2 k} \csc^2\br{\br{\frac{4 \pi^2}{\hbar}}\frac{k}{2}} \re^{- k \br{\frac{2 \pi}{\hbar}} t_F\br{\hbar}}
        \, ,
    \end{multline}
    where $t_F\br{\hbar}$ is related to $\mu$ and $\xi$ via \eqref{eq:QuantumMirrorMapForTheFibre}.
    This can be written in closed form in $t_F \br{\hbar}$ when $\hbar = 2 \pi \br{n / m}$ with $n, m \in \pnaturals$ coprime,
    \begin{multline}
        \rJ_\text{1-loop} \br{\mu , \xi, \hbar}
        =
        \\
        \br{\frac{2 \pi^2 \br{m^2 - n^2} + 3 m^2 t_F^2\br{\hbar}}{12 \pi^2 n m^2}}
        \ln \br{1 - \re^{-m t_F \br{\hbar}}}
        - \frac{t_F\br{\hbar} \Li_2\left( \re^{-m t_F\br{\hbar}}\right)}{2 \pi^2 n m}
        - \frac{\Li_3\left( \re^{-m t_F\br{\hbar}}\right)}{2 \pi^2 n m^2}
        \\
        +
        \sum_{k = 1}^{m-1} \frac{\csc^2\br{\hbar \frac{k}{2}}}{4 \pi k}
        \re^{- k t_F\br{\hbar}}
        \left\{
        \squarebr{\hbar + \br{ \frac{1}{k} + t_F\br{\hbar}} \sin \br{ \hbar k }}
        \pFq{3}{2}{1 , \, \frac{k}{m} , \, \frac{k}{m}}{ 1 + \frac{k}{m} , \,  1 + \frac{k}{m}}{ \re^{- m t_F\br{\hbar}}}
        \right.
        \\
        \left.
        +
        \br{\frac{k / m}{\br{1 + k / m}^2}} \re^{- m t_F\br{\hbar}} \squarebr{\hbar + t_F\br{\hbar} \sin \br{\hbar k }}
        \pFq{3}{2}{2 , \, 1 + \frac{k}{m} , \, 1 + \frac{k}{m}}{2 + \frac{k}{m} , \, 2 + \frac{k}{m}}{ \re^{- m t_F\br{\hbar}}} \right\}
        \\
        - \sum_{k = 1}^{n-1} \frac{\csc^2\br{\br{\frac{4 \pi^2}{\hbar}} \frac{k}{2}}}{2 k}
        \re^{- k \br{\frac{2 \pi}{\hbar}} t_F\br{\hbar}}
        \pFq{2}{1}{1, \,  \frac{k}{n}}{ 1 + \frac{k}{n}}{\re^{ - n \br{\frac{2 \pi}{\hbar}} t_F\br{\hbar}}}
        \, ,
    \end{multline}
    where $\Li_q$ is the polylogarithm of order $q$, and $_pF_q$ is the generalized hypergeometric function.
    It is interesting to note that \eqref{eq:J_Closed_1Loop} also admits the following integral representation \cite[eq.~(3.9)]{bgt}
    \begin{multline}
        \rJ_\text{1-loop} \br{\mu , \xi, \hbar}
        =
        - \frac{\hbar^2}{8 \pi^4} \Li_3\br{\re^{- \br{\frac{2 \pi}{\hbar}} t_F\br{\hbar}}} +
        \\
        2 \Re \int_0^{\infty \re^{\ri 0}} \rd x \ \frac{x}{\re^{2 \pi x} - 1}
        \ln\br{1 - 2 \cosh\br{\br{\frac{4 \pi^2}{\hbar}} x} \re^{- \br{\frac{2 \pi}{\hbar}} t_F\br{\hbar}} + \re^{- 2 \br{\frac{2 \pi}{\hbar}} t_F\br{\hbar}}}
        \, .
    \end{multline}

    \item The instanton part of the grand potential also consists of two parts: one coming from the instanton part of the NS free energy, and one being the instanton part of the GV free energy. Together, they read
    \begin{multline}
    \label{eq:J_Closed_Instanton}
        {\rm J}_{\rm inst}(\mu, \xi, \hbar)
        =
        F^{\rm GV}_{\rm inst} \br{\br{\frac{2 \pi}{\hbar}} t_F \br{\hbar} , \br{\frac{2 \pi}{\hbar}} t_B \br{\hbar}, \frac{4 \pi^2}{\hbar}}
        \\
        +
        \left(-\frac{1}{2 \pi} + \frac{t_F(\hbar)}{2 \pi}\partial_{t_F} + \frac{t_B(\hbar)}{2 \pi}\partial_{t_B} + \frac{\hbar}{2 \pi}\partial_{\hbar} \right)
        F^{\rm NS}_{\rm inst} \left( t_F(\hbar) , t_B(\hbar) , {\hbar}\right)
    \end{multline}
    where $F^{\rm NS}_{\rm inst}$ is the instanton part of the 5d Nekrasov free energy in the NS limit, and $F^{\rm GV}_{\rm inst}$ is similarly the instanton part of the 5d Nekrasov free energy in the GV limit. The leading order reads
    \begin{equation}
    \label{inst2}
        F^{\rm NS}_{\rm inst} \left( t_F , t_B , \hbar \right)
        =
        \squarebr{\frac{\ri \br{1 + \re^{\ri \hbar}}}{\br{1 - \re^{\ri \hbar}} \br{1 -  \re^{\ri \hbar} \re^{- t_F}} \br{1 - \re^{- \ri \hbar} \re^{- t_F}}}} \re^{- t_B}
        + \bigO{\re^{-2 \, t_B}}
        \, ,
    \end{equation}
    \begin{equation}
    \label{eqF_Closed_GV_Instanton}
        F^\mathrm{GV}_{\rm inst}  \left( t_F , t_B , g_s\right)
        =
        \squarebr{\frac{2 \re^{\ri g_s}}{\br{1 - \re^{\ri g_s}}^2 \br{1 - \re^{- t_F}}^2}} \re^{-t_B}
        + \bigO{\re^{- 2 \, t_B}}
        \, ,
    \end{equation}
    and the all order constructions can be found in \autoref{sec:TSfunctions},  equations \eqref{eq:nsef} and \eqref{eq:sdef} respectively.

\end{itemize}
Note that the NS contributions in $\rJ_\text{1-loop}$ \eqref{eq:J_Closed_1Loop} and $\rJ_\mathrm{inst}$ \eqref{eq:J_Closed_Instanton} are perturbative in $\hbar$ for fixed $t_{B,F}$, but non-perturbative in $g_s = 4 \pi^2 / \hbar$ for fixed $t_{B,F}^D = \br{2 \pi / \hbar} t_{B,F}$. Vice versa, the GV contributions in $\rJ_\text{1-loop}$ \eqref{eq:J_Closed_1Loop} and $\rJ_\mathrm{inst}$ \eqref{eq:J_Closed_Instanton} are perturbative in $g_s$ for fixed $t_{B,F}^D$, but non-perturbative in $\hbar = 4 \pi^2 / g_s$ for fixed $t_{B,F} = \br{2 \pi / g_s} t_{B,F}^D$ .
The specific combination of GV and NS free energies in \eqref{eq:closed} provides a genuinely non-perturbative completion of both quantities. Indeed, the NS and GV functions are not well-defined functions of $\hbar$ or $g_s$ when considered separately. Each of them individually exhibits a dense set of poles along the real $\hbar$ or $g_s$ line, making them ill-defined. However, these poles cancel in the combinations \eqref{eq:J_Closed_1Loop} and \eqref{eq:J_Closed_Instanton}, and the resulting function is perfectly well-defined for any value of $\hbar$ or $g_s$.

It is well known that, given an asymptotic series, its non-perturbative completion is not unique; additional conditions are required to fix it. In the context of the TS/ST correspondence, these conditions are provided by the spectral theory of the quantized mirror curve. In particular, the non-perturbative completion defined by the TS/ST correspondence is the unique one that correctly reproduces the spectrum and eigenfunctions of the operator associated with the quantum mirror curve, which in turn is related to an underlying relativistic quantum integrable systems.
The same non-perturbative completion was also used in ABJM theory \cite{hmo3, Hatsuda:2013oxa}, where it successfully captures non-perturbative effects in its corresponding string dual \cite{dmpnp}. Nevertheless, other completions are possible in principle; see for instance \cite{Hatsuda:2015owa,Grassi:2014cla} for an alternative proposal, and \cite{mmhouches} for a broader discussion of non-perturbative completions in topological string theory and their interplay with resurgence.

From the perspective of spectral theory, it is natural to consider the grand-canonical ensemble. One of the key statements of the TS/ST correspondence is then that
\begin{equation}
\label{eq:detimain}
    \det \br{1 + \kappa \rho}
    = \sum_{k \in \IZ} \re^{\rJ\br{\mu + \ri 2 \pi k, \xi, \hbar}}
    \, ,
    \qquad \qquad
    \kappa = \exp\br{\mu}
    \, ,
\end{equation}
where $\rho$ is the operator in \eqref{eq:rhok} and $\rJ$ is the grand potential of \eqref{eq:closed}. An important point is that \eqref{eq:detimain} is entire in the full $\kappa$ plane, therefore all the singularities in the closed string moduli space are smoothed out, and the full quantity is background independent. To extract the partition function around specific points in the moduli space, we need to expand \eqref{eq:detimain} accordingly. For instance, expanding around $\kappa = \infty$ leads to the large radius expansion. Expanding around the orbifold point $\kappa = 0$ gives \cite{ghm, Marino:2015ixa, Kashaev:2015wia}
\begin{equation}
    \det\br{1 + \kappa \rho} = \sum_{N = 0}^{\infty} \kappa^N Z\br{N, \hbar} \, ,
\end{equation}
where $Z(N, \hbar)$ is defined in \eqref{eq:ZNmm}. It is a distinctive feature of this construction that even though one expands the determinant around the orbifold point $\kappa = 0$, the coefficients $Z(N, \hbar)$ actually encode the non-perturbative partition function in the conifold frame.

\subsection{The open string sector and the eigenfunctions}\label{sec:opents}

Let us now turn to the open sector of the topological string. As before, we focus on the example of local $\mathbb{F}_0$. The grand potential for the open topological string in the presence of a toric D-brane on an external leg of the toric diagram was introduced in \cite{Marino:2016rsq, Marino:2017gyg}. Here, we follow the formulation in \cite[app.~A]{Francois:2023trm}, where the resummation in the open string modulus $x$ is also performed.

The open string grand potential is
\begin{equation}
\label{eq:Jopen}
    {\rm J }^{\rm open}(x, \mu, \xi , \hbar) =
    {\rm J }^\mathrm{open}_\rp(x, \xi , \hbar)+{\rm J }^\mathrm{open}_{\text{1-loop}}(x, \mu, \xi , \hbar) + {\rm J}^\mathrm{open}_{\rm inst}\br{ x , \mu , \xi , \hbar }
    \, ,
\end{equation}
where $\mu$ is the closed string modulus, $\xi$ is the mass parameter, and $x$ is the open string modulus. The functions appearing on the right-hand side of \eqref{eq:Jopen} are defined as follows:
\begin{itemize}
    \item  The polynomial part in $x$ is
    \be
        \label{eq:J5DOpenPolynomial}
        \rJ^\mathrm{open}_\rp(x, \xi , \hbar)= - \frac{\ri}{\hbar} 2 \xi x - \frac{\ri}{\hbar} \frac{x^2}{2} + \frac{1}{2} \br{\frac{2 \pi}{\hbar} - 1} x
        \, .
    \ee
    \item The one-loop part is given by \cite[eq.~(A.24)]{Francois:2023trm}
    \begin{multline}
    \label{eq:J5DOpen1Loop}
        \rJ^\mathrm{open}_\text{1-loop}\br{x, \mu, \xi , \hbar}
        =
        \\
        \ln \Phi_\beta\br{\frac{1}{2 \pi \beta} \br{- x - \frac{t_F\br{\hbar}}{2}} + \ri \frac{\beta}{2}}
        +
        \ln \Phi_\beta\br{\frac{1}{2 \pi \beta} \br{- x + \frac{t_F\br{\hbar}}{2}} + \ri \frac{\beta}{2}}
    \end{multline}
    where $\hbar = 2 \pi \beta^2$ and $\ln \Phi_\beta$ is the logarithm of Faddeev's non-compact quantum dilogarithm, see \autoref{sec:NCQuantDiLog}. One can express \eqref{eq:J5DOpen1Loop} in closed form in terms of elementary functions and the classical dilogarithm $\Li_2$ if $\hbar \in 2 \pi \prationals$, by using \eqref{eq:FQDiLogSD}. This 1-loop part of the grand potential consists again of contributions from both the NS and GV free energies, just as in the closed sector. See \cite[app.~A]{Francois:2023trm} for details.

    \item The instanton part ${\rm J}^\mathrm{open}_{\rm inst}$ consists also of a part coming from the NS free energy, and a part coming from the GV free energy
    \begin{multline}
         {\rm J}^\mathrm{open}_{\rm inst}\br{ x , \mu , \xi , \hbar }
         =
         F^\mathrm{open}_\mathrm{NS,inst} \left(x, t_F(\hbar), t_B(\hbar), \hbar \right)
         \\
         +
         F^\mathrm{open}_\mathrm{GV, inst} \left( \br{\frac{2 \pi}{\hbar}} x , \br{\frac{2 \pi}{\hbar}} t_F(\hbar),  \br{\frac{2 \pi}{\hbar}} t_B(\hbar), \frac{4 \pi^2}{\hbar} \right)
         \, .
    \end{multline}
    Here, $F^\mathrm{open}_\mathrm{NS, inst}$ represents the NS limit of the refined open topological string free energy associated with a brane inserted on the outer leg of the toric diagram:
    \begin{multline}
        \label{eq:zns}
        F^\mathrm{open}_\mathrm{NS, inst} \br{ x, t_F, t_B, \hbar}
        =
        \\
        \frac{
        \re^{\ri \hbar} \re^{\frac{t_F}{2} - x} \left(  1 + \re^{-t_F} + \re^{\ri \hbar} \left( 1 +\re^{\ri \hbar }\right) \re^{-\frac{t_F}{2}-x}  \right)}
        {\left( 1 - \re^{\ri \hbar} \right) \left( 1 - \re^{\ri \hbar} \re^{- t_F}\right) \left( 1 - \re^{- \ri \hbar } \re^{-t_F}\right) \left( 1 + \re^{\ri \hbar} \re^{\frac{t_F}{2} - x}\right) \left( 1 + \re^{\ri \hbar} \re^{-\frac{t_F}{2} - x}\right)} \re^{- t_B}
        + \bigO{\re^{- 2 \, t_B}}
        ,
    \end{multline}
    and the all order definition is given in \autoref{sec:TSfunctions}, equation \eqref{eq:nsef}.
    Similarly, $F^\mathrm{open}_\mathrm{GV, inst}$ is the open topological string free energy corresponding to a brane inserted in the outer leg of the toric diagram:
    \begin{multline}
        \label{eq:zgv}
        F^\mathrm{open}_\mathrm{GV, inst} \left( x , t_F, t_B, g_s \right)
        =
        \\
        -
        \frac{\re^{\ri \frac{g_s}{2}} \re^{\frac{t_F}{2} - x} \br{1 + \re^{-t_F} - 2 \re^{\ri \frac{g_s}{2}} \re^{- \frac{t_F}{2} - x}}}{\br{1 - \re^{\ri g_s}} \br{1 - \re^{-t_F}}^2 \br{1 - \re^{\ri \frac{g_s}{2}} \re^{\frac{t_F}{2} - x}} \br{1 - \re^{\ri \frac{g_s}{2}} \re^{- \frac{t_F}{2} - x}}} \re^{- t_B}
        + \bigO{\re^{-2 t_B}}
        \, ,
    \end{multline}
    and the all order expression for \eqref{eq:zgv} can be found in \autoref{sec:TSfunctions}, equation \eqref{eq:sdef}.
\end{itemize}
Similar to what was observed in the closed string sector, the role of the NS partition function in \eqref{eq:Jopen} in the open sector is also purely non-perturbative in \( g_s \), and it plays a crucial role in cancelling the poles in $g_s$ of the GV part and making the full expression well-defined. Hence, \eqref{eq:Jopen} provides a well-defined non-perturbative completion for the open topological string partition function around the large radius frame.
On the other side, from the spectral theory perspective, the perturbative contributions in $\hbar = 4 \pi^2 / g_s$ are captured by the NS partition function in \eqref{eq:Jopen}, whereas the GV part remains purely non-perturbative in \( \hbar \). We refer to \cite{Marino:2016rsq, Marino:2017gyg} for further details.

Note that $\exp\br{J^\mathrm{open}_\text{1-loop}}$ has poles at
\begin{equation}
    x = \pm \frac{t_F \br{\hbar}}{2} - \ri 2 \pi \br{n - \frac{1}{2}} - \ri \hbar m
    \, ,
    \qquad \qquad
    n \in \pnaturals ,
    \,
    m \in \naturals
    \, ,
\end{equation}
and similarly,
$\exp\br{J^\mathrm{open}_\mathrm{inst}}$ has poles coming from $F^\mathrm{open}_\mathrm{NS, inst}$ and $F^\mathrm{open}_\mathrm{GV, inst}$ respectively when
\begin{equation}
    \begin{aligned}
    \label{polesNS}
        x & = \pm \frac{t_F \br{\hbar}}{2} + \ri 2 \pi \br{m - \frac{1}{2}} + \ri \hbar n
        \, ,
        \qquad \qquad
        n \in \pnaturals,
        \,
        m \in \integers
        \, ,
        \\
        x & = \pm \frac{t_F \br{\hbar}}{2} + \ri 2 \pi \br{n - \frac{1}{2}} + \ri \hbar m
        \, ,
        \qquad \qquad
        n \in \pnaturals,
        \,
        m \in \integers
        \, .
    \end{aligned}
\end{equation}
Note that only the poles with $n \leqslant N$ occur at order $\exp\br{- N \, t_B}$ in $F^\mathrm{open}_\mathrm{NS, inst}$ or $F^\mathrm{open}_\mathrm{GV, inst}$.
These poles should be related to the transition from the external to the internal leg of the toric diagram, and they
do not disappear in the open string grand potential ${\rm J }^{\rm open}(x, \mu, \xi, \hbar) $.

Now we want to relate \eqref{eq:Jopen} to the eigenfunctions of the quantum mirror curve \eqref{eq:DifferenceEquation}. It is important to emphasize that there are numerous ways to construct formal solutions to \eqref{eq:DifferenceEquation}. For instance, consider
\begin{equation}
    \exp\Bigl[{\rm J }^\mathrm{open}_{\rm p}(x, \xi , \hbar) + {\rm J }^\mathrm{open}_{\text{1-loop}}(x, \mu, \xi , \hbar) + F^\mathrm{open}_\mathrm{NS,inst} \left(x, t_F(\hbar), t_B(\hbar), \hbar \right) \Bigr]
    \, .
\end{equation}
While this expression formally satisfies \eqref{eq:DifferenceEquation}, it is not a well-defined function for $\hbar \in \preals$ due to the dense set of poles at $\hbar \in \pi \rationals$, and it fails to satisfy the analytic properties required by a proper eigenfunction as discussed below \eqref{eq:DifferenceEquation}.
The exponential of \eqref{eq:Jopen} on the other hand, as well as each individual term in \eqref{tsstopenf}, would give a well-defined solution to the difference equation for all $\hbar \in \preals$. However, it does not have the correct analytical properties or asymptotic behaviour to be in the domain of the quantized mirror curve. Hence, it does not qualify as an eigenfunction either.

\begin{figure}[t]
    \centering
    \includegraphics[width=0.49\linewidth]{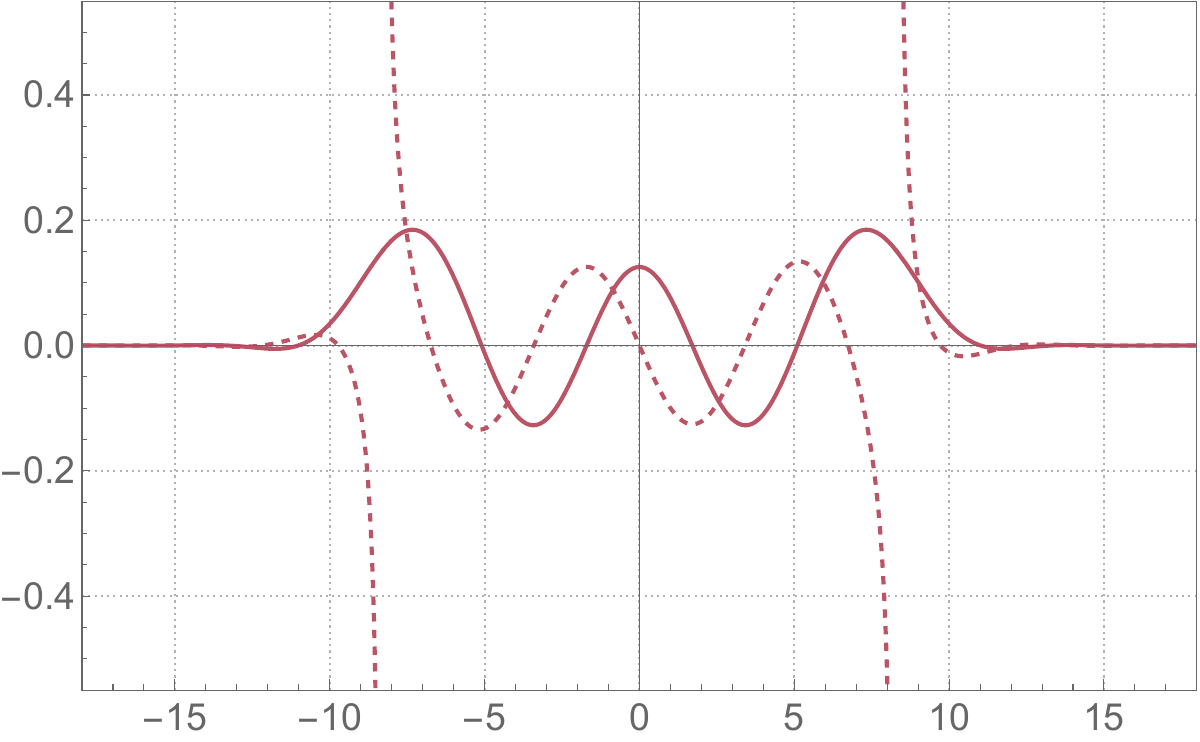}
    \includegraphics[width=0.49\linewidth]{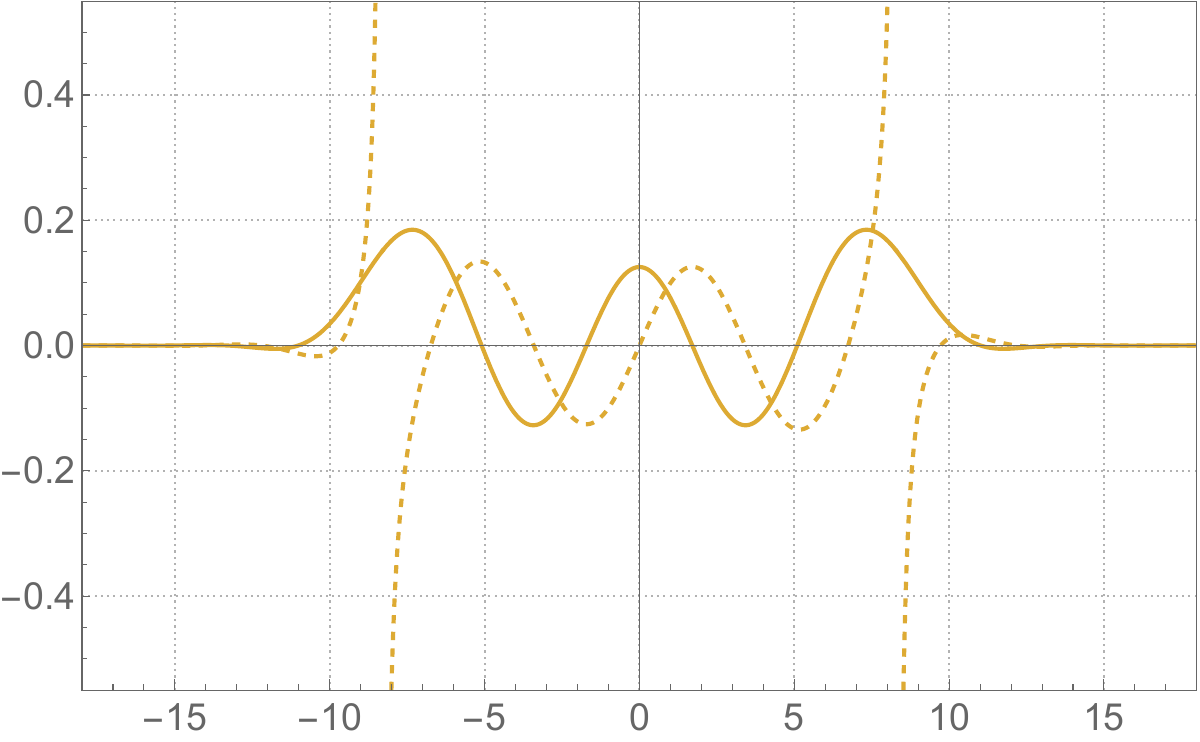}
    \includegraphics[width=0.49\linewidth]{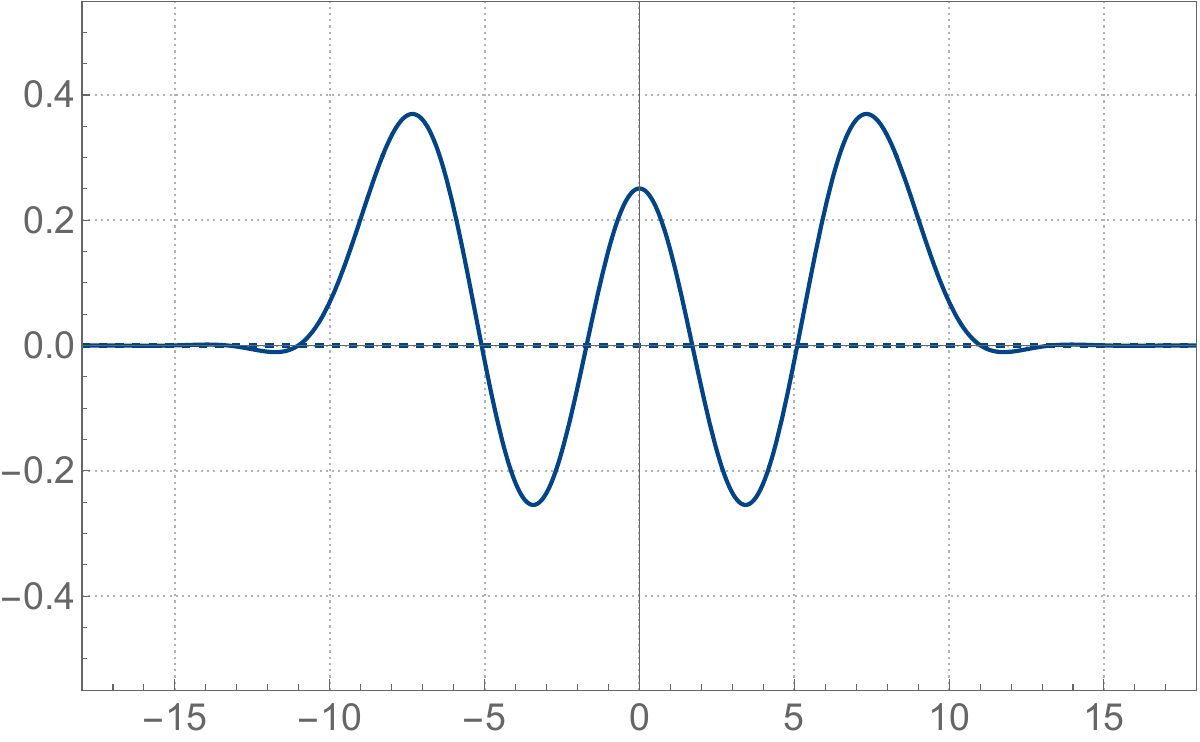}
    \includegraphics[width=0.49\linewidth]{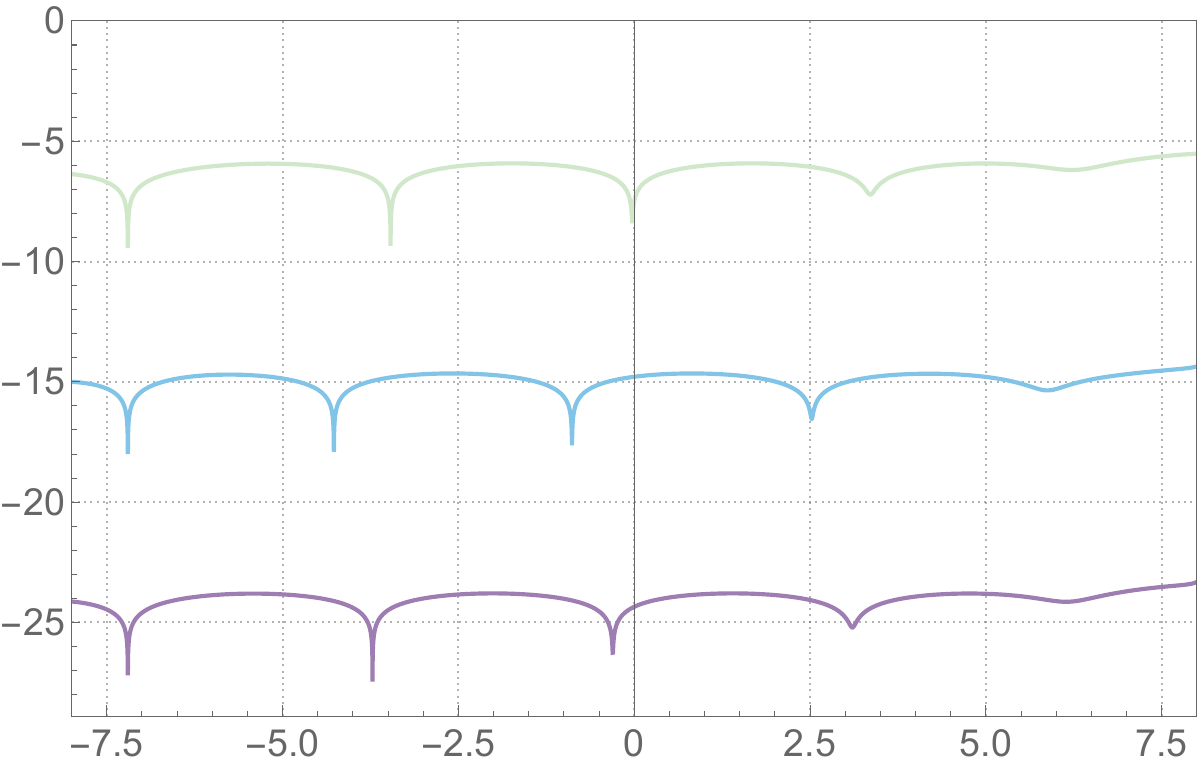}
    \caption{The on-shell eigenfunction from topological strings for $\xi = - 16 / 53$ and $\hbar = 8 \pi / 3$ at energy $E = E_4 \approx 7.69$. Top left: the $\sigma = 1$ term in \eqref{eq:sigmas}. Top right: the $\sigma = 2$ term in \eqref{eq:sigmas}. Bottom left: the full eigenfunction $\psi$ as given in \eqref{tsstopenf}, after normalization. Bottom right: the absolute difference between the analytical and numerical eigenfunctions, including 0 (green), 2 (blue), and 4 (violet) instantons in the small $\exp\br{-t_B}$ expansion of $\psi$. The plot on the bottom right uses a logarithmic scale with base 10. The solid and dashed lines correspond to the real and imaginary parts, respectively.}
    \label{fig:test1}
\end{figure}

\begin{figure}[t]
    \centering
    \includegraphics[width=0.49\linewidth]{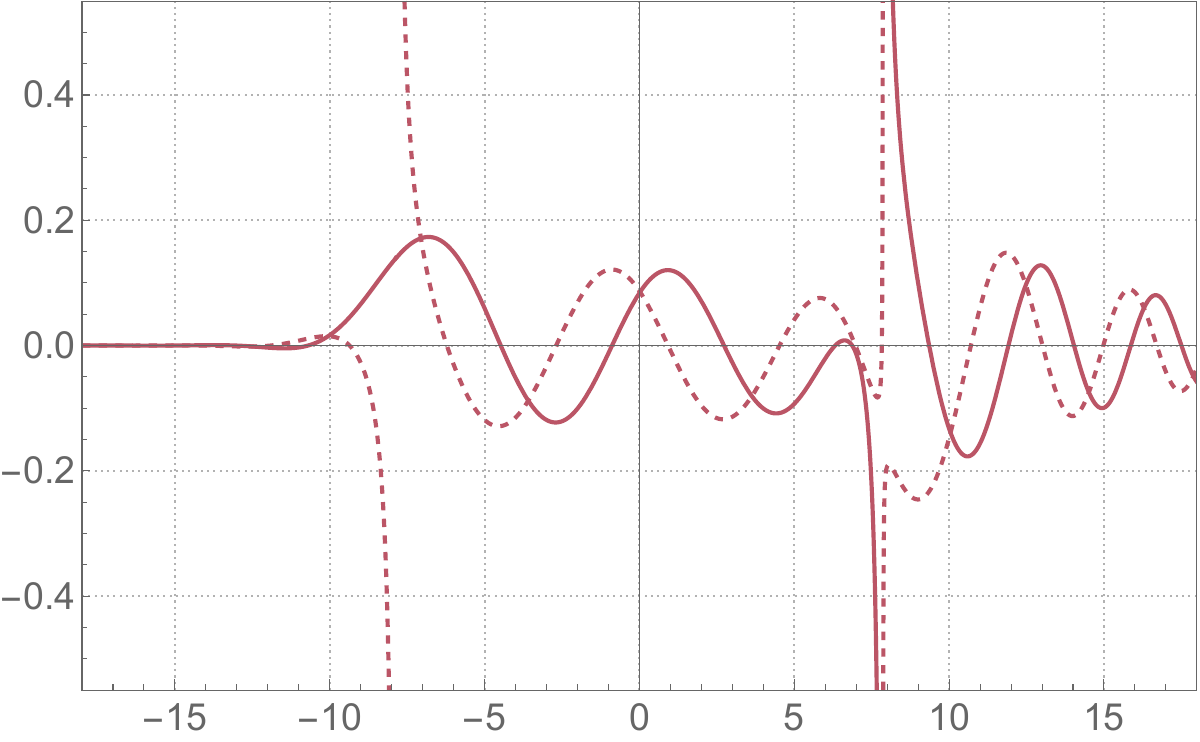}
    \includegraphics[width=0.49\linewidth]{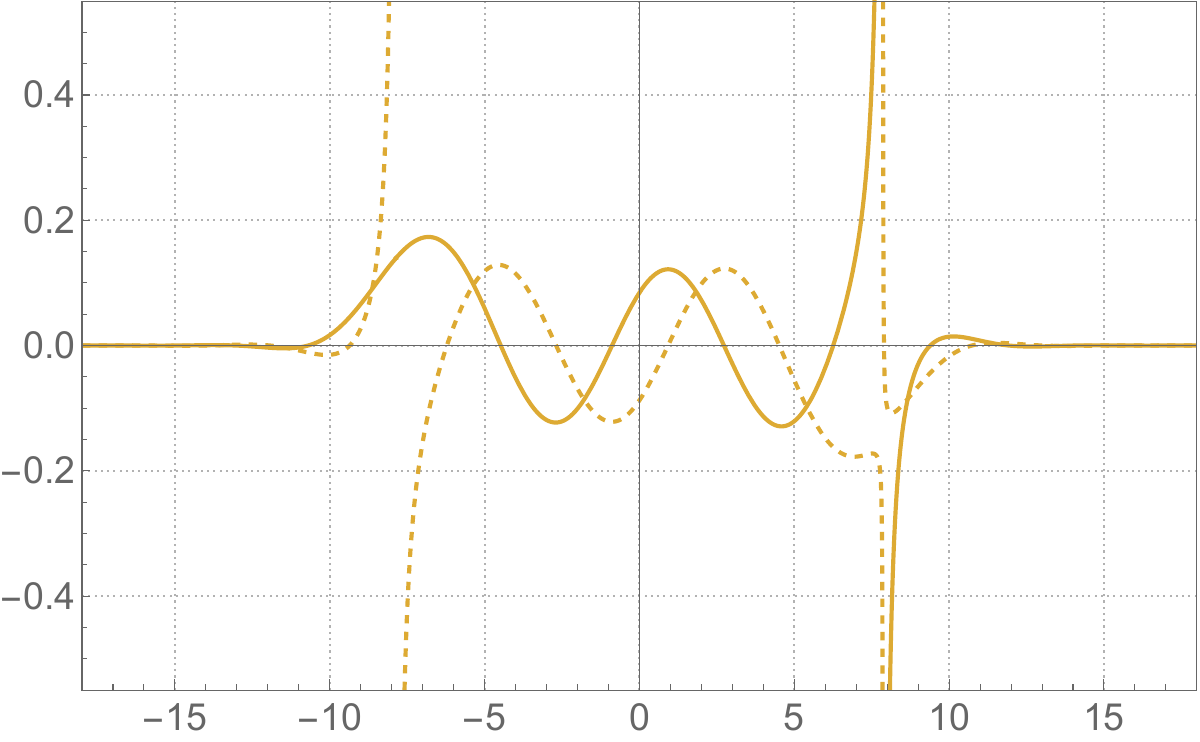}
    \includegraphics[width=0.49\linewidth]{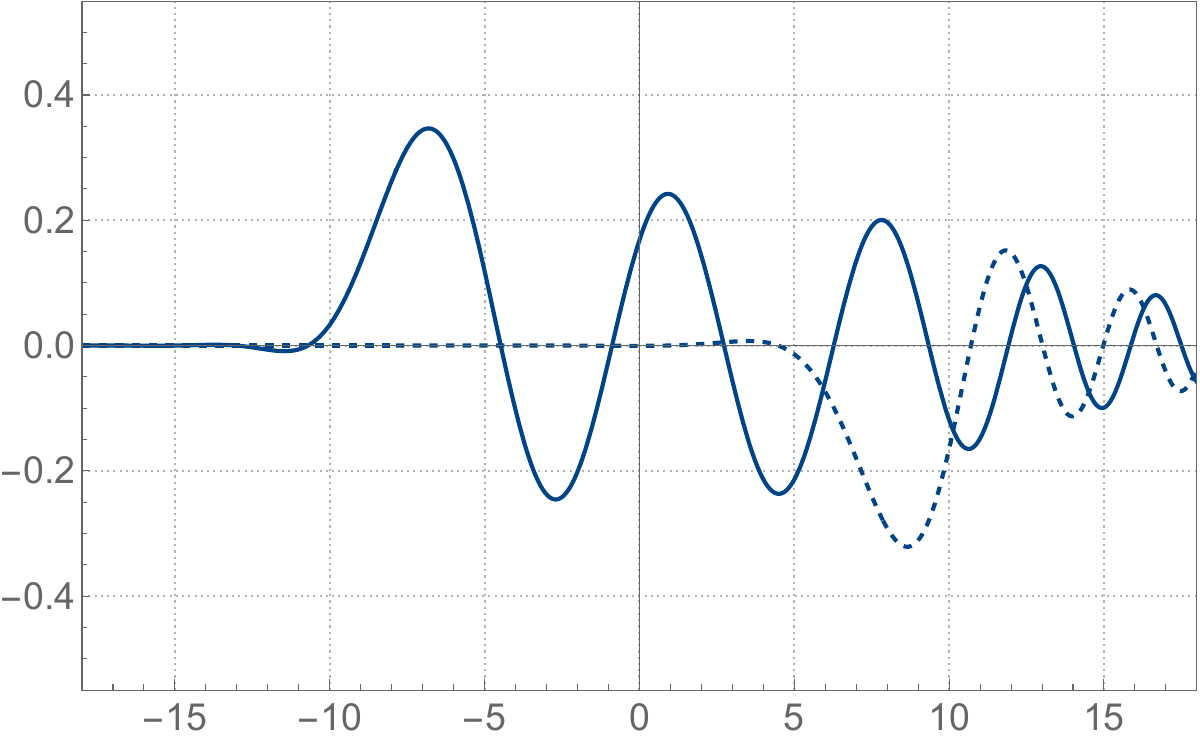}
    \caption{The off-shell eigenfunction from topological strings for $\xi = - 16 / 53$ and $\hbar = 8 \pi / 3$ at energy $E = 225/31$, which is between $E_3$ and $E_4$. From top left to bottom: the $\sigma = 1$ term in \eqref{eq:sigmas}, the $\sigma = 2$ term in \eqref{eq:sigmas}, and the complete eigenfunction $\psi$ in \eqref{tsstopenf}, after normalization. The solid and dashed lines correspond to the real and imaginary parts, respectively.}
    \label{fig:test2}
\end{figure}

To construct proper eigenfunctions, it is useful to introduce the full grand potential, which is defined as
\be\label{eq:Jfull} {\rm J}(x, \mu, \xi, \hbar)=  {\rm J}(\mu, \xi, \hbar)+{\rm J }^{\rm open}(x, \mu, \xi, \hbar) \,. \ee
It was conjectured in \cite{Marino:2016rsq,Marino:2017gyg} that one of the  terms in the sum \eqref{eq:sigmaint} should be given by
\be
\label{eq:fpts}
\sum_{k \in \mathbb{Z}} \re^{{\rm J}(x, \mu + \ri 2 \pi k, \xi, \hbar)} \, .
\ee
Building on the structure found in \eqref{eq:ProposedStructureEigenFun} for the second term, we can now express the full eigenfunctions of the quantum mirror curve \eqref{eq:QuantizedMirrorCurve} in a compact form:
\begin{equation}
\label{eq:sigmas}
    \psi(x, \kappa) = \sum_{k \in \mathbb{Z}} \sum_{\sigma \in \cbr{ 1, 2 }} \re^{{\rm J_\sigma}(x, \mu + \ri  2 \pi k, \xi, \hbar)} \, ,
    \qquad \qquad
    \kappa = \re^{\mu} \, ,
\end{equation}
\begin{equation}
    {\rm J}_1(x, \mu, \xi, \hbar) = {\rm J}(x, \mu, \xi, \hbar) \, , \qquad
    {\rm J}_2(x, \mu, \xi, \hbar) = \frac{\ri}{\hbar} \frac{\pi^2}{2} + \frac{\pi x}{\hbar} + {\rm J}(-x - \ri \pi, \mu + \ri \pi, \xi, \hbar) \, .
\end{equation}
Hence, the eigenfunctions are simply
\begin{equation}
\label{tsstopenf}
    \boxed{ \psi (x, \kappa) = \sum_{k \in \IZ}\left(\re^{ {\rm J}(x, \mu + \ri 2 \pi k, \xi, \hbar)} + \re^{ \frac{\ri}{\hbar} \frac{\pi^2}{2}+ \frac{\pi x}{\hbar} + {\rm J}(-x-\ri \pi, \mu + \ri \pi + \ri 2 \pi k, \xi, \hbar)} \right)}
    \, .
\end{equation}
According to our construction, these are well-defined functions for all $x, \kappa \in \complexes$, $\xi \in \reals$ and $\hbar \in \preals$, which are \textit{entire} in $x$ and solve the difference equation \eqref{eq:DifferenceEquation} for any value of the parameters, and which are the true, square-integrable eigenfunctions of the quantum mirror curve when $\kappa = - \exp\br{E_n}$ coincides with a value in the spectrum. We do not have a rigorous proof of this, but there are many non-trivial tests.
See \autoref{fig:test1}, \autoref{fig:test2}, and \autoref{sec:plots} for some graphical representations of \eqref{tsstopenf}.

Let us make some comments on this result:
\begin{itemize}

    \item The grand potential ${\rm J}(x, \mu, \xi, \hbar)$ is defined in the large-radius frame, corresponding to placing a brane on the outer leg of the toric diagram. It is therefore expected that ${\rm J}(x, \mu, \xi, \hbar)$ has poles in $x$: it is associated with a specific patch of the open-string moduli space and is not background independent. Starting from ${\rm J}(x, \mu, \xi, \hbar)$ and moving to a different patch in the moduli space requires performing a modular transformation followed by an analytic continuation.

    In contrast, the non-perturbative completion that we propose \textemdash{} given in equation \eqref{tsstopenf} \textemdash{} is an entire function of both the open modulus $x$ and the closed moduli. In this sense, equation \eqref{tsstopenf} is background independent: we can simply expand it around any desired point in the moduli space, such as the large-radius or orbifold regions, and directly recover the corresponding open-string partition function valid in that particular frame.

    At the technical level, such background independence is obtained by the summation over $k$ and the summation over $\sigma$ in \eqref{eq:sigmas} and \eqref{tsstopenf}. Indeed, as we reviewed in the introduction, the summation over $k$ in equation \eqref{eq:detimain} smooths all the singularities in the closed-string moduli space, parametrized by $\kappa$, yielding an entire function in $\kappa$. The sum over $\sigma$ in equation \eqref{tsstopenf} plays an analogous role for the open-string modulus $x$: each of the two terms in \eqref{eq:sigmas} and \eqref{tsstopenf} individually has singularities in $x$, but these are smoothed out once the two contributions are combined and the sum over $k$ is performed. This property holds even off-shell, i.e.~for $\kappa \neq -\re^{E_n}$.  Geometrically, it was argued in \cite{Marino:2017gyg,Marino:2016rsq} that such a sum over $\sigma$ should correspond to a sum over the two sheets of the mirror curve.

    \item As can be seen in \autoref{fig:test1} and in \autoref{sec:plots}, when evaluated on-shell, the two terms in \eqref{tsstopenf} have the same real part along the real line, which is pole-free, while having opposite imaginary parts, with poles that cancel in the sum.

    \item An important part of the statement is that the sums in \eqref{tsstopenf} are convergent. The instanton expansion of $\rJ_\text{inst}$ and the sum over the shifts $k \in \integers$ are both essentially expansions in $\exp\br{-t_B\br{\hbar}}$, and the mirror map $t_B\br{\hbar}$ is itself given as an expansion in large $\kappa$. Hence, one can think of this as a convergent semi-classical expansion. For $\hbar \in 2 \pi \prationals$ numerical evidence suggests that the convergence of this series is compact, that is uniform on every compact subset of the complex $x$-plane, but only pointwise near complex infinity.
    Similar properties seem to hold for $\hbar \in 2 \pi \br{\preals \setminus \prationals}$ as well.

\end{itemize}

Let us end with a few words about the evidence in favour of \eqref{tsstopenf}. Many tests that do not directly involve the $\sigma = 2$ term as written in \eqref{eq:sigmas} were done in \cite{Marino:2016rsq, Marino:2017gyg}. Let us in particular mention the closed form solutions they wrote down for $\xi = 0$, $\hbar = 2 \pi$, for which one can explicitly check the pole cancellation. Strong numerical evidence for our proposal \eqref{tsstopenf} comes from diagonalizing the quantum mirror curve in the basis of the harmonic oscillator as in \cite{Huang:2014eha}, and comparing the resulting numerical eigenfunctions with \eqref{tsstopenf}. See also the explanation at the beginning of \autoref{sec:plots}. Another non-trivial test of \eqref{tsstopenf} comes from the so-called standard and dual four-dimensional limits. These are interesting stories in their own right, which we will discuss next.

\section{Four-dimensional limits}
\label{sec:sec4}

It is well-known that topological string theory on local CY manifolds can be used to geometrically engineer four-dimensional $\mathcal{N}=2$ theories \cite{kkv, Klemm:1996bj}.
In the case of a local $\mathbb{F}_0$ geometry, the corresponding four-dimensional gauge theory is $\mathcal{N}=2$, SU(2) SYM. In this section, we study the four-dimensional limit of the open TS/ST correspondence. There are two distinct four-dimensional limits one can implement: the standard limit, discussed in \autoref{sec:sd4d}, and the dual limit, presented in \autoref{sec:dual4d}.

\subsection{The standard four-dimensional limit}\label{sec:sd4d}

Let us consider the mirror curve for local $\mathbb{F}_0$ \eqref{eq:MC}. In the standard 4d limit, the parameters of the curve scale as \cite{kkv}
\be\label{eq:4d}
x= R {x}_{\rm 4d}\, ,
\qquad \re^{2 \xi}= \frac{1}{\sqrt{t} R^2} \, ,
\qquad \kappa = \frac{1}{\sqrt{t}R^2} \left(-2-R^2 E\right) \, ,
\qquad \hbar = R \epsilon \, ,
\ee
and the limit is taken as $R \to 0$. In this limit, \eqref{eq:QuantizedMirrorCurve} becomes the Fourier transformed modified Mathieu operator
\be
\label{mat1o}
{\rm O}_{\rm FMa}= \sqrt{t} \left(\re^{\hat y} + \re^{-\hat y }\right)+\hat{x}^2
\, ,
\qquad \qquad
[\hat x, \hat y]=\ri \epsilon
\, ,
\qquad \qquad
t, \epsilon > 0
\, ,
\ee
where we omit the subscripts $\rm 4d$ in the variable $x$ for the sake of notation. The corresponding eigenvalue equation reads
\be  \label{eq:math}
\sqrt{t} \left(\phi(x - \ri \epsilon, E)+\phi(x + \ri \epsilon, E)\right)+{x}^2\phi(x, E)- E \phi(x, E) = 0 \, .
\ee
If we perform a Fourier transform on \eqref{mat1o}, i.e.~we exchange position and momentum operator, we obtain the modified Mathieu operator in the standard form
\be  \label{eq:mathN} \left(\sqrt{t} \left(\re^{ q} +\re^{- q }\right)-\epsilon^2\partial_q^2- E\right)\widehat{\phi}(q, E)=0 \, .\ee
The eigenfunctions of \eqref{eq:math} and \eqref{eq:mathN} are related by a Fourier transform,
\be \label{eigft}\widehat{\phi}(q, E) = \int_{\IR} \rd x \, \re^{\ri q x / \epsilon} \, \phi(x, E) \, . \ee
The quantization condition for the spectrum of \eqref{mat1o} is derived analogously to the one for \eqref{eq:QuantizedMirrorCurve}, i.e.~by imposing analytic continuation within the strip \eqref{qctopco}, and requiring square integrability of $\phi(x, E)$. On the other hand, for the Fourier transformed operator \eqref{eq:mathN}, this corresponds to square integrability of $\widehat{\phi}(q, E)$. The energy spectrum is then discrete and the same for both operators.

For the sake of notation, we work at  $\epsilon=1$. The $\epsilon$-dependence can be reinstated by appropriately shifting the variable as follows:
\begin{itemize}
    \item For the modified Mathieu operator \eqref{eq:mathN}, we can work at $\epsilon=1$ and then re-install the $\epsilon$ dependence by shifting
    \begin{equation}
        t \to t / \epsilon^4
        \, ,
        \qquad \qquad
        E \to E / \epsilon^2
        \, .
    \end{equation}
    \item For the Fourier transformed modified Mathieu \eqref{mat1o} \eqref{eq:math}, we can work at $\epsilon=1$ and then re-install the $\epsilon$ dependence by shifting
    \begin{equation}
    \label{eq:epsidep}
        x \to  x / \epsilon
        \, ,
        \qquad \qquad
        t \to t / \epsilon^4
        \, ,
        \qquad \qquad
        E \to E / \epsilon^2
        \, .
    \end{equation}
\end{itemize}

\subsubsection{Result}

The standard four-dimensional limit \eqref{eq:4d} was examined in the context of the closed TS/ST correspondence in \cite{Hatsuda:2015qzx, ggm, Grassi:2018bci}.
In \cite{ggm} it was shown that in this limit \eqref{eq:detimain} gives
\begin{equation}
    \label{eq:matdet}
    \det\left(1 - E \, \mathrm{O}^{-1}_{\rm FMa} \right) = A(t) \br{\frac{\sinh{\br{\frac{\ri}{2} \partial_\sigma F_\mathrm{NS}^{\rm 4d}{\br{\sigma, t}}}}}{\ri \sinh{\br{2 \pi \sigma}}}} \, ,
\end{equation}
where $A(t)$ is a normalization constant independent of $E$, chosen such that the left-hand side, when evaluated at $E = 0$, equals $1$. In \eqref{eq:matdet}, we denote by $F_\mathrm{NS}^{\rm 4d}$ the full, four-dimensional NS free energy associated with $\mathcal{N}=2$, $\mathrm{SU}\br{2}$ SYM
\begin{multline}
    \label{eq:FreeEnergy4dNS}
    F_\mathrm{NS}^{\rm 4d}\br{\sigma, t} =
    - \psi^{(-2)}{\br{1 - \ri 2 \sigma}}
    - \psi^{(-2)}{\br{1 + \ri 2 \sigma}}
    - \sigma^2 \ln{\br{t}}
    \\
    - \br{\frac{2}{4 \sigma^2 + 1}} t
    - \br{\frac{20 \sigma^2 - 7}{4\br{4 \sigma^2 + 1}^3 \br{\sigma^2 + 1}}} t^2
    + \bigO{t^3}
    \, ,
\end{multline}
where $\psi^{\br{-2}}$ is the polygamma function of order $-2$. Higher orders in the $t$ expansion can be found in \eqref{eq:F_4d_NS_NLO} and according to the all order expression \eqref{eq:F_4d_NS}.
The variable $\sigma$ and the energy $E$ are related via the quantum Matone relation \cite{matone,francisco}
\be
\label{eq:quantumm}
E = -t \partial_t F_{\rm NS}^{\rm 4d}\left(\sigma, t \right)
\, ,
\qquad \qquad
{2 \sigma \notin \ri \IZ \setminus \cbr{0}} \, .
\ee
This relation is the 4d limit of the Wilson loop \eqref{WL}, and is hence the 4d equivalent of the quantum mirror map, relating the gauge theory and spectral parameters to each other.
The quantization condition for the operator spectrum, determined by the vanishing of the determinant \eqref{eq:matdet}, exactly reproduces the NS quantization condition \cite{ns}:
\be
\label{eq:qc4d}
\partial_\sigma F_{\rm NS}^{\rm 4d}(\sigma, t) = 2\pi (n+1)
\, ,
\qquad \qquad
n \in \mathbb{N}
\, ,
\ee
see also \cite{Hollands:2019wbr,Hollands:2021itj} and reference therein.
For a fixed value of $t$, we denote by  \(\left\{\sigma_n\right\}_{n \geqslant 0}\) the solution to \eqref{eq:qc4d}. These give the energy spectrum of the operator \eqref{eq:math} via the quantum Matone relation \eqref{eq:quantumm}.

Implementing the four-dimensional limit \eqref{eq:4d} on the eigenfunction expression appearing on the right-hand side of \eqref{tsstopenf} gives (see \autoref{sec:deriv4d})
\begin{equation}
\label{eq:4dlimeig}
    \sum_{k \in \IZ}
    \br{\re^{ {\rm J}(x,\mu + \ri 2 \pi k , \xi,\hbar)}
    +  \re^{
    \frac{\ri}{\hbar} \frac{\pi^2}{2} + \frac{\pi x}{\hbar}
    + {\rm J}(-x - \ri \pi, \mu + \ri \pi + \ri 2 \pi k , \xi , \hbar)}}
    \to  \phi_1(x, \sigma,t)+\phi_2(x,\sigma,t)
    \, ,
\end{equation}
where\footnote{The overall normalization $ - \ri \exp\left(\frac{\ri}{4} \partial_\sigma F_\mathrm{NS}^{\mathrm{4d}} \left( \sigma, t \right)\right)$  was added so that the on-shell eigenfunctions are real.}
\begin{equation}
\label{eq:s12}
    \boxed{
    \phi_1(x,\sigma,t) =
    - \ri \exp\left(\frac{\ri}{4} \partial_\sigma F_\mathrm{NS}^{\rm 4d}\left( \sigma, t \right)\right)
    t^{- \ri \frac{x}{2}} \Gamma\left( \ri \left( x + \sigma \right) \right) \Gamma\left( \ri \left( x - \sigma \right) \right) Z^{\rm 2d/4d}_\mathrm{NS, inst} \left( - x , \sigma, t \right)
    \, , }
\end{equation}
\begin{equation}
\label{eq:s11}\boxed{
    \phi_2(x, \sigma, t)=
    \phi_1(-x, \sigma, t)
    \left[ \frac{ \re^{- \frac{\ri}{2} \partial_\sigma F_\mathrm{NS}^{\rm 4d}\left( \sigma, t \right) }\left( \re^{2 \pi x} - \re^{2 \pi \sigma}  \right) - \re^{\frac{\ri}{2} \partial_\sigma F_\mathrm{NS}^{\rm 4d}\left( \sigma, t \right) }\left( \re^{2 \pi x} - \re^{- 2 \pi \sigma}  \right)}{\re^{2 \pi \sigma} - \re^{-2 \pi \sigma}} \right]
    \, ,}
\end{equation}
with $Z^{\mathrm{2d/4d}}_\mathrm{NS, inst}$ denoting the instanton part of the NS function in the presence of a surface defect
\begin{equation}
    Z^\mathrm{2d/4d}_\mathrm{NS, inst}\br{x,\sigma,t}
    =
    1 + \squarebr{\frac{ 1 + 2 (\ri x+1)}{\left(1+4 \sigma ^2\right) \left((\ri x+1)^2+\sigma ^2\right)}} t
    + \bigO{t^2}
    \, .
\end{equation}
Higher-order terms are provided in \eqref{eq:Z_2d-Defect_NS_NLO}, while the full all-order definition is given in \eqref{eq:Z_Defect_NS}.
We obtain then the following eigenfunction of \eqref{eq:math}
\begin{equation}
\label{eq:matsum}
    \boxed{\phi(x, E, t) = \phi_1(x, \sigma, t)+\phi_2(x, \sigma, t) \, , }
\end{equation}
where $E$ and $\sigma$ are related as in \eqref{eq:quantumm}.
Let us make some comments on the above result.
\begin{itemize}
    \item The finite difference equation \eqref{eq:math} has extensive families of formal solutions. For instance, both functions \eqref{eq:s11} and \eqref{eq:s12} are solutions to the Fourier transform Mathieu equation \eqref{eq:math}. Each of these functions is meromorphic, with poles located at \( x = \pm \sigma + \ri  \ell \), where  $\ell \in \mathbb{Z}$. However, what makes the solution \eqref{eq:matsum} special is that in the summation, all poles cancel, yielding a final expression that is entire in $x$, even when evaluated at generic values of the energy. The factor in the square brackets in \eqref{eq:s11} is crucial for this to happen.
    \item Although the symmetric structure of the two contributions from
    \eqref{eq:ProposedStructureEigenFun} is lost in the 4d limit, the key feature that remains is
    that only the sum \eqref{eq:matsum} of these two contributions is  entire
    in $x$.
    \item When we evaluate the above eigenfunctions \eqref{eq:matsum}, \eqref{eq:s11}, \eqref{eq:s12} on-shell, i.e.~on the locus \eqref{eq:qc4d}, we obtain
    \begin{multline}
    \label{eq:4dons}
        \phi(x, E_n, t) =
        \br{-1}^n \re^{\ri \frac{\pi}{2} n}
        t^\frac{\ri x}{2} \Gamma\left( \ri \left( - x + \sigma_n \right) \right) \Gamma\left( \ri \left( - x - \sigma_n \right) \right)  Z^{\mathrm{2d/4d}}_\mathrm{NS, inst} \left( x , \sigma_n, t \right)
        \\
        + \re^{\ri \frac{\pi}{2} n} t^{-\frac{\ri x}{2}}  \Gamma\left( \ri \left( x + \sigma_n \right) \right) \Gamma\left( \ri \left( x - \sigma_n \right) \right) Z^{\mathrm{2d/4d}}_\mathrm{NS, inst} \left( - x , \sigma_n, t \right)
        \, ,
    \end{multline}
    which is the well-known form of the eigenfunctions obtained in \cite{Kozlowski:2010tv, Sciarappa:2017hds, Kanno:2011fw, Jeong:2018qpc, Jeong:2017pai}. Notice that the relation between the two terms in \eqref{eq:4dons} once again exhibits the same symmetry structure as in \eqref{eq:ProposedStructureEigenFun}.

    \item Let us now consider the asymptotic behaviour of \eqref{eq:s11} and \eqref{eq:s12}   for $\real{x} \to \pm \infty$ with constant $\imaginary{x}$. The non-trivial asymptotics are determined by the part involving $\Gamma$ functions as well as the term inside the square brackets in \eqref{eq:s11}.
    For $\phi_1$, we get an exponential decay in both directions
    \begin{equation}
          \abs{\phi_1\br{x + \ri y, \sigma, t}}
          \propto
          \re^{\mp \pi x - \br{1 + 2 y} \ln\br{\pm x}}
          \br{1 + \bigO{\frac{1}{x}}} \, ,
          \qquad \qquad
          x \to \pm \infty
          \, ,
    \end{equation}
    for any constant $y \in \reals$.
    For $\phi_2$, we have instead
    \begin{multline}
        \label{eq:asy2}
        \abs{\phi_2\br{x + \ri y, \sigma, t}}
        \propto
        \\
        \begin{dcases}
            \abs*{\frac{\sinh[\frac{\ri}{2} \partial_\sigma F_\mathrm{NS}^{\rm 4d}{\br{\sigma, t}}]}{\ri \sinh[2 \pi \sigma]}}
            \, \re^{+ \pi x - \br{1 - 2 y} \ln\br{x}}
            \br{1 + \bigO{\frac{1}{x}}} \, ,
            &
            x \to + \infty
            \\
            \abs*{\frac{\sinh[\frac{\ri}{2} \partial_\sigma F_\mathrm{NS}^{\rm 4d}{\br{\sigma, t}} - 2 \pi \sigma]}{\ri \sinh[2 \pi \sigma]}}
            \, \re^{+ \pi x - \br{1 - 2 y} \ln\br{-x}}
            \br{1 + \bigO{\frac{1}{x}}} \, ,
            &
            x \to - \infty
        \end{dcases}
    \end{multline}
    for any constant $y \in \reals$. The overall trigonometric terms come  from
    the factor in square brackets in \eqref{eq:s11}.
    Hence, the complete eigenfunction is square-integrable if and only if
    \begin{equation}
        \frac{\sinh[\frac{\ri}{2}\partial_\sigma F_\mathrm{NS}^{\rm 4d}\br{\sigma, t}]}{\ri \sinh[2 \pi \sigma]}=0
        \, ,
    \end{equation}
    which is precisely the vanishing of the Fredholm determinant \eqref{eq:matdet}.\footnote{Note that each of the two terms themselves are never square-integrable for all $y \in \reals$ due to the simple poles at $x + \ri y = \pm \sigma + \ri k$ for $k \in \mathbb{Z}$.}
\end{itemize}

\subsubsection{Derivation}\label{sec:deriv4d}

Here we derive \eqref{eq:4dlimeig}.
We are interested in the standard 4d limit \eqref{eq:4d} of the eigenfunctions as they appear in terms of the gauge theory grand potential in \eqref{tsstopenf}. The scaling of the complex structure parameters as given in \eqref{eq:4d} corresponds to the following scaling of the Kähler parameters
\begin{equation}
\label{eq:Standard4DLimit_KahlerParameters_Scaling}
    t_B\br{\hbar} = - {\ln\br{t \hbar^4}} + 2 \sigma \hbar
    \, ,
    \qquad \qquad
    t_F\br{\hbar} = 2 \sigma \hbar
    \, ,
\end{equation}
and in addition, the negative sign of $\kappa$ in \eqref{eq:4d} is reflected in a shift
\begin{equation}
\label{eq:Standard4DLimit_KahlerParameters_Shifts}
    t_{B,F}\br{\hbar} \to t_{B,F}\br{\hbar} - \ri 2 \pi
    \, .
\end{equation}
The standard 4d limit is then taking $\hbar \to 0$ from above while keeping $x_{\rm 4d}$, $\sigma$ and $t$ fixed with $\real{\sigma}, \, t > 0$.

We will not keep track of the overall normalization of the 5d eigenfunctions, since we didn't fix it in the first place. Hence, we can freely choose an overall normalization, and it turns out that normalizing the eigenfunctions \eqref{tsstopenf} as
\begin{equation}
\label{eq:Eigenfunctions5DForStandard4DLimit}
    \re^{-\rJ\br{\mu, \xi, \hbar}}
    \sum_{k \in \integers} \br{\re^{{\rJ\br{x, \mu + \ri \pi \br{2 k - 1}, \xi, \hbar}}} +
    \re^{\frac{\ri}{\hbar} \frac{\pi^2}{2} + \frac{\pi x}{\hbar} + {\rJ\br{- x - \ri \pi, \mu + \ri \pi \br{2 k}, \xi, \hbar}}}}
    \, ,
\end{equation}
will be a convenient choice.\footnote{Interchanging the sum over $k$ with the limit is quite subtle, since the sum is not always uniformly converging in $x$. However, we will not go into this issue and simply note that interchanging the sum and limit in this case gives a perfectly well-behaved answer in line with our expectations.}

Let us first look at the closed sector, closely following \cite{Hatsuda:2015qzx,ggm}.
One gets for the polynomial part of the closed grand potential \eqref{eq:J5DClosedPolynomial}
\begin{multline}
    \rJ_\rp\br{\mu + \ri \pi \ell, \xi , \hbar}
     -  \rJ_\rp{\br{\mu, \xi ,  \hbar}}
    =
    \\
    \frac{\pi \ell^2 \ln{\br{t \hbar^4}}}{2 \hbar}
    - \ri \frac{\pi^2 \ell \br{2 \ell^2 - 1}}{3 \hbar}
    - \ri \ell \sigma \ln{\br{t \hbar^4}}
    - 2 \pi \ell^2 \sigma
    + \bigO{\hbar}
    \, ,
\end{multline}
where $\ell = 2 k - 1$ for the first term and $\ell = 2 k$ for the second term.
Hence, one can see that the dominating contributions in the sum over the shifts $k \in \mathbb{Z}$ are given by $k = 0, 1$ for the first term and by $k = 0$ term for the second term. One can already note that this gives a trivial closed sector contribution for the second term, due to our normalization.
The 4d limit for the 1-loop and instanton part of the closed grand potential can be dealt with as done in \cite{Hatsuda:2015qzx,ggm} and one finds
\begin{multline}
    \rJ{\br{\mu \pm \ri \pi, \xi, \hbar}}
    - \rJ{\br{\mu , \xi , \hbar}}
    =
    \\
    \frac{\pi}{2} \frac{\ln( t \hbar^4 )}{\hbar} \mp \ri \frac{\pi}{2}
    \pm \frac{\ri}{2} \partial_\sigma F_\mathrm{NS}^{\rm 4d}{\br{t , \sigma}}
    - \ln{\br{\re^{2 \pi \sigma} - \re^{- 2 \pi \sigma}}}
    + \bigO{\hbar}
    \, .
\end{multline}
where the 4d NS free energy is given in \eqref{eq:FreeEnergy4dNS}.

Let us now turn our attention to the open sector. The polynomial part of the open grand potential \eqref{eq:J5DOpenPolynomial} can be dealt with straightforwardly.
To take the standard 4d limit on the 1-loop part \eqref{eq:J5DOpen1Loop} it is useful to first use the quasi-periodicity of the quantum dilogarithm \eqref{eq:NCQDiLogQuasiPerRepeated} to write\footnote{When implementing the limit we assume for simplicity \( 2 \abs{\imaginary{x \pm \sigma}} < 1 \). However, the final result extends to all values of $x$ in the complex plane.}
\begin{multline}\label{eq:lim1}
    \rJ^\mathrm{open}_\text{1-loop}{\br{x, \mu \pm \ri \pi , \xi , \hbar}}
    =
    - 2 \ln\br{\hbar}
    - \ln{\br{x_{\rm 4d}^2 - \sigma^2}}
    - 2 \pi x_{\rm 4d}
    + \ln{\br{\re^{2 \pi x_{\rm 4d}} - \re^{\mp 2 \pi \sigma}}} +
    \\
    \ln{\Phi_\beta\squarebr{\beta \br{\sigma - x_{\rm 4d}} + \frac{\ri}{2} \br{\beta^{-1} - \beta}}}
    + \ln{\Phi_\beta\squarebr{\beta \br{- \sigma - x_{\rm 4d}} + \frac{\ri}{2} \br{\beta^{-1} - \beta}}} + \bigO{\hbar} \, ,
\end{multline}
\begin{multline}\label{eq:lim2}
    \rJ^\mathrm{open}_\text{1-loop}{\br{- x - \ri \pi, \mu , \xi, \hbar}}
    =
    - 2 \ln\br{\hbar}
    - \ln{\br{x_{\rm 4d}^2 - \sigma^2}} +
    \\
    \ln{\Phi_\beta\squarebr{\beta \br{\sigma + x_{\rm 4d}} + \frac{\ri}{2} \br{\beta^{-1} - \beta}}}
    + \ln{\Phi_\beta\squarebr{ \beta \br{- \sigma + x_{\rm 4d}} + \frac{\ri}{2} \br{\beta^{-1} - \beta}}}  + \bigO{\hbar} \, ,
\end{multline}
with $\beta = \sqrt{\hbar / 2 \pi}$. In the end, we are interested in the exponential of the grand potential, so all equalities are modulo integer multiples of $\ri 2 \pi$.
Let us introduce a variable $z$ which is
\begin{equation}
    z = \frac{1}{2} \pm \ri \br{x_{\rm 4d} \pm \sigma} \, ,
\end{equation}
with the signs chosen independently and $2 \abs{\imaginary{x_{\mathrm{4d}} \pm \sigma}} < 1$ or $0 < \real{z} < 1$.
To compute the expansion of the quantum dilogarithms in \eqref{eq:lim1} and \eqref{eq:lim2}, we use the integral representations in \eqref{eq:NonCompQuantDiLog:IntRepsBLimit} and \eqref{eq:LogGamma}.
We find that
\begin{multline}
    \ln{\Phi_\beta{\br{\frac{\ri}{2} \beta^{-1} - \ri \beta z + \bigO{\beta^3} }}}
    =
    \\
    - \ri \frac{\pi}{12\beta^2} + z \ln{\br{2\pi \beta^2}} - \frac{\ln{\br{2 \pi}}}{2} + \ri \frac{\pi}{2} z + \ln{\Gamma{\br{z + \frac{1}{2}}}}
    + \bigO{\beta^2} \, .
\end{multline}
Combining the polynomial and 1-loop parts of the grand potential one finds
\begin{multline}
    \rJ^\mathrm{open}_\mathrm{p}{\br{x, \xi, \hbar}}
    + \rJ^\mathrm{open}_\text{1-loop}\left(x, \mu \pm \ri \pi , \xi, \hbar\right)
    =
    - \frac{\ri \pi^2}{3 \hbar} - \ln \left( 2 \pi \hbar \right)
    - \ri \frac{\pi}{2}
    \\
    + \ln\Gamma\left( \ri \left( - x_{\rm 4d} + \sigma \right) \right) + \ln\Gamma\left( \ri \left( - x_{\rm 4d} - \sigma \right) \right) + \ri \frac{x_{\rm 4d}}{2} \ln{\br{t}} + \ln{\br{\re^{2 \pi x_{\rm 4d}} - \re^{\mp 2 \pi \sigma}}}
    + \bigO{\hbar}
\end{multline}
\begin{multline}
    \frac{\ri}{\hbar} \frac{\pi^2}{2} + \frac{\pi}{\hbar} x + \rJ^\mathrm{open}_\mathrm{p}{\br{  - x - \ri \pi , \xi , \hbar}}
    + \rJ^\mathrm{open}_\text{1-loop}\left( - x - \ri \pi , \mu , \xi, \hbar \right)
    =
    \frac{\pi}{2}\frac{\ln{\br{t \hbar^4}}}{\hbar}- \frac{\ri \pi^2}{3 \hbar} - \ln \left( 2 \pi \hbar \right)
    \\
    + \ln\Gamma\left( \ri \left( x_{\rm 4d} + \sigma \right) \right) + \ln\Gamma\left( \ri \left( x_{\rm 4d} - \sigma \right) \right) - \ri \frac{x_\mathrm{4d}}{2} \ln{\br{t}}
    + \bigO{\hbar} \, .
\end{multline}
Regarding the instanton part for the open sector, one can see that the NS part is a rational function of $\exp\br{x - t_F / 2}$, $\exp{\br{-t_F}}$ and $\exp{\br{-t_B}}$. Hence, the shifts of $t_{B, F}$ by $\ri 4 \pi k$ act trivially, and the only difference between the first and second term is a change in the sign of $x$. The GV part, on the other hand, vanishes in the standard 4d limit. The resulting defect instanton partition function can be found in \eqref{eq:Z_2d-Defect_NS_NLO}.

Hence, putting all parts of the grand potential for both the closed and open sectors together gives an overall divergent factor
\begin{equation}
    \frac{\pi}{2} \frac{\ln\left(t \hbar^4\right)}{\hbar} - \frac{\ri \pi^2}{3 \hbar}  - \ln\left( 2 \pi \hbar \right) \, ,
\end{equation}
so that, after appropriate normalization, \eqref{eq:Eigenfunctions5DForStandard4DLimit} reduces to \eqref{eq:s11}, \eqref{eq:s12} and \eqref{eq:matsum} in the standard 4d limit \eqref{eq:Standard4DLimit_KahlerParameters_Scaling}.

\subsection{The dual four-dimensional limit}\label{sec:dual4d}

A few years ago, \cite{bgt,bgt2} showed that, starting from the quantum mirror curve \eqref{eq:spectralc}, one can implement another scaling limit to connect with the four-dimensional gauge theory. In this limit, we take
\be
\label{eq:dual4d}
4\xi = 4 \pi \ri \sigma - \frac{2\pi}{R} \log \left(R^4 t\right)
\, ,
\qquad \qquad
\re^{-2 \xi} \kappa = 2 \cos(2\pi \sigma)
\, ,
\qquad \qquad
\hbar = \frac{4\pi^2}{R}
\, ,
\ee
and send $R\to 0$ from above.
One of the key differences between the limits \eqref{eq:4d} and \eqref{eq:dual4d} is that in \eqref{eq:4d}, we take $\hbar \to 0$, whereas in \eqref{eq:dual4d}, we take $\hbar \to \infty$. Hence, we refer to \eqref{eq:dual4d} as the ``dual'' four-dimensional limit.

Applying the scaling  \eqref{eq:dual4d} and taking the limit $R\to 0$ on the operator kernel \eqref{eq:rhok} yields the  integral operator $\rho_{\rm GV }:~L^2(\IR)\to L^2(\IR)$ with kernel \cite{bgt}
\begin{equation}
\label{eq:4dop}
    \rho_{\rm GV}\br{q, p} = \frac{\re^{-4 t^{1/4}\cosh(q)} \, \re^{-4 t^{1/4}\cosh(p)}}{\cosh\br{\frac{q-p}{2}}} \, ,
\end{equation}
and we look for square-integrable eigenfunctions
\be
\label{eq:speceig} \int_{\IR }\rd p \ \rho_{\rm GV}(q, p)\widehat{\varphi}_{n}(p, \widehat{E}_n, t)=\widehat{E}_n \widehat{\varphi}_n(q, \widehat{E}_n, t) \, .
\ee
Interestingly, \eqref{eq:4dop} first appeared in the literature in the 1970s in studies of the 2d Ising model and the theory of Painlev\'{e} equations, see \cite{wu1, zamo, mccoy2011romance, gm}. We will refer to it as  McCoy-Tracy-Wu operator. The connection between the quantum mirror curve \eqref{eq:spectralc} and the Painlev\'{e} kernel \eqref{eq:4dop} made it possible to prove the TS/ST correspondence for local $\IF_0$ in this particular dual 4d limit \cite{bgt}. See also \cite{Gavrylenko:2023ewx} for the generalization to all $Y^{N,0}$ geometries.

The spectral problem \eqref{eq:speceig} was solved in \cite{Francois:2023trm}, where it was shown that the eigenfunctions can be explicitly computed using the partition function of four-dimensional, $\mathcal{N} = 2$, $\mathrm{SU}\br{2}$ supersymmetric Yang-Mills theory in the GV (or self-dual) phase of the $\Omega$-background ($- \epsilon_1 = \epsilon_2 = 1$), with the inclusion of a surface defect. More precisely we have \cite{Francois:2023trm}
\begin{multline}
    \label{eiged}
    \widehat \varphi(q, \widehat E_n  , t)
    = \int_{\IR} \rd x \ \re^{\ri {2} q x}\sum_{k \in \IZ} \left( Z^{\text{2d/4d}}_{\rm GV}\left(x , k + \frac{1}{2} + \ri\widehat\sigma_n  , t  \right) Z^{\rm 4d}_{\rm GV}\br{ k + \frac{1}{2} +  \ri \widehat\sigma_n  , t} \right.
    \\
    \left.+ Z^{\text{2d/4d}}_{\rm  GV}\left(- x - \frac{\ri}{2}, k + \ri \widehat\sigma_n , t \right)  Z^{\rm 4d}_{\rm GV}\br{ k + \ri \widehat\sigma_n  , t } \right)
    \, ,
\end{multline}
\be
\label{eq:McCoyTracyWu:RelationSpectrumSigma}
\widehat{E}_n = 2 \pi \sech[2 \pi \widehat{\sigma}_n]
\, ,
\ee
where the gauge theory partition function of the defect is given by
\be \label{ZIItot} Z^{\mathrm{2d/4d}}_{\rm GV}(x, \sigma  , t) =  t^{\ri \frac{x}{2}}
\Gamma \left(- \mathrm{i} x-\sigma + \frac{1}{2}\right) \Gamma \left(- \mathrm{i} x + \sigma + \frac{1}{2}\right) Z^{\rm 2d/4d}_{\rm GV, inst}(x, \sigma, t) \, ,
\ee
\begin{multline}
\label{def21}
    Z^{\rm 2d/4d}_{\rm GV, inst} \left( x, \sigma  , t \right)
    =
    1 - \left[ \frac{\widetilde{x}}{2 \sigma^2 \left(\widetilde{x}^2 - \sigma^2 \right)}\right] t
    \\
    + \left[ \frac{\widetilde{x} \left( \widetilde{x} + 1 \right)^2 - \widetilde{x} \left( 10 \widetilde{x}^2 + 19 \widetilde{x} + 10 \right) \sigma^2 + \left(  8 \widetilde{x}^2 + 30 \widetilde{x} + 9 \right) \sigma ^4}{4 \sigma^4 \left( 4 \sigma^2 - 1 \right)^2 \left(\widetilde{x}^2 - \sigma^2\right) \left( \left( \widetilde{x} + 1 \right)^2 - \sigma^2\right)} \right] t^2
   +\mathcal{O}\left(t^3\right) \, ,
\end{multline}
with $\widetilde{x}=\ri x + 1/2$ and higher orders in the $t$ expansion can be found from the definition \eqref{eq:Z_2d4d_SelfDual}. The
$\widehat\sigma_n \in \mathbb{R} \setminus \left\{ 0 \right\}$ are solutions to
\be\label{QC2} \sum_{k \in \IZ} Z^{\rm 4d}_{\rm GV}\left(k + \frac{1}{2} + \ri \widehat\sigma_n  , t \right) = 0
\, ,
\ee
and $Z^{\rm 4d}_{\rm GV}(\sigma, t)$ is the Nekrasov function in the GV-phase of the $\Omega$-background:
\be \label{nek4d} Z^{\rm 4d}_{\rm GV}(\sigma, t) = \frac{t^{\sigma^2}}{G(1-2\sigma)G(1+2\sigma)} \left(1+\frac{t}{2 \sigma ^2}+ \frac{\left(8 \sigma ^2+1\right) t^2}{4 \sigma ^2 \left(4 \sigma ^2-1\right)^2}+\mathcal{O}(t^3)\right) \, , \ee
with $G$ the Barnes G-function, and higher orders in the instanton expansion can be found in \eqref{eq:Z_4d_SD_NLO} and according to \eqref{eq:Z_2d4d_SelfDual}.
The Fourier transform in \eqref{eiged} can be interpreted exactly as in \eqref{eigft}. Indeed, we can represent the kernel \eqref{eq:4dop} in operator form as
\be {\re^{-4 t^{1/4}\cosh \hat q}\frac{1}{\cosh\left(\frac{\hat p}{2}\right)}\re^{-4 t^{1/4}\cosh \hat q}}
\, ,
\qquad \qquad
[\hat q, \hat p]=\ri 2\pi \, .\ee
By exchanging momentum and position operator according to the transformation
\begin{equation}
\label{eq:candual}
    \begin{pmatrix} \hat{x} \\ \hat{y} \end{pmatrix} =
    \begin{pmatrix} 0 & \frac{1}{4\pi} \\ -{4 \pi} & 0 \end{pmatrix}
    \begin{pmatrix} \hat{q} \\ \hat{p} \end{pmatrix}
    \, ,
\end{equation}
we obtain the Fourier transformed operator whose eigenfunctions are given by
\begin{multline}
    \label{eq:phisdft}
    {\varphi}(x, \widehat{E}_n  , t) =
    \sum_{k \in \IZ} \left( Z^{\mathrm{2d/4d}}_{\rm GV}\left(x, k + \frac{1}{2} +  \ri \widehat\sigma_n , t \right) Z^{\rm 4d}_{\rm GV}\br{k + \frac{1}{2} + \ri\widehat\sigma_n , t } \right.
    \\
    \left.+ Z^{\mathrm{2d/4d}}_{\rm  GV}\left(- x - \frac{\ri}{2}, k + \ri \widehat\sigma_n , t \right)  Z^{\rm 4d}_{\rm GV}\br{ k + \ri \widehat\sigma_n  , t } \right)
    \, .
\end{multline}

\subsection{Relating modified Mathieu to McCoy-Tracy-Wu} \label{sec:relation}

From the preceding discussion, we see that:
\begin{enumerate}
    \item The eigenfunctions and spectrum of the (Fourier transformed) modified Mathieu operator \eqref{eq:mathN} are determined by gauge theory partition functions in the NS phase of the $\Omega$ background.
    \item The eigenfunctions and spectrum of the (Fourier transformed) McCoy–Tracy–Wu operator \eqref{eq:4dop} are determined by gauge theory partition functions in the GV phase of the $\Omega$ background.
\end{enumerate}
It was first shown in \cite{ggu}, based on \cite{huang1606},\footnote{In \cite{huang1606}, the relationship between GV invariants and the NS phase of the $\Omega$-background was used to express the quantization condition of \cite{ghm} in a more symmetric form \cite{Wang:2015wdy}.} that these two phases of the  $\Omega$ background can be related by using the Nakajima-Yoshioka blowup equations \cite{naga}. This relation was extended to partition functions in the presence of surface defects in \cite{Jeong:2020uxz, Nekrasov:2020qcq}.
It is therefore natural to ask whether we can relate the eigenfunctions and the spectra of \eqref{eq:mathN} and \eqref{eq:4dop}.

\subsubsection{The spectrum}

For the modified Mathieu equation the spectrum $E_n$ is given in terms of $\sigma_n$ \eqref{eq:quantumm}, which is a solution of \eqref{eq:qc4d} \cite{ns}, while for the McCoy-Tracy-Wu operator the spectrum $\widehat{E}_n$ is given in terms of $\widehat{\sigma}_n$ \eqref{eq:McCoyTracyWu:RelationSpectrumSigma}, which is a solution of \eqref{QC2} \cite{bgt}.
Using the NS limit of blowup equations without defects, it follows that solutions of \eqref{eq:qc4d} are mapped to solutions of \eqref{QC2}, that is \cite{ggu, Bershtein:2021uts}
\begin{equation}
    \sigma_n = \widehat{\sigma}_n \, .
\end{equation}
This gives a direct, but non-trivial relation between the spectra of the two operators above, namely
\be\label{eq:energyrel}\boxed{\ba
\text{modified Mathieu:} \qquad E_n&= -t \partial_t F^\mathrm{4d}_{\rm NS}\left(\sigma_n, t \right) \\
\text{McCoy-Tracy-Wu:}\qquad \widehat E_n & = 2 \pi \sech[2 \pi  \sigma_n]
\ea} \, . \ee
Note that the relation between the energy and $\sigma$ is much simpler in the McCoy-Tracy-Wu case. The spectra of the two operators are hence related by
\be\label{eq:relationsp} E_n = -t \partial_t F^\mathrm{4d}_{\rm NS}\left(\frac{1}{2\pi} \arcsech[\frac{\widehat{E}_n}{2 \pi}], t \right) \, . \ee
Let us emphasize that much of the derivation of \eqref{eq:relationsp} from \cite{bgt,ggm,Bershtein:2021uts} is fully rigorous, as it is based on several well-established results: the AGT correspondence \cite{Alday:2009aq}, which was later proven in \cite{Fateev:2009aw,Maulik:2012wi}; the Kyiv formula for the Painlev\'{e} tau function \cite{ilt1}, subsequently proven in \cite{Bershtein:2016aef,ilte,Gavrylenko:2016zlf}; the Nakajima–Yoshioka blowup relation derived and proven in \cite{naga}; the special solution to the Painlev\'{e} ${\rm III}_3$ equation constructed in \cite{wu1,Widom:1996ff}; and the convergence of the NS function \cite{Desiraju:2024fmo}. While many of these results were originally conjectural, they have all been proven by now. What is not rigorously established yet is that $\widehat{\sigma}_n = \arcsech(\widehat{E}_n/2\pi)/ 2 \pi$ with $\widehat{E}_n \in  \, ]0, 2 \pi[ $ lies within the radius of convergence of $t \partial_t F^{\mathrm{4d}}_{\mathrm{NS}}\br{\widehat{\sigma}_n, t}$.\footnote{
We recall that \( F^{\mathrm{4d}}_{\mathrm{NS}}(\sigma, t) \) is defined by a power series in \( t \), whose radius of convergence is finite and may depend on \( \sigma \).}

\subsubsection{The eigenfunctions}

One can verify that the modified Mathieu and the McCoy–Tracy–Wu operators commute. Since the modified Mathieu operator possesses a self-adjoint, trace-class inverse, and the McCoy–Tracy–Wu operator is itself self-adjoint and trace-class, it follows that they admit a common $L^2$-orthonormal basis of eigenfunctions.
Indeed, we can also check numerically, that
\be\label{eigen} \boxed{ \frac{\phi (x, E_n, t)}{\phi(x_0, E_n, t)} = \frac{\varphi(x, \widehat E_n, t)}{\varphi(x_0, \widehat{E}_n, t)}}\ee
where $\varphi( x, \widehat{E}_n, t )$ are the Fourier transformed eigenfunctions of the McCoy-Tracy-Wu operator in \eqref{eq:phisdft},  $\phi( x, E_n, t ) $ are the Fourier transformed eigenfunctions of the modified Mathieu operator \eqref{eq:matsum} and  $x_0$ is some arbitrary point which is not a zero of the eigenfunctions.
The equality \eqref{eigen} should follow from taking the NS limit of the blowup equations in the presence of surface defect \cite{Jeong:2020uxz}. A detailed study will appear elsewhere.
Note that for \eqref{eigen} to hold, we must evaluate both sides on-shell, i.e.~at \eqref{eq:energyrel}, \eqref{eq:qc4d}. For generic values of $E$, $\widehat{E}$ \eqref{eigen} does not hold.

\subsubsection{The operators}

Let us now use the relation between the spectrum and the eigenfunctions of the modified Mathieu operator \eqref{eq:mathN} and McCoy-Tracy-Wu operator \eqref{eq:4dop} to find a relation between the operators themselves. To avoid subtleties related to the domain of definition, it is convenient to work with bounded operators.

We denote by $\rho_\mathrm{NS}$ the trace class inverse of the modified Mathieu operator, where the latter is in our local conventions\footnote{There is a factor of 2 multiplying $q$ in \eqref{eq:manor} compared to \eqref{eq:mathN}, which arises from the factor of 2 in the unitary transformation of the eigenfunctions in \eqref{eiged}.}
\begin{equation}
\label{eq:manor}
    - \frac{1}{4}\partial_q^2+\sqrt{t} \br{\re^{2q}+\re^{-2q}} \, .
\end{equation}
As discussed previously, the modified Mathieu operator and $\rho_{\mathrm{GV}}$  admit a common basis of eigenfunctions. Consequently, the same holds for $\rho_{\mathrm{NS}}$  and $\rho_{\mathrm{GV}}$. Hence, by employing their spectral decompositions together with the relation between their spectra given in \eqref{eq:relationsp}, it follows immediately that
\be \label{eq:functional}
\boxed{ \rho_\mathrm{NS} = {\mathfrak{F}}\br{\rho_\mathrm{GV}}
\, ,
}
\ee
\begin{equation}
    \boxed{
    \mathfrak{F}
    \
    :
    \
    \Sigma\br{\rho_\mathrm{GV}} \to \Sigma\br{\rho_\mathrm{NS}}
    \
    :
    \
    E \mapsto \br{- t \partial_t F^\mathrm{4d}_\mathrm{NS}\br{\frac{1}{2 \pi} \arcsech[\frac{E}{2\pi}] , t}}^{-1}
    \, ,
    }
\end{equation}
where $F^\mathrm{4d}_\mathrm{NS}$ is defined by \eqref{eq:FreeEnergy4dNS} and \eqref{eq:F_4d_NS}, $\Sigma\br{\rho} \subset \nnreals$ is the spectrum of $\rho$, and $\mathfrak{F}\br{\rho}$ should be understood in the functional calculus sense \cite[def.~7.13]{Hall_2013}.
Note that $\mathfrak{F}$ is strictly increasing on $\Sigma\br{\rho_\mathrm{GV}}$, hence $\mathfrak{F}^{-1}$ is also well-defined on $\Sigma\br{\rho_\mathrm{NS}}$.

\section{Conclusion and outlook}\label{sec:conc}

In this paper, we formulated the open topological string/spectral theory correspondence for local $\IF_0$, by generalizing \cite{Marino:2016rsq, Marino:2017gyg} away from the self-dual point. Focusing on local $\mathbb{F}_0$, our main result is encapsulated in \eqref{tsstopenf}. From the perspective of topological string theory, the right-hand side provides a non-perturbative, background-independent formulation of the open topological string partition function, which is entire in both the closed string modulus $\kappa$ and the open string modulus $x$. From the viewpoint of the quantum mirror curve, what makes \eqref{tsstopenf} particularly significant is that it provides a solution to the corresponding difference equation \eqref{eq:intro} which is entire, even off-shell. When evaluated on-shell, this solution gives the eigenfunctions of the relativistic two-particle Toda lattice.

We explored the implications of our construction \eqref{tsstopenf} in both the standard \cite{kkv} and dual \cite{bgt} four-dimensional limits, where the quantum mirror curves reduce to the (Fourier-transformed) Mathieu operator \eqref{mat1o} and the McCoy-Tracy-Wu operator \eqref{eq:4dop}, respectively. In the standard 4d limit, our construction provides entire off-shell eigenfunctions of the Fourier-transformed Mathieu operator \eqref{eq:intro}, expressed as special combinations of NS functions in the presence of 2d/4d surface defects, see \eqref{eq:matsum}. When evaluated on-shell, these solutions reproduce the known results \cite{Kozlowski:2010tv, Alday:2010vg, Kanno:2011fw, Jeong:2021rll, Jeong:2018qpc, Jeong:2017pai, Alday:2009fs, Drukker:2009id, Sciarappa:2017hds}.
On the other hand, in the dual 4d limit, our result \eqref{tsstopenf} reproduces the expression of \cite{Francois:2023trm}, where the eigenfunctions of the McCoy-Tracy-Wu operator are obtained through a special combination of 2d/4d surface defects in the GV phase of the background $\Omega$. Notably, we find that the eigenfunctions of the Mathieu and the McCoy-Tracy-Wu operators, when evaluated on-shell, are the same. This gives an explicit functional relation between the two operators, see \eqref{eq:functional}.

Many open questions remain; we summarize some of them below.
\begin{itemize}
    \item  It would be important to understand the geometric meaning of the second term in \eqref{tsstopenf}. This insight would enable a straightforward generalization to all other toric CY threefolds.

    \item Non-perturbative effects in the context of the closed TS/ST correspondence have been analysed from a resurgence perspective in
    \cite{Couso-Santamaria:2016vwq, Hatsuda:2015owa, Gu:2021ize, Alim:2021mhp, Rella:2022bwn, Alim:2022oll, Grassi:2022zuk,Gu:2022fss}, see \cite{mmrev2} for a review and a more exhaustive list of references. It would be interesting to explore the open version of the TS/ST correspondence through the lens of resurgence, particularly the role that the special combinations \eqref{tsstopenf} and \eqref{eq:matsum} may play in the context of exact WKB  \cite{DingleMorgan1968I, DingleMorgan1968II, Kashani-Poor:2016edc, Alim:2022oll, Grassi:2022zuk, DelMonte:2024dcr}, as well as the connection with quantum modularity \cite{Fantini:2024snx}.

    \item In \autoref{sec:relation}, we numerically demonstrated an explicit relation between the on-shell eigenfunctions of the modified Mathieu operator and the McCoy-Tracy-Wu operator. It should be possible to derive this relation analytically using blowup equations in the presence of surface defects. The proof will appear elsewhere.

    \item Over the years, many formal solutions to the functional difference equation \eqref{eq:DifferenceEquation} have been constructed using topological string/gauge theory partition functions. However, most of these proposals are not well-defined for $\hbar\in \preals$ and, moreover, they do not satisfy the analytic properties required for the eigenfunctions of the relativistic Toda lattice, discussed below \eqref{eq:DifferenceEquation}. To our knowledge, the only exceptions are \cite{Marino:2016rsq, Marino:2017gyg, Sciarappa:2017hds}.

    The constructions in \cite{Marino:2016rsq, Marino:2017gyg} are specific to the self-dual point $\hbar = 2\pi$ with $\xi = 0$. Our proposal naturally reduces to theirs when these parameter values are imposed.

    On the other hand, the connection to \cite{Sciarappa:2017hds} is less straightforward.
    A key distinction between \eqref{tsstopenf} and the eigenfunctions in \cite{Sciarappa:2017hds} is that our functions in \eqref{tsstopenf} remain entire even off-shell, whereas those in \cite{Sciarappa:2017hds} exhibit poles at specific values of $x$.
    It would be interesting to understand this better, e.g. via blowup equations.

    \item In the closed version of the TS/ST correspondence, the sum over integers on the right-hand side of \eqref{eq:detintro} has a direct interpretation in the context of  $q$-isomonodromic tau functions \cite{Bonelli:2017gdk}. It would be interesting to explore whether the special combinations of the two terms in \eqref{tsstopenf} carry any particular meaning from the perspective of  $q$-isomonodromic deformations.

    \item Another point for future investigation is the relation to fibre-base duality, i.e. invariance under exchange of \( t_B \) and \( t_F \). This corresponds to the transformation $ \xi, \kappa \to -\xi, \re^{-2 \xi} \kappa$ at the level of the complex moduli. One can verify that the Fredholm determinant \eqref{eq:detintro} remains invariant under this duality.
    It would be interesting to explore how this duality manifests at the level of the special eigenfunctions \eqref{tsstopenf} we constructed.

    \item Finally, it is important to establish a rigorous analytic proof of our results, such as demonstrating that \eqref{tsstopenf} is entire in $x$ for generic values of $\kappa \in \complexes$, $\xi \in \reals$, and $\hbar \in \preals$.

\end{itemize}
We hope to report on some of these topics in the future.

\clearpage

\appendix

\addtocontents{toc}{\protect\setcounter{tocdepth}{1}}

\section{Faddeev's non-compact quantum dilogarithm}
\label{sec:NCQuantDiLog}

Good summaries of the properties of Faddeev's non-compact quantum dilogarithm $\Phi_b \br{z}$ can be found in the appendices of \cite{Garoufalidis:2014ifa,Kashaev:2015kha} and the more comprehensive \cite[app.~A]{EllegaardAndersen:2011vps}.

\subsection{Definition and general properties}

The defining representation of the quantum dilogarithm is often taken to be \cite[eq.~(42)]{EllegaardAndersen:2011vps}
\begin{equation}
    {\Phi_b\br{z}} =
    \exp\br{\frac{1}{4} \int_{\reals + \ri 0} \frac{\re^{- \ri 2 z u}}{{\sinh\br{b u}} {\sinh\br{b^{-1}u}}} \frac{\rd u}{u}}
    \, ,
    \qquad \qquad
    2 \abs{\imaginary{z}} < \abs{\real{b + b^{-1}}}
    \, .
\end{equation}
It can be analytically continued to a meromorphic function of $z$ on the whole complex plane with poles and roots at \cite[eq.~(45)]{EllegaardAndersen:2011vps}
\begin{equation}
\label{eq:poles&rootsNCQDiLog}
    z =
    \begin{cases}
        + \ri \squarebr{\br{k + \frac{1}{2}} b + \br{\ell + \frac{1}{2}} b^{-1}} & \text{poles}
        \\
        - \ri \squarebr{\br{k + \frac{1}{2}} b + \br{\ell + \frac{1}{2}} b^{-1}} & \text{roots}
    \end{cases}
    \, ,
    \qquad \qquad
    k , \ell \in \mathbb{N} \, ,
\end{equation}
and with an essential singularity at complex infinity \cite[p.~34]{EllegaardAndersen:2011vps}. One can determine the order of the poles and roots when $b^2 \in \prationals$, based on \cite[eq.~(21)]{Garoufalidis:2014ifa}, as we do around \eqref{eq:FaddeevQuantumDilogOrderOfRationalPoles}.
The parameter $b$ is in general such that $b^2 \in \mathbb{C} \setminus \mathbb{R}_{\leqslant 0}$, but we are mostly interested in $b^2 > 0$, since this corresponds to $\hbar, g_s > 0$.
The asymptotic behaviour of the quantum dilogarithm is given by \cite[eq.~(46)]{EllegaardAndersen:2011vps}
\begin{equation}
\label{eq:NCQDiLogAsymp}
    \Phi_b(z) \simeq
    \begin{cases}
        \Phi_b^2(0) \exp\left( \ri \pi z^2 \right) \qquad & \Re(z) \gg 1
        \\
        1 \qquad & \Re(z) \ll -1
    \end{cases} \, ,
    \qquad \qquad
    \Re(b) > 0 \, ,
\end{equation}
and the asymptotic behaviour elsewhere in the complex $z$-plane can be found in \cite[eq.~(46)]{EllegaardAndersen:2011vps}.

The non-compact quantum dilogarithm has some important symmetries.
There is an inversion and parity symmetry in $b$ \cite[p.~(34)]{EllegaardAndersen:2011vps},
\begin{equation}
    \Phi_b\left(z\right) = \Phi_{b^{-1}}\left(z\right) = \Phi_{-b}\left(z\right)
\end{equation}
and the following behaviour under a parity transformation for $z$ \cite[eq.~(47)]{EllegaardAndersen:2011vps} \cite[p.~16]{Garoufalidis:2014ifa},
\begin{equation}
\label{eq:NCQuantDiLogParity}
    \Phi_b\left(z\right) \Phi_b\left(-z\right) = \Phi_b^2\left(0\right) \exp\left( \ri \pi z^2 \right) \, ,
    \qquad \qquad
    \Phi_b\left(0\right) = \exp \left( \ri \frac{\pi}{24} \left( b^2 + b^{-2} \right) \right) \, .
\end{equation}
One has furthermore that $\overline{\Phi_b(z)} = 1 / \Phi_b(\overline{z})$ whenever $b \in \mathbb{R}\setminus\left\{0\right\}$ or $|b| = 1$.
Another important property for us is the quasi-periodicity in $z$ \cite[eq.~(48)]{EllegaardAndersen:2011vps} \cite[eq.~(77)]{Garoufalidis:2014ifa},
\begin{equation}
\label{eq:NCQDiLogQuasiPer}
\Phi_b(z + s \ri b^{\pm}) = \left( 1 + \re^{s \ri \pi b^{\pm 2}} \re^{2 \pi b^{\pm} z} \right)^{-s} \Phi_b(z) \, ,
\end{equation}
where $s \in \left\{ \pm 1 \right\}$.
One has similarly
\begin{equation}
\label{eq:NCQDiLogQuasiPerRepeated}
    \Phi_b(z + \ri k b^{\pm}) = \left\{ \prod_{\ell=0}^{\left| k \right| - 1} \left( 1 + \re^{\sgn[k] \ri \pi (2 \ell + 1) b^{\pm 2}} \re^{2 \pi b^{\pm} z} \right)^{- \sgn[k]} \right\} \Phi_b(z) \, ,
    \qquad
    k \in \mathbb{Z}
    \, ,
\end{equation}
by applying $\abs{k}$ times \eqref{eq:NCQDiLogQuasiPer}.

\subsection{Representations and expansions of Faddeev's quantum dilogarithm}

One can express the non-compact quantum dilogarithm in terms of elementary functions and the classical dilogarithm when $b^2 \in \mathbb{Q}_{>0}$ \cite[eqs.~(9),~(11),~(21)]{Garoufalidis:2014ifa},
\begin{equation}
\label{eq:FQDiLogSD}
    {\Phi_{b}\br{z}} =
    \frac{{\exp\squarebr{\frac{\ri}{2 \pi n m} {\dilog{\exp\br{\widetilde{z}}}} + \br{1 + \frac{\ri}{2 \pi n m} \widetilde{z}} {\ln\br{1 - \exp\br{\widetilde{z}}}}}}}
    {{D_m\br{\exp\br{\frac{\widetilde{z}}{m}}; \exp\br{\ri 2 \pi \frac{n}{m}}}} {D_n\br{\exp\br{\frac{\widetilde{z}}{n}}; \exp\br{\ri 2 \pi \frac{m}{n}}}}}
    \, ,
\end{equation}
\begin{equation}
    {D_k\br{X; q}} = {\prod_{\ell = 1}^{k-1} \br{1 - q^\ell X}^{\ell/k}}
    \, ,
    \qquad \quad
    \widetilde{z} = 2 \pi \sqrt{n m} z + \ri \pi \br{n + m} \, ,
    \qquad \quad
    b^2 = \frac{n}{m} \, ,
\end{equation}
where $n , m \in \pnaturals$ are coprime.
It should be noted that this is a representation of the meromorphic quantum dilogarithm in terms of multivalued functions. One can use \eqref{eq:FQDiLogSD} to check that the poles and roots of the quantum dilogarithm \eqref{eq:poles&rootsNCQDiLog} located at
\begin{equation}
\label{eq:FaddeevQuantumDilogOrderOfRationalPoles}
    z =
    \pm \ri \squarebr{\br{k m + r + \frac{1}{2}} b + \br{\ell n + s + \frac{1}{2}} b^{-1}}
    \, ,
    \qquad \qquad
    b^2 = \frac{n}{m}
    \, ,
\end{equation}
are of order $1 + k + \ell$, where $k, \ell \in \naturals$ and $r \in \naturals_{< m}$, $ s \in \naturals_{< n}$, and the upper, plus sign is for the poles and the lower, minus sign for the roots.

When taking the 4d limits the following representation of Faddeev's quantum dilogarithm comes in useful \cite[eqs.~(3.2)-(3.8)]{Hatsuda:2015owa}\footnote{There appears to be a constant term missing in \cite[eqs.~(3.2)-(3.8)]{Hatsuda:2015owa}.}
\begin{equation}
\label{eq:NonCompQuantDiLog:IntRepsBLimit}
    \ln\Phi_b\left( z \right) = - \frac{\ri}{2 \pi b^2} \Li_2 \left( - \re^{2 \pi b z} \right) - \ri \int_0^{+\infty} \frac{\rd u}{1 + \re^{2 \pi u}} \ln \left( \frac{1 + \re^{2 \pi b z - 2 \pi b^2 u}}{1 + \re^{2 \pi b z + 2 \pi b^2 u}} \right)
    \, ,
\end{equation}
under the condition that $2 \abs{\imaginary{z}} < b^{-1}$ when $b > 0$.\footnote{One can note from the behaviour under shifts of $z$ by $\ri b^{-1}$ that the representation above has a limited domain of validity: the quantum dilogarithm is quasi-periodic under such shifts while the same shifts act trivially on the right-hand side above. Some numerical checks seem to suggest that the representation above is only valid for $ 2 \abs{\imaginary{z}} < b^{-1}$ when $\Re(z) \geqslant 0$.}
In taking the standard or dual 4d limit on the integral representation above, one may use the following integral representation of the log gamma function \cite[p.~8]{adamchik2003contributions},
\begin{equation}
\label{eq:LogGamma}
    \ln \Gamma \br{z + \frac{1}{2}} = \frac{\ln \br{2 \pi}}{2} - z + z \ln \br{z} - 2 \int_{0}^{+ \infty} \frac{ \rd u}{1 + \re^{2 \pi u}} \arctan \br{\frac{u}{z}}
    \, ,
    \qquad
    \real{z} > 0
    \, ,
\end{equation}
which can be obtained from Binet's second formula for the log gamma function.

From \cite[eqs.~(65),~(67)]{EllegaardAndersen:2011vps} one also has the following asymptotic expansion
\begin{equation}
\label{eq:NonCompQuantDiLog_QuasiClassicalExpansion}
    {\ln \Phi_b \br{\frac{z}{2 \pi b}}}
    \simeq
    \sum_{n = 0}^{\infty} \br{\ri 2 \pi b^2}^{2 n - 1} \frac{{B_{2n}\br{1/2}}}{\br{2n}!} \partial_z^{2n} \dilog{-\re^z}
    =
    - \frac{\ri}{2 \pi b^2} \dilog{-\re^z} + \bigO{b^2}
    \, ,
\end{equation}
when $b \to 0$ and where ${B_n\br{1/2}}$ are the Bernoulli polynomials evaluated at $1/2$.

\section{Special functions from topological strings}
\label{sec:TSfunctions}

Here, we summarize the all-order construction of the instanton partition function or instanton free energy for the relevant gauge theories.

One can go from the instanton partition function to the free energy by
\begin{equation}
    {Z\br{Q}} = \sum_{n \in \naturals} \frac{Z_n}{n!} Q^n
    \, ,
    \qquad \qquad
    {F\br{Q}} = \ln\br{Z\br{Q}} = \sum_{n \in \pnaturals} \frac{F_n}{n!} Q^n
    \, ,
\end{equation}
where $F_n$ is given by Fa\`{a} di Bruno's formula
\begin{equation}
    F_n = \sum_{m = 1}^{n} \br{-1}^{m-1} \ \br{m - 1}! \ B_{n, m}\br{ Z_1 , \cdots , Z_{n - m + 1}} \, ,
\end{equation}
and the $B_{n,m}$ are the partial or incomplete, exponential Bell polynomials, \texttt{BellY} in \textit{Wolfram Mathematica}.

\subsection{\texorpdfstring{Topological vertex for local $\mathbb{F}_0$ / five-dimensional, $\mathcal{N} = 1$, SU(2) SYM}{Topological vertex for local F0 / five-dimensional, N = 1, SU(2) SYM}}

We summarize the refined open topological vertex for local $\mathbb{F}_0$ as presented in \cite[app.~A.1]{Francois:2023trm}, which follows \cite{Cheng:2021nex}. See \cite{akmv2,ikv} for earlier works.

A Young diagram, or partition, $\mu$ is given by
\begin{equation}
    \mu = \left\{ \mu_1, \mu_2, \mu_3, \cdots \ | \ \forall \ k, \ell \in \pnaturals : \left( \mu_k \in \mathbb{N} \right) \land \left(  k \leqslant \ell \Rightarrow \mu_k \geqslant \mu_\ell \right) \right\} \, .
\end{equation}
We denote by $\mu^t$  the transposed Young diagram
\begin{equation}
    \mu^t =\left\{ \mu_1^t, \mu_2^t, \mu_3^t, \cdots \ | \ \forall \ k \in \pnaturals : \mu_k^t =\left| \left\{ \ell \in \pnaturals \ | \ \mu_\ell \geqslant k \right\} \right|\right\} \, .
\end{equation}
For any pair $k, \ell \in \pnaturals$ we say that $\left(k, \ell \right) \in \mu$ if $1 \leqslant \ell \leqslant \mu_k$ and we define
\begin{equation}
    \abs{\mu} =\sum_{k=1}^{+\infty} \mu_k \, ,
    \qquad \qquad
    \norm{\mu}^2 =\sum_{k=1}^{+\infty} \mu_k^2 \, .
\end{equation}
Let $\mu, \nu$ be two Young diagrams and define
\begin{equation}
\label{eq:defZtilde3}
    Z_\mu \left( r_1, r_2 \right) =\prod_{\left( k , \ell \right) \in \mu} \left( 1 - r_2^{\mu_k - \ell} r_1^{\mu_\ell^t - k + 1} \right)^{-1}
    \, ,
    \qquad
    \norm{Z_\mu \left( r_1, r_2 \right)}^2 =Z_{\mu^t} \left( r_1, r_2 \right) Z_\mu \left( r_2, r_1 \right)
    \, ,
\end{equation}
where $r_1 = \exp\br{\ri \epsilon_1}$ and $r_2 = \exp\br{- \ri \epsilon_2}$, with $\epsilon_{1,2}$ the $\Omega$-background parameters. Note the different sign in the definition of $r_{1,2}$.
Define then the Nekrasov factors as \cite[eq.~(A.7)]{Cheng:2021nex}
\begin{equation}
    \label{eq:NekrasovFactors}
    N_{\mu, \nu} \left( Q; r_1 , r_2 \right)
    =
    \prod_{\left(k, \ell\right) \in \nu} \left( 1 - Q \ r_2^{\nu_k - \ell} r_1^{\mu_\ell^t-k+1} \right) \prod_{\left(k, \ell\right) \in \mu} \left( 1 - Q \ r_2^{-\mu_k + \ell - 1} r_1^{-\nu_\ell^t+k} \right)
    \, ,
\end{equation}
as well as the combination
\begin{multline}
        C_{\mu, \nu} \left( Q_X, Q_F, r_1, r_2 \right)
        =
        \br{ N_{\mu^t,\nu}\left( Q_F; r_1^{-1}, r_2^{-1} \right) N_{\mu^t,\nu}\left( Q_F \frac{r_1}{r_2}; r_1^{-1}, r_2^{-1} \right)}^{-1}
        \\
        \frac{N_{\emptyset, \mu^t} \left( Q_X \frac{r_1^2}{r_2}; r_1^{-1}, r_2^{-1} \right)}{N_{\emptyset, \mu^t} \left( Q_X \frac{r_1}{r_2}; r_1^{-1}, r_2^{-1} \right)}
        \frac{N_{\emptyset, \nu} \left( Q_X Q_F \frac{r_1^2}{r_2}; r_1^{-1}, r_2^{-1} \right)}{N_{\emptyset, \nu} \left( Q_X Q_F \frac{r_1}{r_2}; r_1^{-1}, r_2^{-1} \right)}
        \, ,
\end{multline}
where $\emptyset$ is the empty partition and $Q_{X, F}$ is related to the Kähler parameter of the brane and the fibre respectively.
The ``\textit{open-closed t-brane partition function}'' for local $\mathbb{F}_0$ is then given by \cite[p.~50, eq.~(5.4)]{Cheng:2021nex}
\begin{multline}
\label{eq:RefinedOpenTopologicalVertexLocalF0PartitionFunction}
    Z^\mathrm{open-closed}_\mathrm{inst} \left(  Q_X , Q_F , Q_B , r_1, r_2 \right) =
    \\
    \sum_{n \in \mathbb{N}} Q_B^n
    \sum_{m = 0}^n \sum_{\substack{\abs{\mu} = m \\ \abs{\nu} = n - m}} r_1^{\norm{\nu^t}^2} r_2^{\norm{\mu^t}^2}
    \norm{Z_\mu \left( r_1, r_2 \right)}^2 \norm{Z_\nu \left( r_2, r_1 \right)}^2 C_{\mu, \nu} \left(  Q_X , Q_F, r_1, r_2 \right) \, ,
\end{multline}
where $Q_B$ is related to the Kähler parameter of the base.
The closed partition function is then obtained by setting $Q_X = 0$, that is
\begin{equation}
    Z_\mathrm{inst} \left( Q_F, Q_B, r_1 , r_2 \right) = Z^\mathrm{open-closed}_\mathrm{inst} \left( 0 , Q_F, Q_B, r_1 , r_2 \right) \, .
\end{equation}
Finally we define the t-brane instanton partition function for local $\mathbb{F}_0$ by \cite[p.~50, eq.~(5.7)]{Cheng:2021nex}
\begin{equation}
\label{eq:T-BraneInstanton}
        Z^\mathrm{open}_\mathrm{inst} \left(  Q_X , Q_F, Q_B , r_1, r_2 \right) =\frac{Z^\mathrm{open-closed}_\mathrm{inst} \left(  Q_X , Q_F, Q_B, r_1, r_2 \right)}{Z_\mathrm{inst} \left( Q_F, Q_B, r_1, r_2 \right)} \, .
\end{equation}

Two particular phases of the refined topological vertex are of interest to us: the self-dual or Gopakumar-Vafa (GV) phase $\epsilon_1 + \epsilon_2 = 0$ or $r_1 = r_2$, and the Nekrasov-Shatashvili (NS) phase $\epsilon_1 \to 0$ or $r_1 \to 1$. For the NS phase we have
\begin{equation}\label{eq:nsef}
    \begin{split}
        F^\mathrm{NS}_\mathrm{inst} \br{t_F, t_B, \hbar} & = \lim_{\epsilon_1 \to 0} \br{- \epsilon_1} F_\mathrm{inst} \br{ \re^{-t_F} , \re^{-t_B} , \re^{\ri \epsilon_1}, \re^{- \ri \hbar}} \, ,
        \\
        F^\mathrm{open}_\mathrm{NS, inst} \br{ x, t_F, t_B, \hbar} & = \lim_{\epsilon_1 \to 0} F_\mathrm{inst}^\mathrm{open} \br{ -  \re^{- \ri \epsilon_1 / 2} \re^{t_F / 2} \re^{-x},  \re^{-t_F}, \re^{-t_B}, \re^{\ri \epsilon_1}, \re^{- \ri \hbar}}
        \, .
    \end{split}
\end{equation}
Numerically, one can see that the series expansions in $t_B$ of the above quantities are convergent,\footnote{\label{footnoteconv}There are some issues when $\hbar$ is real, but these are resolved once  $F_\mathrm{NS}$ and $F_\mathrm{GV}$ are combined in the definition of topological string grand potential $\rm J$, as discussed in the main text.} but, to our knowledge, at present there is not a rigorous mathematical proof.
At leading order we have
\begin{equation}
\label{inst2_appendix}
    F^\mathrm{NS}_\mathrm{inst}\left( t_F , t_B , \hbar \right)
    =
    \squarebr{\frac{\ri \br{1 + \re^{\ri \hbar}}}{\br{1 - \re^{\ri \hbar}} \br{1 -  \re^{- \ri \hbar} \re^{- t_F}} \br{1 - \re^{\ri \hbar} \re^{- t_F}}}} \re^{- t_B}
    + \bigO{\re^{-2 \, t_B}}
    \, ,
\end{equation}
 \begin{multline}
    \label{eq:zns_appendix}
    F^\mathrm{open}_\mathrm{NS, inst} \br{ x, t_F, t_B, \hbar}
    =
    \\
    \squarebr{\frac{
    \re^{\ri 2 \hbar} \re^{\frac{t_F}{2} - x} \left(  1 + \re^{-t_F} + \re^{\ri \hbar} \left( 1 + \re^{\ri \hbar }\right) \re^{-t_F} \re^{\frac{t_F}{2}-x}  \right)}
    {\left( 1 - \re^{\ri \hbar} \right) \left( 1 - \re^{\ri \hbar} \re^{- t_F}\right) \left( \re^{\ri \hbar } - \re^{-t_F}\right) \left( 1 + \re^{\ri \hbar} \re^{-\frac{t_F}{2} - x}\right) \left( 1 + \re^{\ri \hbar} \re^{\frac{t_F}{2} - x}\right)}} \re^{- t_B}
    + \bigO{\re^{- 2 \, t_B}}
\end{multline}
For the GV phase we have similarly
\begin{equation}\label{eq:sdef}
    \begin{split}
        F^\mathrm{GV}_{\rm inst}\br{t_F , t_B , g_s} & = F_\mathrm{inst} \left( \re^{-t_F },  \re^{-t_B }, \re^{- \ri g_s}, \re^{- \ri g_s} \right)
        \, ,
        \\
        F^\mathrm{open}_\mathrm{GV, inst} \br{x  , t_F ,  t_B , g_s} & = F_\mathrm{inst}^\mathrm{open} \left( \re^{\ri g_s / 2}\re^{t_F  / 2} \re^{-x  }, \re^{-t_F },  \re^{-t_B }, \re^{- \ri g_s}, \re^{- \ri g_s} \right)
        \, .
    \end{split}
\end{equation}
The series expansion in $t_B$ of the above quantities is convergent. In the case of  $F^\mathrm{GV}_{\rm inst}$, this was rigorously demonstrated in \cite{Bershtein:2016aef}.\footnote{There are some issues when $g_s$ is real, but these are resolved once  $F_\mathrm{NS}$ and $F_\mathrm{GV}$ are combined in the definition of topological string grand potential $\rm J$, as discussed in the main text.} The proof for  $F^\mathrm{open}_\mathrm{GV, inst}$ follows analogously.
The leading order reads then
\begin{equation}
\label{eqF_Closed_GV_Instanton_appendix}
    F^\mathrm{GV}_{\rm inst} \left( t_F, t_B , g_s \right)
    =
    \squarebr{\frac{2 \re^{\ri g_s}}{\br{1 - \re^{\ri g_s}}^2 \br{1 - \re^{- t_F }}^2}} \re^{-t_B }
    + \bigO{\re^{- 2 \, t_B }}
    \, ,
\end{equation}
\begin{multline}
    \label{eq:app:zgv}
    F^\mathrm{open}_\mathrm{GV, inst} \left( x, t_F , t_B , g_s \right)
    =
    \\
    \squarebr{\frac{\re^{\ri \frac{g_s}{2}} \re^{\frac{t_F }{2} - x  } \br{2 \re^{\ri \frac{g_s}{2}} \re^{- \frac{t_F }{2} - x  }- 1 - \re^{-t_F } }}{\br{1 - \re^{\ri g_s}} \br{1 - \re^{-t_F }}^2 \br{1 - \re^{\ri \frac{g_s}{2}} \re^{\frac{t_F }{2} - x  }} \br{1 - \re^{\ri \frac{g_s}{2}} \re^{- \frac{t_F }{2} - x  }}}} \re^{- t_B }
    + \bigO{\re^{-2 \, t_B }}
    \, ,
\end{multline}
for the closed and open parts respectively.

\subsection{\texorpdfstring{Four-dimensional, $\mathcal{N} = 2$, SU(2) SYM}{Four-dimensional, N = 2, SU(2) SYM}}

Consider the following four-dimensional limit for the variables in the previous section \cite{Iqbal:2003ix, kkv}
\begin{equation}
    Q_X = \exp\br{- R \br{z - a + \ri \epsilon_1^\mathrm{4d}}}
    \qquad \qquad
    Q_F = \exp\br{- 2 R a}
    \qquad \qquad
    Q_B = R^4 t
\end{equation}
\begin{equation}
   \epsilon_{1,2} = R \epsilon_{1,2}^\mathrm{4d} \, ,
    \qquad \qquad
    R \to 0 \, .
\end{equation}
This is the same scaling as for the standard variables in the standard 4d limit, \eqref{eq:4d}, \eqref{eq:Standard4DLimit_KahlerParameters_Scaling}, and \eqref{eq:Standard4DLimit_KahlerParameters_Shifts}, and the same scaling as for the dual variables in the dual 4d limit, \eqref{eq:dual4d}.
This gives for \eqref{eq:defZtilde3}
\begin{equation}
    Z_\mu^\mathrm{4d} \left( \epsilon_1 , \epsilon_2 \right)
    =
    \prod_{\left( k , \ell \right) \in \mu}
    \br{\br{\mu_k - \ell} \epsilon_2 - \br{\mu_\ell^t - k + 1} \epsilon_1}^{-1} \, ,
\end{equation}
\begin{equation}
    \norm{Z_\mu^\mathrm{4d} \left( \epsilon_1, \epsilon_2 \right)}^2 = Z_{\mu^t}^\mathrm{4d} \left( \epsilon_1, \epsilon_2 \right) Z_\mu^\mathrm{4d} \left( \epsilon_2, \epsilon_1 \right) \, ,
\end{equation}
while the Nekrasov factors \eqref{eq:NekrasovFactors} become
\begin{multline}
    N_{\mu, \nu}^\mathrm{4d} \left( \alpha ; \epsilon_1, \epsilon_2 \right)
    =
    \prod_{\left(k, \ell\right) \in \nu} \left( \br{\nu_k - \ell} \epsilon_2 - \br{\mu_\ell^t-k+1} \epsilon_1 + \ri \alpha \right)
    \\
    \prod_{\left(k, \ell\right) \in \mu} \left( \br{ - \mu_k + \ell - 1} \epsilon_2 - \br{ - \nu_\ell^t + k} \epsilon_1 + \ri \alpha \right) \, ,
\end{multline}
and similarly
\begin{equation}
     C_{\mu, \nu}^\mathrm{4d} \left(a, \epsilon_1, \epsilon_2 \right)
     =
     \br{N_{\mu^t,\nu}^\mathrm{4d} \left( 2 a ; \epsilon_1, \epsilon_2 \right) N_{\mu^t,\nu}^\mathrm{4d} \left( 2 a - \ri \br{\epsilon_1 + \epsilon_2};  \epsilon_1, \epsilon_2 \right)}^{-1}
     \, ,
\end{equation}
\begin{multline}
    C_{\mu, \nu}^\mathrm{2d/4d} \left( z, a, \epsilon_1, \epsilon_2 \right)
    =
    C_{\mu, \nu}^\mathrm{4d} \left( a , \epsilon_1, \epsilon_2 \right)
    \\
    \frac{N_{\emptyset, \mu^t}^\mathrm{4d} \left( z - a - \ri \br{\epsilon_1 + \epsilon_2} ; \epsilon_1, \epsilon_2 \right)}{N_{\emptyset, \mu^t}^\mathrm{4d} \left( z - a - \ri \epsilon_2 ; \epsilon_1, \epsilon_2 \right)}
    \frac{N_{\emptyset, \nu}^\mathrm{4d} \left( z + a - \ri \br{\epsilon_1 + \epsilon_2} ; \epsilon_1, \epsilon_2 \right)}{N_{\emptyset, \nu}^\mathrm{4d} \left( z + a - \ri \epsilon_2; \epsilon_1, \epsilon_2 \right)}
    \, .
\end{multline}
Hence we find for the complete 2d/4d defect instanton partition function in a generic phase of the $\Omega$-background
\begin{equation}
    Z^\mathrm{4d}_{\rm inst} \left( a , t, \epsilon_1, \epsilon_2 \right) =
    \sum_{n \in \mathbb{N}} t^n
    \sum_{m = 0}^n \sum_{\substack{\abs{\mu} = m \\ \abs{\nu} = n - m}}
    \br{-1}^n
    \norm{Z_\mu^\mathrm{4d} \left( \epsilon_1, \epsilon_2 \right)}^2 \norm{Z_\nu^\mathrm{4d} \left( \epsilon_2, \epsilon_1 \right)}^2 C_{\mu, \nu}^\mathrm{4d} \left( a , \epsilon_1, \epsilon_2 \right) \, ,
\end{equation}
\begin{multline}
    Z^\mathrm{D}_{\rm inst} \left(  z , a, t, \epsilon_1, \epsilon_2 \right) =
    \\
    \sum_{n \in \mathbb{N}} t^n
    \sum_{m = 0}^n \sum_{\substack{\abs{\mu} = m \\ \abs{\nu} = n - m}}
    \br{-1}^n
    \norm{Z_\mu^\mathrm{4d} \left( \epsilon_1, \epsilon_2 \right)}^2 \norm{Z_\nu^\mathrm{4d} \left( \epsilon_2, \epsilon_1 \right)}^2 C_{\mu, \nu}^\mathrm{2d/4d} \left( z , a , \epsilon_1, \epsilon_2 \right) \, ,
\end{multline}
\begin{equation}
     Z^\mathrm{2d/4d}_\mathrm{inst} \left( z, a , t , \epsilon_1, \epsilon_2 \right) = \frac{Z^\mathrm{D}_\mathrm{inst} \left( z, a , t , \epsilon_1, \epsilon_2 \right)}
    {Z^\mathrm{4d}_\mathrm{inst} \left( a , t , \epsilon_1, \epsilon_2 \right)} \, .
\end{equation}
The NS limit as used in \autoref{sec:sd4d} is then given by
\begin{equation}
\label{eq:F_4d_NS}
    F^\mathrm{4d}_\mathrm{NS, inst} \br{\sigma, t}
    =
    \lim_{\epsilon_1 \to 0} \br{ - \epsilon_1} F^\mathrm{4d}_\mathrm{inst} \br{ \sigma, t, \epsilon_1, 1}
    \, ,
\end{equation}
\begin{equation}
\label{eq:Z_Defect_NS}
     Z^\mathrm{2d/4d}_\mathrm{NS, inst} \br{x, \sigma, t} = \lim_{\epsilon_1 \to 0} Z^\mathrm{2d / 4d}_\mathrm{inst} \left( x - \ri \frac{\epsilon_1}{2}, \sigma, t , \epsilon_1, 1 \right)
     \, .
\end{equation}
The convergence of \eqref{eq:F_4d_NS} as series in $t$ for $2\sigma\neq\epsilon_1 \IZ$ follows from  \cite{Desiraju:2024fmo}. Similar arguments should hold for \eqref{eq:Z_Defect_NS} as well.
The first few terms of the NS limit are
\begin{multline}
\label{eq:F_4d_NS_NLO}
    F^\mathrm{4d}_\mathrm{NS, inst} \br{\sigma, t}
    =
    - \squarebr{\frac{2}{4 \sigma^2 + 1}} t
    - \squarebr{\frac{20 \sigma^2 - 7}{4  \br{4 \sigma^2 + 1}^3 \br{\sigma^2 + 1}}} t^2
    \\
    - \squarebr{\frac{4 \br{ 144 \sigma^4 - 232 \sigma^2 + 29 }}{3 \br{ 4 \sigma^2 + 1 }^5 \br{\sigma^2 + 1} \br{4 \sigma^2 + 9}}} t^3
    + \bigO{t^4}
    \, ,
\end{multline}
\begin{multline}
\label{eq:Z_2d-Defect_NS_NLO}
    Z^\mathrm{2d/4d}_\mathrm{NS, inst} \br{x,\sigma,t}
    =
    1 + \squarebr{\frac{ 1 - 2 \widetilde{x}}{\left(1+4 \sigma ^2\right) \left(\widetilde{x}^2+\sigma ^2\right)}} t
    +
    \\
    \squarebr{\frac{5+9 \widetilde{x}-27 \widetilde{x}^2+14 \widetilde{x}^3+(91-10 \widetilde{x} (17+2 \widetilde{x} (-7+2 \widetilde{x}))) \sigma ^2+4 (-7+2 \widetilde{x}) (-5+4 \widetilde{x}) \sigma^4}{4 \left(1+4 \sigma ^2\right)^3 \left(1+\sigma ^2\right)  \left(\widetilde{x}^2+\sigma^2\right) \left((\widetilde{x} - 1)^2+\sigma ^2\right) }} t^2
    \\
    + \bigO{t^3}
\end{multline}
where $\widetilde{x} = - \ri x - 1$.

For the GV phase, we define
\be\ba
\label{eq:Z_2d4d_SelfDual}
    Z^\mathrm{4d}_\mathrm{GV, inst}\br{\sigma, t} = & Z^\mathrm{4d}_\mathrm{inst} \br{ \ri \sigma, t, -1, 1}
     \, ,
     \\
    Z^\mathrm{2d/4d}_\mathrm{GV, inst}\br{x, \sigma, t} = & Z^\mathrm{2d/4d}_\mathrm{inst} \left(x + \frac{\ri}{2}, \ri \sigma, t, -1, 1 \right)
    \, .
\ea\ee
The convergence of \eqref{eq:Z_2d4d_SelfDual} as a series in  $t$  for $ 2\ri\sigma \notin \mathbb{Z} $ follows from \cite{ilt}.
The first few orders of the GV phase \eqref{eq:Z_2d4d_SelfDual} are given by
\begin{multline}
\label{eq:Z_4d_SD_NLO}
    Z^\mathrm{4d}_\mathrm{GV, inst} \left( \sigma, t \right)
    =
    1 + \squarebr{\frac{1}{2 \sigma^2}} t
    + \squarebr{\frac{ 8 \sigma^2 + 1 }{4 \sigma^2 \left( 4 \sigma^2 - 1 \right)^2}} t^2
    + \squarebr{\frac{8 \sigma^4 - 5 \sigma^2 + 3}{24 \sigma^2 \left( 4 \sigma^2 - 1 \right)^2 \left( \sigma^2 - 1 \right)^2}} t^3
    \\
    + \squarebr{\frac{ 256 \sigma^8 - 832 \sigma^6 + 972 \sigma^4 - 177 \sigma^2 + 81 }{384 \sigma^4 \left( 4 \sigma^2 - 1 \right)^2 \left( \sigma^2 - 1 \right)^2 \left( 4 \sigma^2 - 9 \right)^2}} t^4
    + \bigO{t^5}
    \, ,
\end{multline}
\begin{multline}
\label{eq:Z_2d-Defect_SD_NLO}
    Z^\mathrm{2d/4d}_{\rm GV, inst} \left( x, \sigma  , t \right)
    =
    1 - \left[ \frac{\widetilde{x}}{2 \sigma^2 \left(\widetilde{x}^2 - \sigma^2 \right)}\right] t
    \\
    + \left[ \frac{\widetilde{x} \left( \widetilde{x} + 1 \right)^2 - \widetilde{x} \left( 10 \widetilde{x}^2 + 19 \widetilde{x} + 10 \right) \sigma^2 + \left(  8 \widetilde{x}^2 + 30 \widetilde{x} + 9 \right) \sigma ^4}{4 \sigma^4 \left( 4 \sigma^2 - 1 \right)^2 \left(\widetilde{x}^2 - \sigma^2\right) \left( \left( \widetilde{x} + 1 \right)^2 - \sigma^2\right)} \right] t^2
   +\mathcal{O}\left(t^3\right) \, ,
\end{multline}
where we used $\widetilde{x} = \ri x + 1/2$ for convenience.

\section{\texorpdfstring{Comment on the case $\xi = 0$ and $\hbar = 2 \pi$}{Comment on the case \pdfxi~=~0 and \pdfhbar~=~2\pdfpi}}
\label{app:mass1}

The eigenfunctions of \eqref{eq:spectralc} for $\xi = 0$ and $\hbar = 2 \pi$ were studied extensively in \cite{Marino:2016rsq, Marino:2017gyg}. The explicit expression for $\psi_0(x)$ can be found in \cite[eqs.~(2.95), (2.96)]{Marino:2016rsq}. However, when written in the form given in \cite[eqs.~(2.95), (2.96)]{Marino:2016rsq}, the structure presented in \eqref{eq:ProposedStructureEigenFun2} is not immediately apparent. In this appendix, we clarify the reasons for this.

Let us consider the $\xi \to 0$ limit when $\hbar = 2 \pi$ of \eqref{eq:Psi0x_RationalB_Result}. In this case one gets
\begin{equation}
    \sqrt{2} q_{\pm, k}\br{x}
    =
    \ln \squarebr{\re^{x} \br{\frac{\re^{x} + \re^{- x}
    \pm \sgn[\arg[x + \ri \frac{\pi}{2}]] \sqrt{\br{\re^{x} - \re^{- x} }^2}}{2}}} + \ri 2 \pi k
    \, ,
\end{equation}
and it should be noted that the important symmetry $q_{\pm, k}\br{x} = -q_{\mp, - k}\br{- x - \ri \pi} \, \mathrm{mod} \, \ri 2 \pi$ is still present.
Let us now write the two terms \eqref{eq:Psi0x_RationalB_Result} explicitly for $x \in \reals$ and $\hbar = 2 \pi$. We find that
\begin{equation}
\label{eq:cco}
    \begin{split}
        \omega_{0,+}(x)
        & =  \ri 2 \pi \lim_{\xi\to 0} \br{\mathrm{Res}_+(x) \sum_{k = 0}^1 U(x, q_{+,k}(x)) E(q_{+, k}(x))}
        \\
        &
        =
        \begin{dcases}
            - \frac{1}{\sqrt{\pi}} \re^{\ri x^2 / 4 \pi}  \br{\frac{\re^{x}}{\re^{2 x} - 1}}
            & \text{if} \, x < 0
            \\
            \frac{\re^{\ri \pi / 4}}{\sqrt{2 \pi}} \re^{- \ri x^2 / 4 \pi}  \br{\frac{\re^{x}\br{\re^x - \ri}}{\re^{2 x} - 1}}
            & \text{if} \, x > 0
        \end{dcases}
        \quad ,
    \end{split}
\end{equation}
\begin{equation}
    \begin{split}
        \omega_{0,-}(x)
        & =  \ri 2 \pi \lim_{\xi\to 0} \br{\mathrm{Res}_-(x) \sum_{k = 0}^1 U(x, q_{-,k}(x)) E(q_{-, k}(x))}
        \\
        &
        =
        \begin{dcases}
            \frac{\re^{\ri \pi / 4}}{\sqrt{2 \pi}} \re^{- \ri x^2 / 4 \pi}  \br{\frac{\re^{x}\br{\re^x - \ri}}{\re^{2 x} - 1}}
            & \text{if} \, x < 0
            \\
            - \frac{1}{\sqrt{\pi}} \re^{\ri x^2 / 4 \pi}  \br{\frac{\re^{x}}{\re^{2 x} - 1}}
            & \text{if} \, x > 0
        \end{dcases}
        \quad ,
    \end{split}
\end{equation}
and $\omega_{0, \pm}\br{x}$ diverges when $x = 0$ but the sum $\psi_0\br{x}$ is still well-defined. One finds for all
$x \in \reals$ that
\begin{equation}
\label{fin}
    \psi_0(x) = \omega_{0,+}(x)+\omega_{0,-}(x) =
    \frac{\re^{\ri \pi / 4}}{\sqrt{2 \pi}} \re^{- \ri x^2 / 4 \pi}  \br{\frac{\re^{x}\br{\re^x - \ri}}{\re^{2 x} - 1}}
    - \frac{1}{\sqrt{\pi}} \re^{\ri x^2 / 4 \pi}  \br{\frac{\re^{x}}{\re^{2 x} - 1}}
    \, ,
\end{equation}
which is indeed the same expression as in \cite[eq.~(2.95)-(2.96)]{Marino:2016rsq}. However, if we were to start from the final expression on the right-hand side of \eqref{fin}, we would not see the general structure \eqref{eq:ProposedStructureEigenFun2} which is instead manifest when $\xi\neq 0$.

\section{\texorpdfstring{Selected plots of the eigenfunctions at various $\xi$ and $\hbar$}{Selected plots of the eigenfunctions at various \pdfxi~ and \pdfhbar}}
\label{sec:plots}

Below, we plot \eqref{tsstopenf} for various values of the parameters.
For the on-shell eigenfunctions, we have always from top left to bottom right: the $\sigma = 1$ term in \eqref{eq:sigmas}, the $\sigma = 2$ term in \eqref{eq:sigmas}, the complete eigenfunction $\psi$ in \eqref{tsstopenf}, after normalization, and the absolute difference between the analytical and numerical eigenfunctions for 0, 2 and 4 instantons, using a logarithmic scale with base 10. With ``0, 2 and 4 instantons'' we mean that we truncate the large $t_B$ expansion of $\psi$ up to terms of order $\exp \br{- n \min\br{1, \br{2 \pi / \hbar}} t_B}$ where $n \in \cbr{0, 2, 4}$.
The numerical eigenfunctions are obtained by diagonalizing in the basis of the harmonic oscillator, as in \cite{Huang:2014eha}.
The peaks visible in the bottom-right plots of the figures occur at $x = \pm \real{t_F\br{\hbar}} / 2$, where the $\sigma = 1$ and $\sigma = 2$ terms of \eqref{eq:sigmas} have poles.
The convergence of the sum over $k$ in \eqref{tsstopenf} is slower near these points, leading to those peaks.
For the off-shell eigenfunctions, we have similarly from top left to bottom: the $\sigma = 1$ term in \eqref{eq:sigmas}, the $\sigma = 2$ term in \eqref{eq:sigmas}, and the complete eigenfunction $\psi$ in \eqref{tsstopenf}, after normalization.
The solid and dashed lines correspond to the real and imaginary parts, respectively.

\clearpage

\subsection{\texorpdfstring{The case $\xi = - \ln \br{3} / 3$ and $\hbar = 2 \pi / 3$}{The case \pdfxi~=~-ln(3)/3 and \pdfhbar~=~2\pdfpi/3}}

\begin{figure}[h!]
    \centering
    \includegraphics[width=0.44\textwidth]{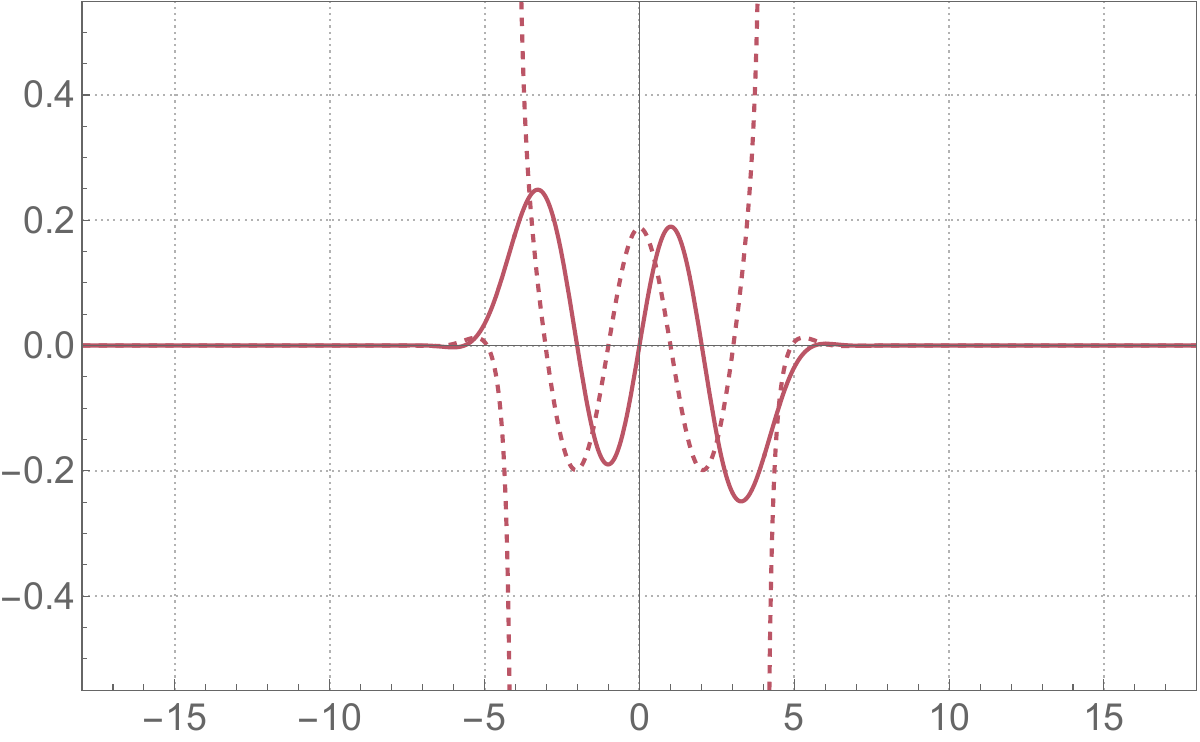}
    \includegraphics[width=0.44\textwidth]{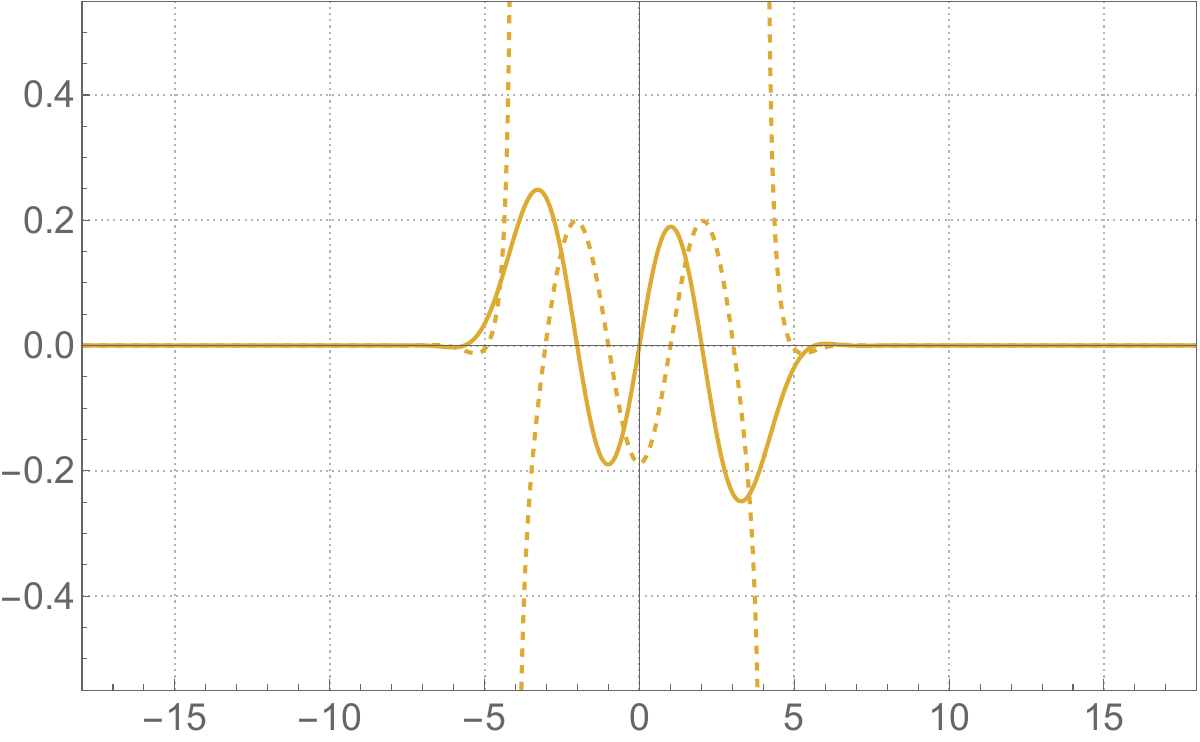}
    \includegraphics[width=0.44\textwidth]{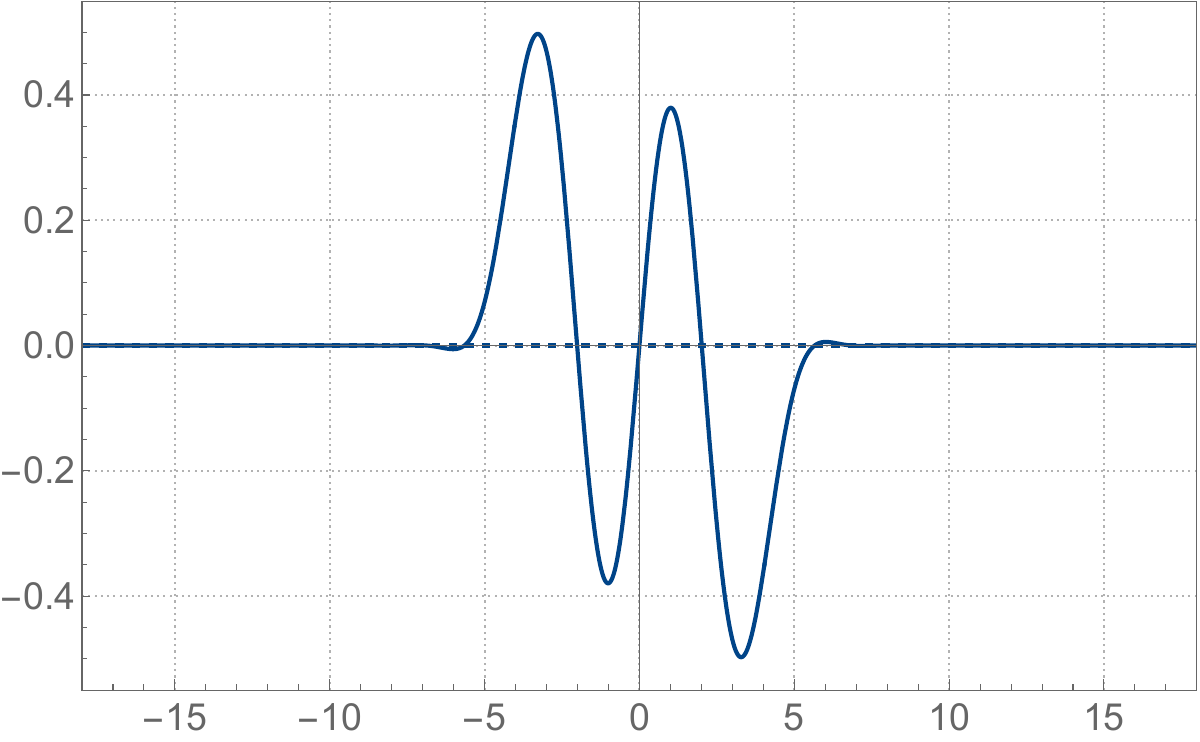}
    \includegraphics[width=0.44\textwidth]{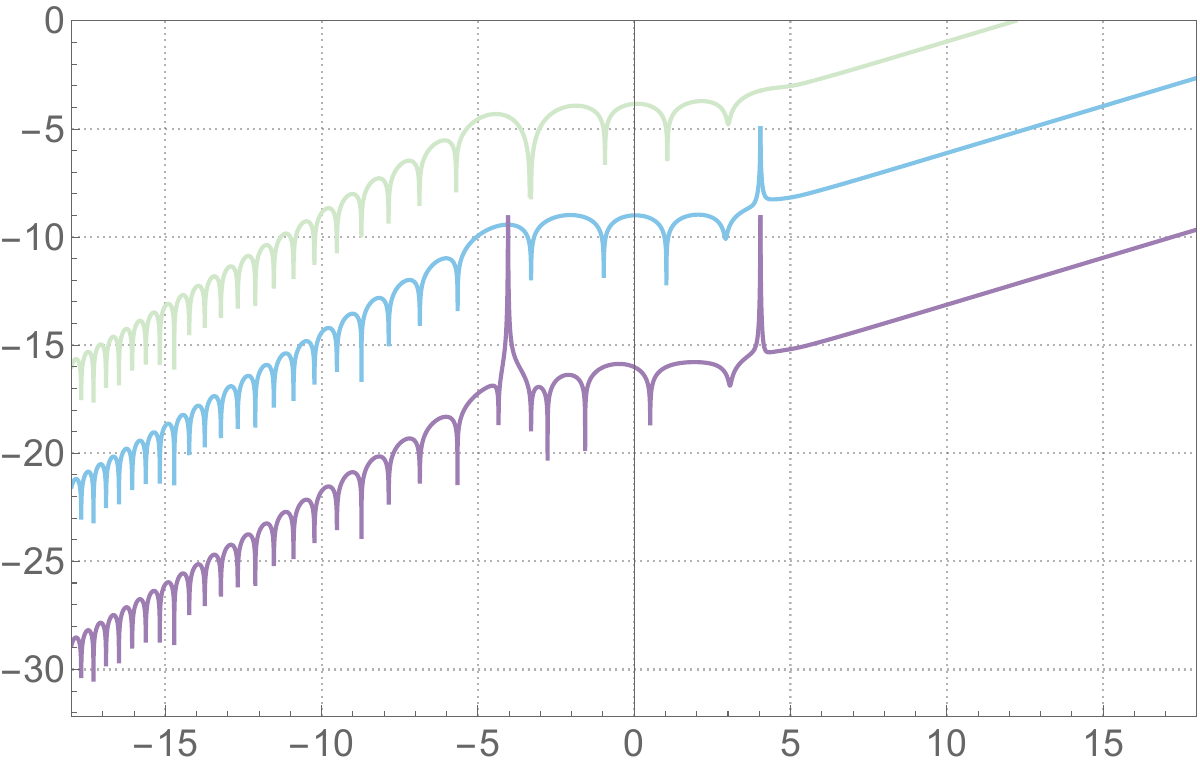}
    \caption{The on-shell eigenfunction from topological strings for $\xi = - \ln \br{3} / 3$ and $\hbar = 2 \pi / 3$ at energy $E = E_3 \approx 3.31$\,. See the explanation at the beginning of \autoref{sec:plots}.}
\end{figure}

\begin{figure}[h!]
    \centering
    \includegraphics[width=0.44\textwidth]{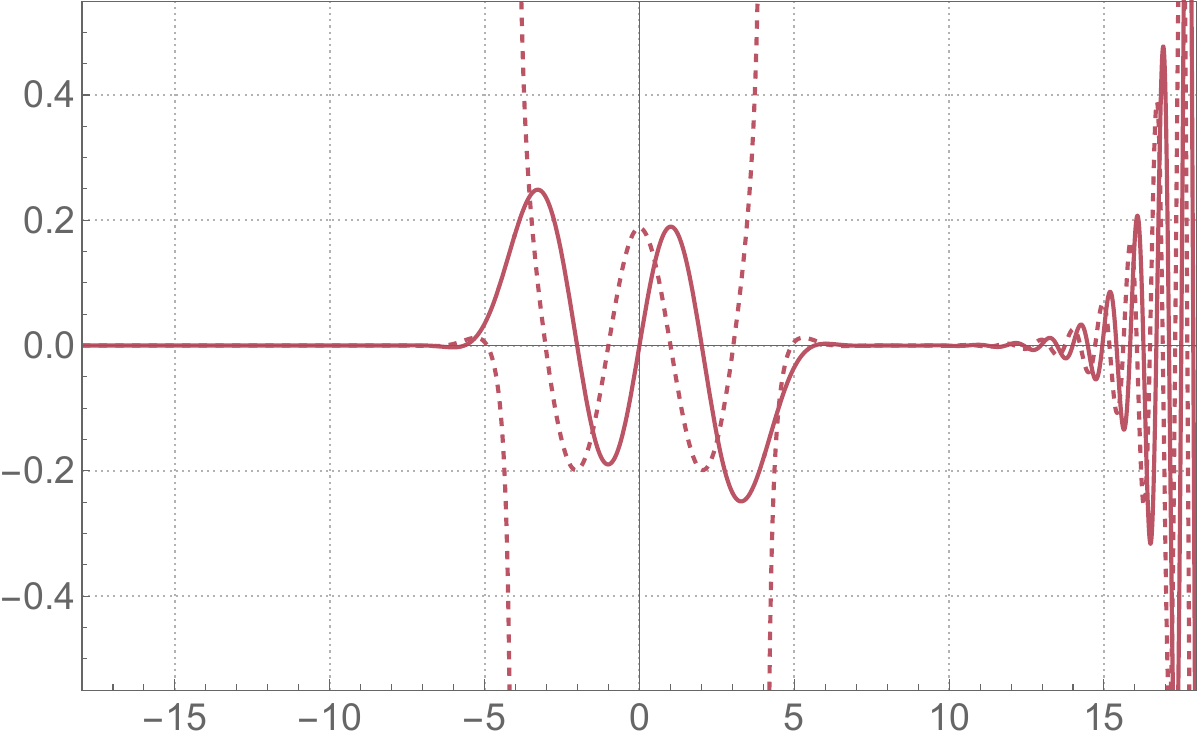}
    \includegraphics[width=0.44\textwidth]{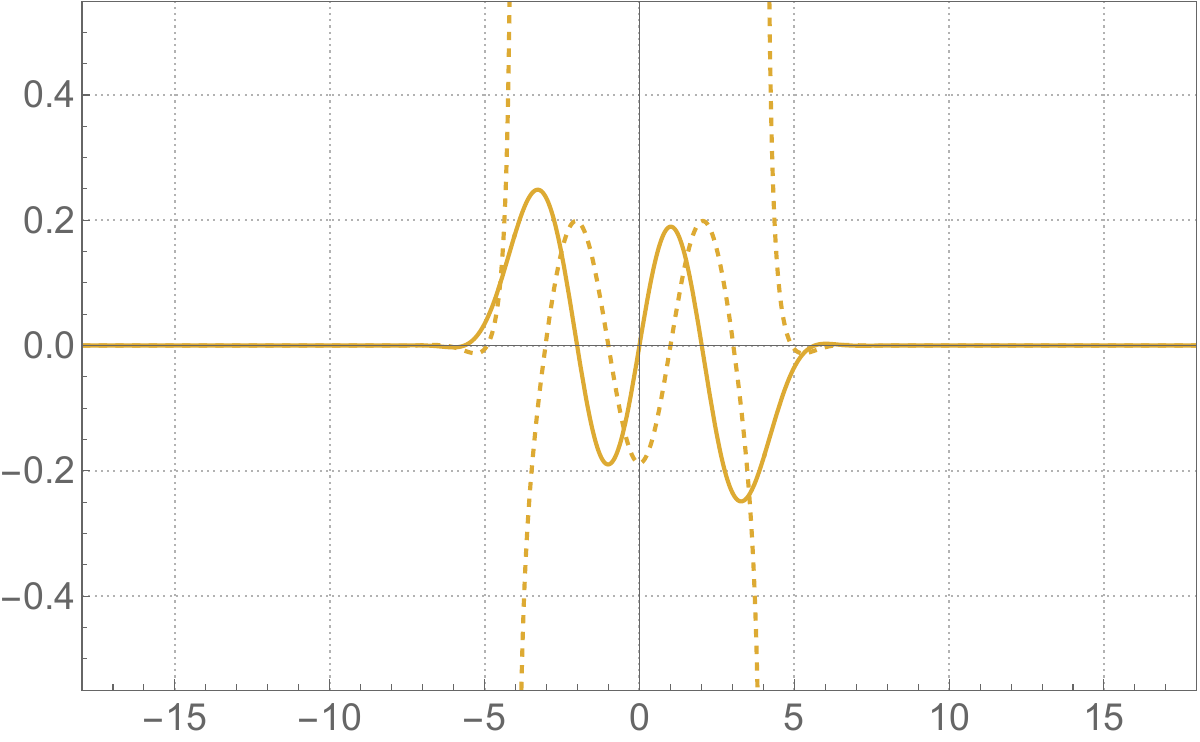}
    \includegraphics[width=0.44\textwidth]{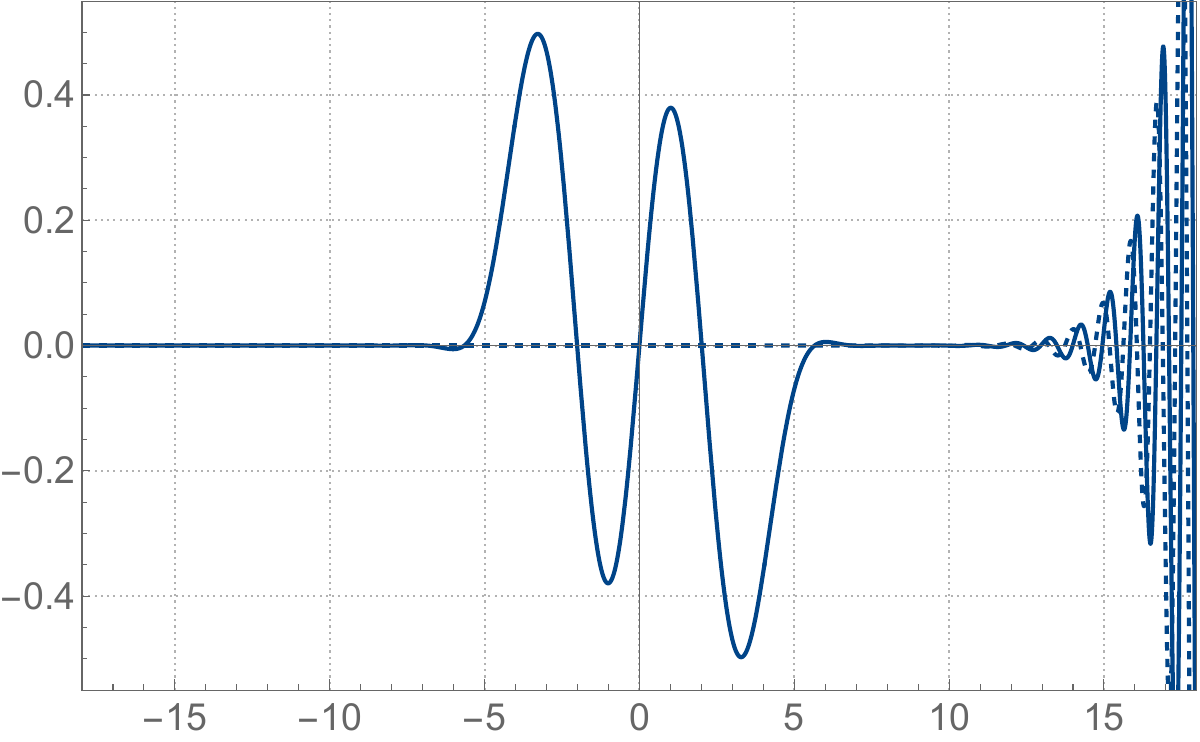}
    \caption{The off-shell eigenfunction from topological strings for $\xi = - \ln \br{3} / 3$ and $\hbar = 2 \pi / 3$ at energy $E = 40367 / 12209$, just below $E_3$. See the explanation at the beginning of \autoref{sec:plots}.}
\end{figure}

\clearpage

\subsection{\texorpdfstring{The case $\xi = \sqrt{7} / 4$ and $\hbar = 3 \pi$}{The case \pdfxi~=~\pdfsqrt7/4 and \pdfhbar~=~3\pdfpi}}

\begin{figure}[h!]
    \centering
    \includegraphics[width=0.44\textwidth]{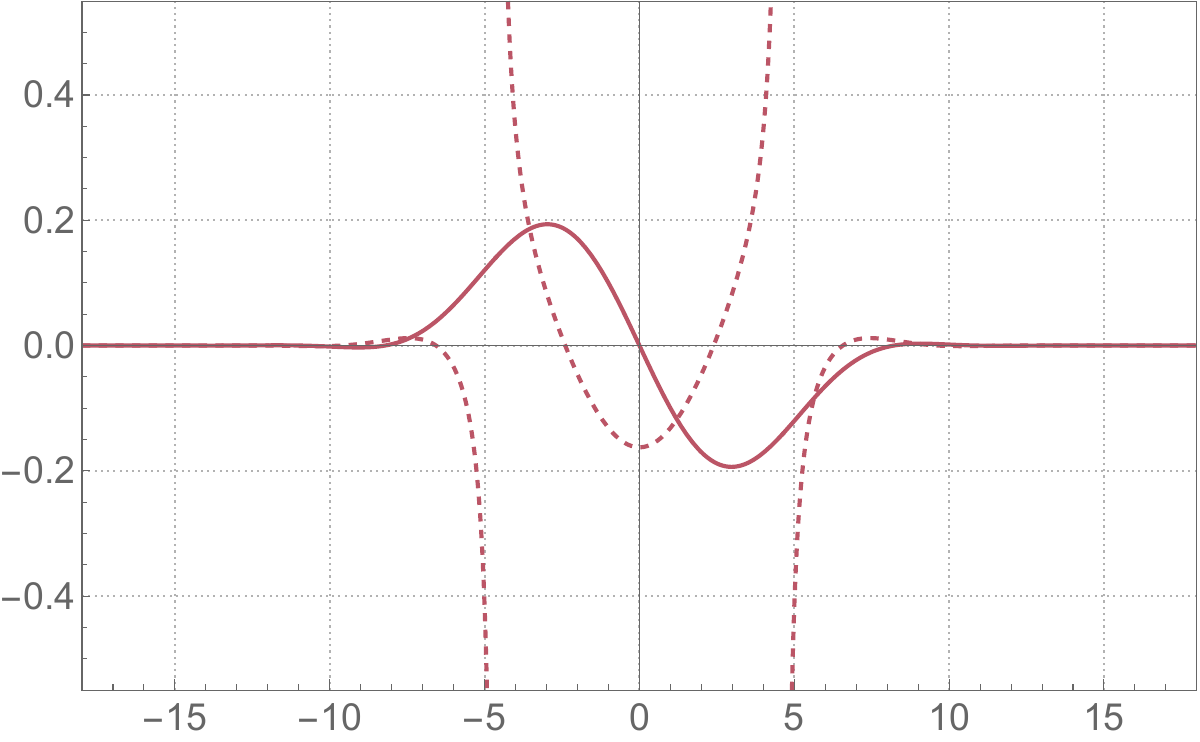}
    \includegraphics[width=0.44\textwidth]{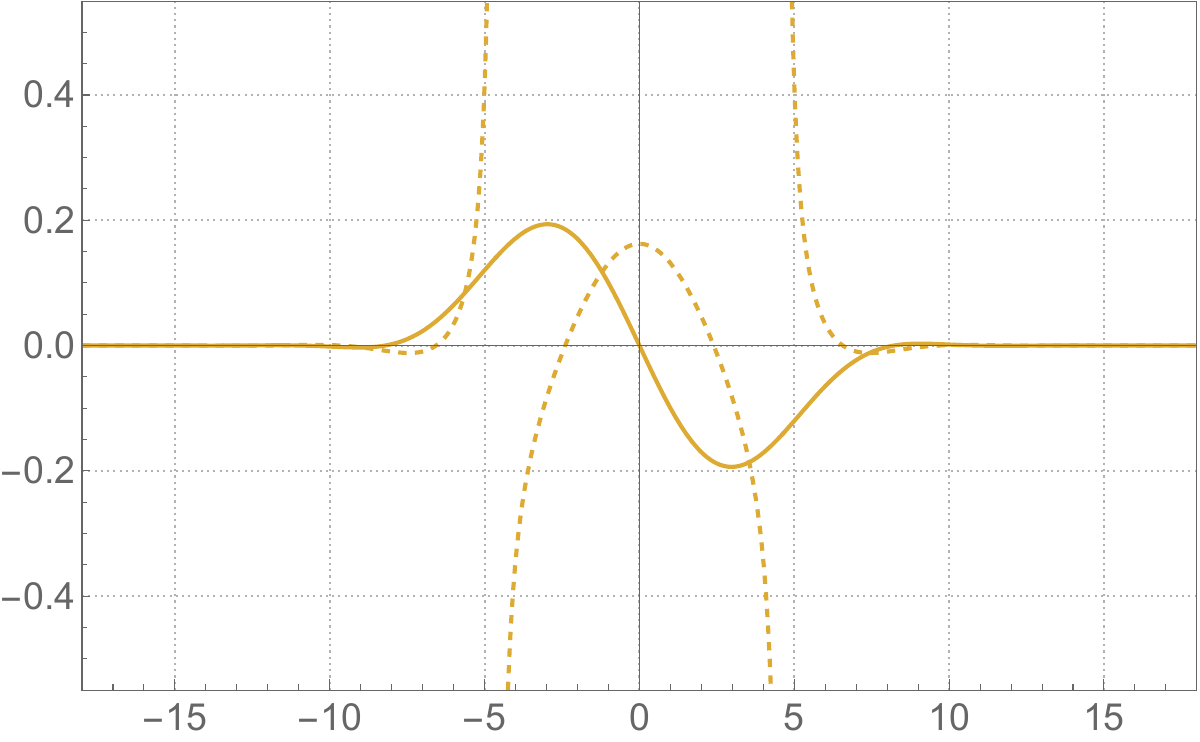}
    \includegraphics[width=0.44\textwidth]{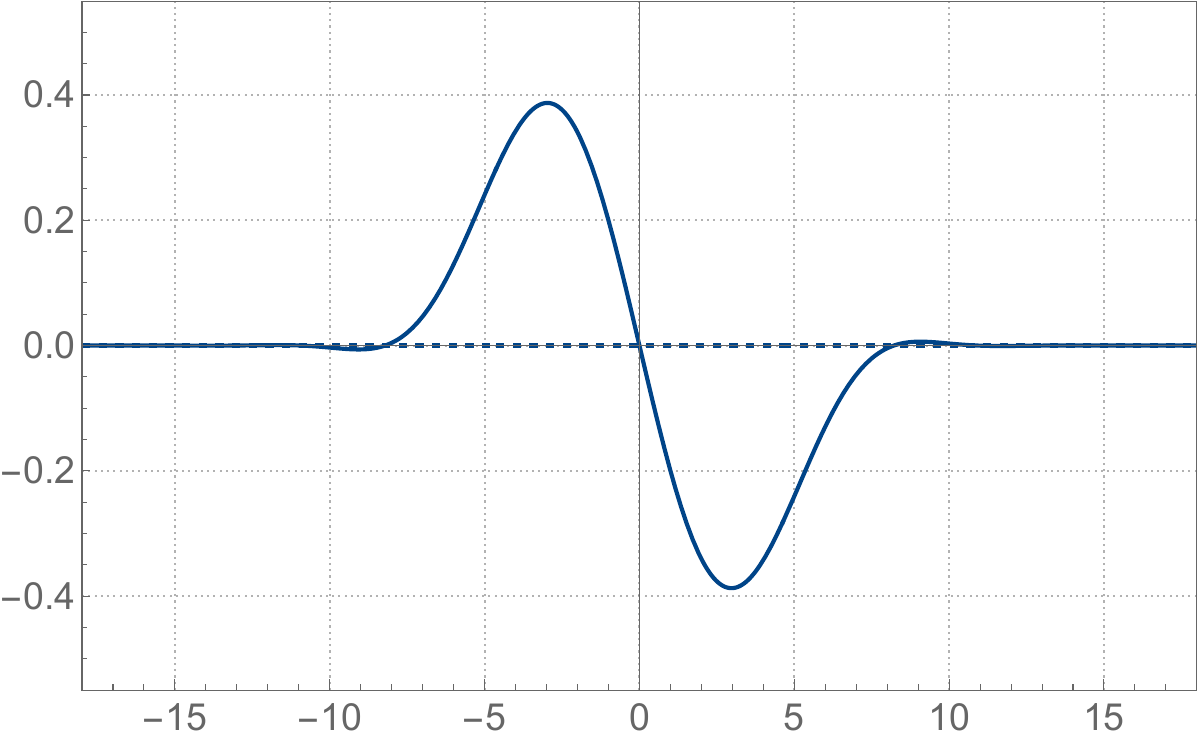}
    \includegraphics[width=0.44\textwidth]{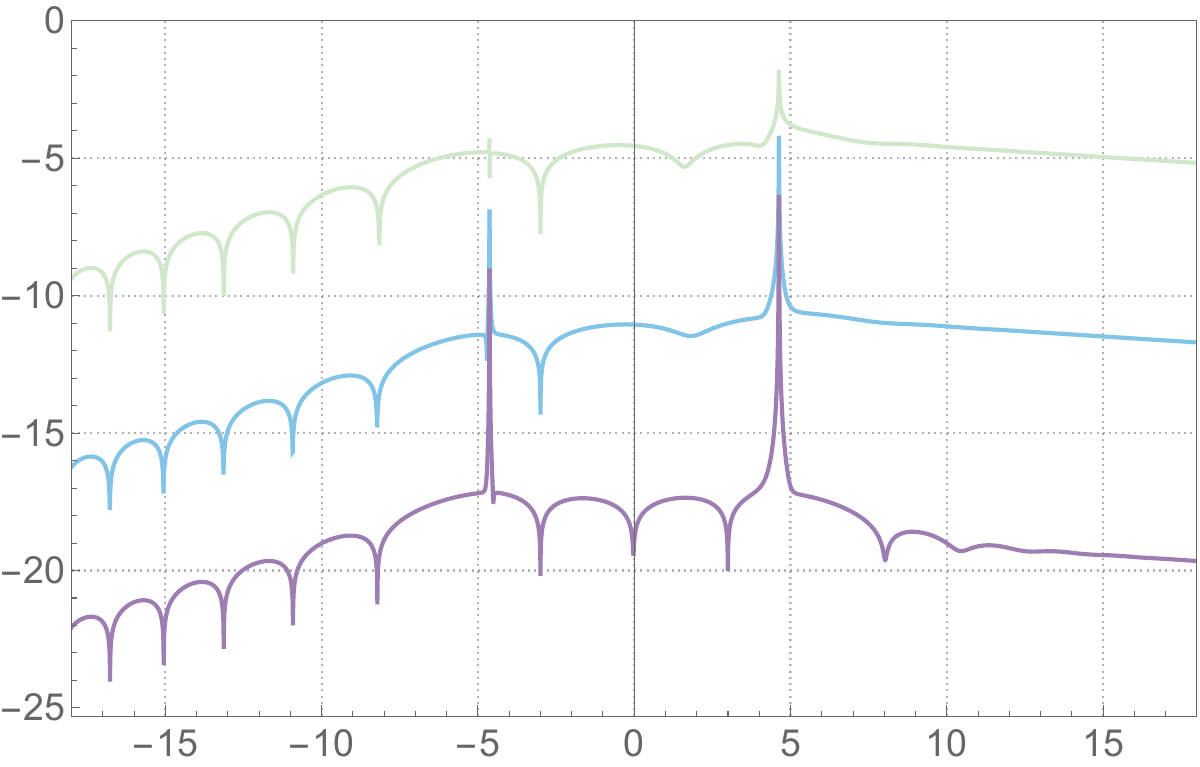}
    \caption{The on-shell eigenfunction from topological strings for $\xi = \sqrt{7} / 4$ and $\hbar = 3 \pi$ at energy $E = E_1 \approx 5.95$\,. See the explanation at the beginning of \autoref{sec:plots}.}
\end{figure}

\begin{figure}[h!]
    \centering
    \includegraphics[width=0.44\textwidth]{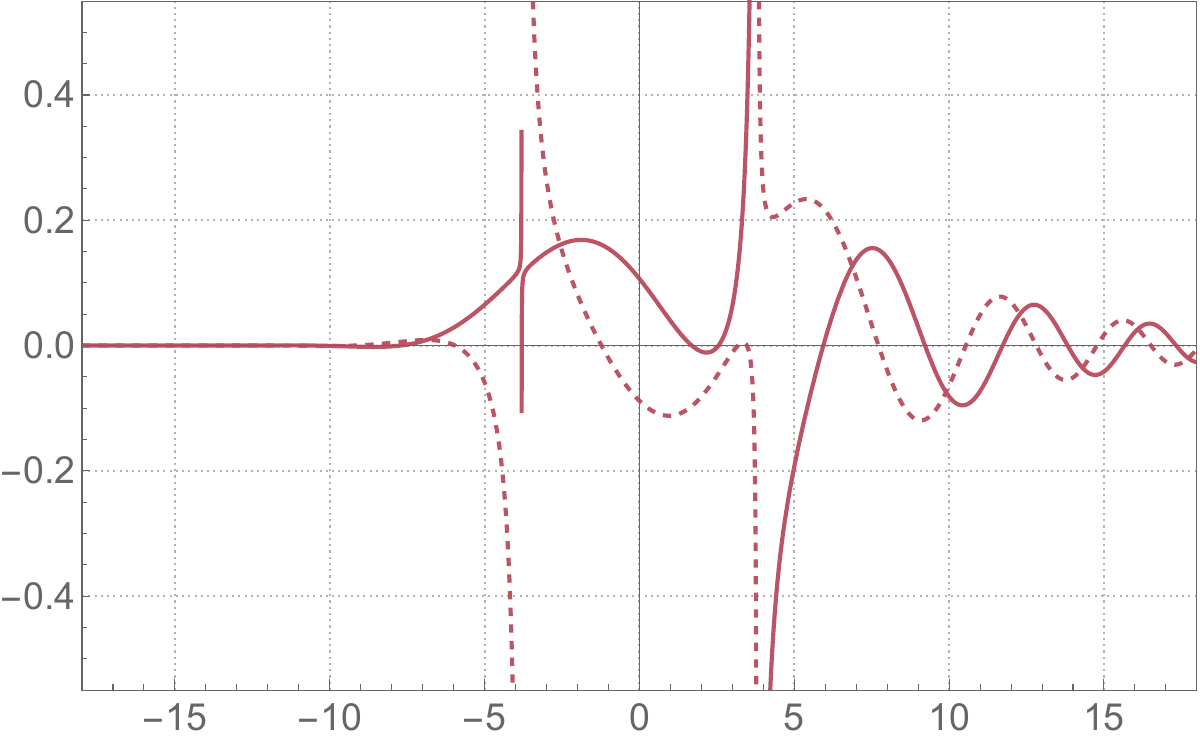}
    \includegraphics[width=0.44\textwidth]{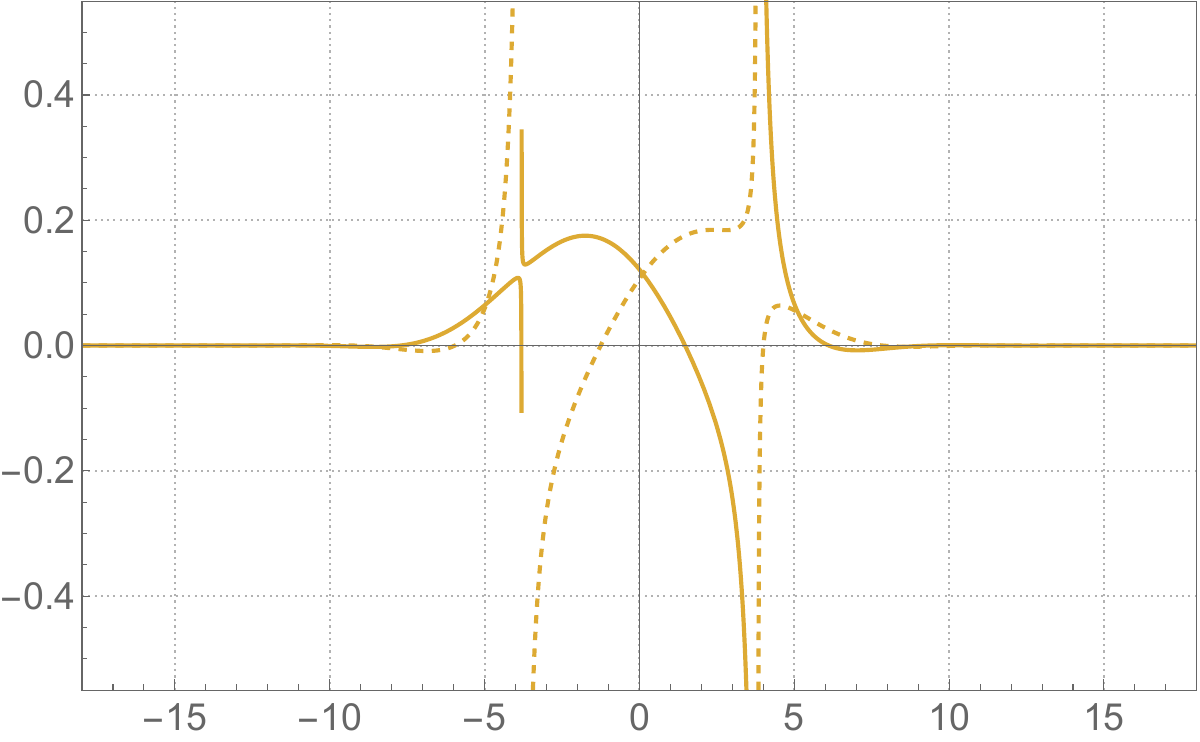}
    \includegraphics[width=0.44\textwidth]{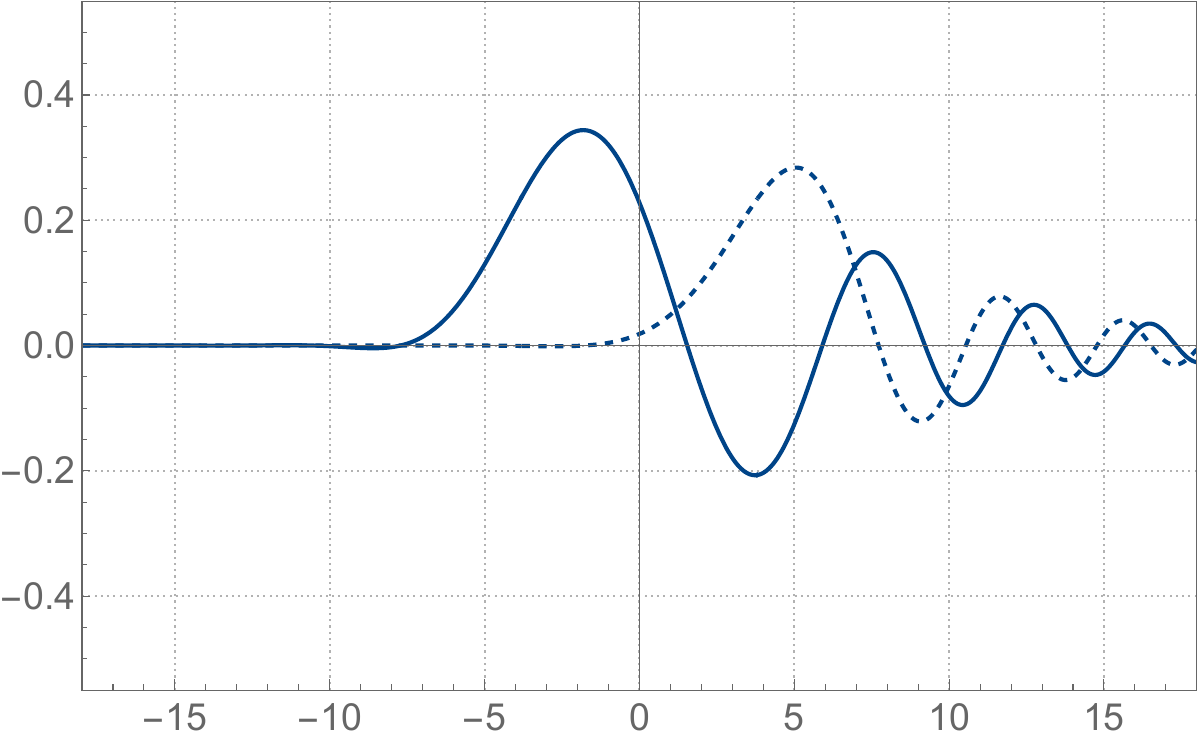}
    \caption{The off-shell eigenfunction from topological strings for $\xi = \sqrt{7} / 4$ and $\hbar = 3 \pi$ at energy $E = 118/23$\,, which is between $E_0$ and $E_1$. See the explanation at the beginning of \autoref{sec:plots}.}
\end{figure}

\clearpage

\subsection{\texorpdfstring{The case $\xi = 1/7$ and $\hbar = 5 \pi / 2$}{The case \pdfxi~=~1/7 and \pdfhbar~=~5\pdfpi/2}}

\begin{figure}[h!]
    \centering
    \includegraphics[width=0.44\textwidth]{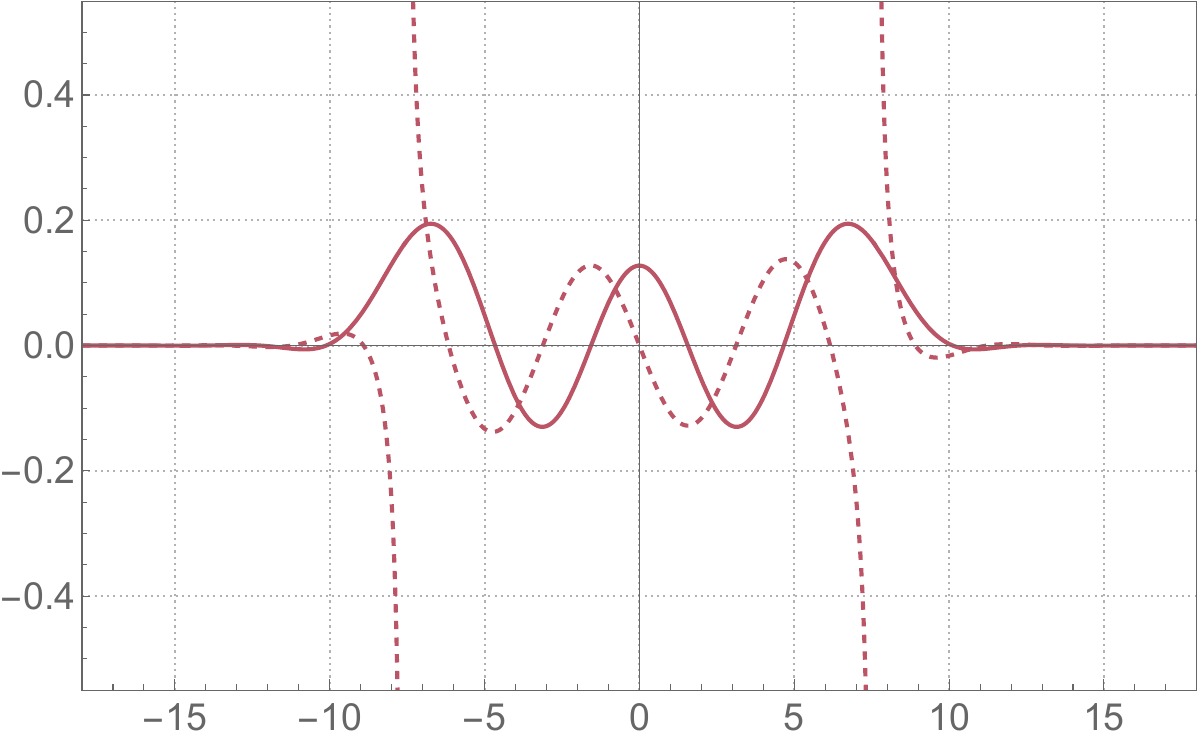}
    \includegraphics[width=0.44\textwidth]{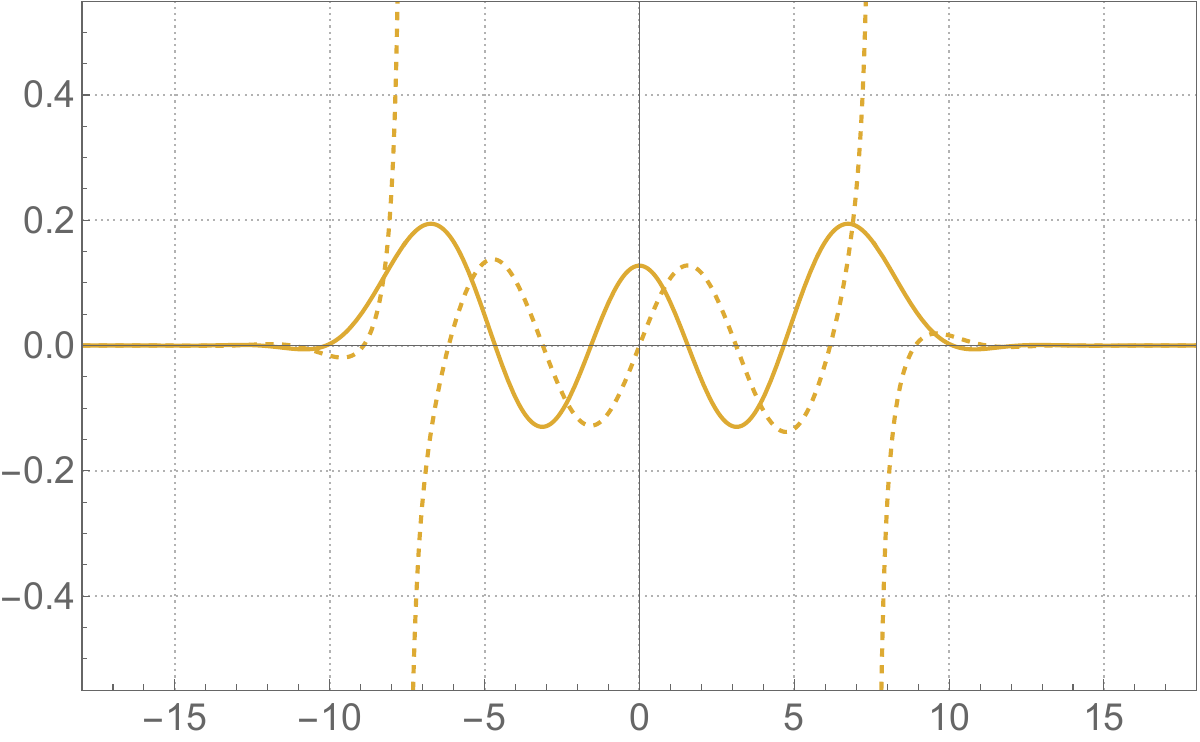}
    \includegraphics[width=0.44\textwidth]{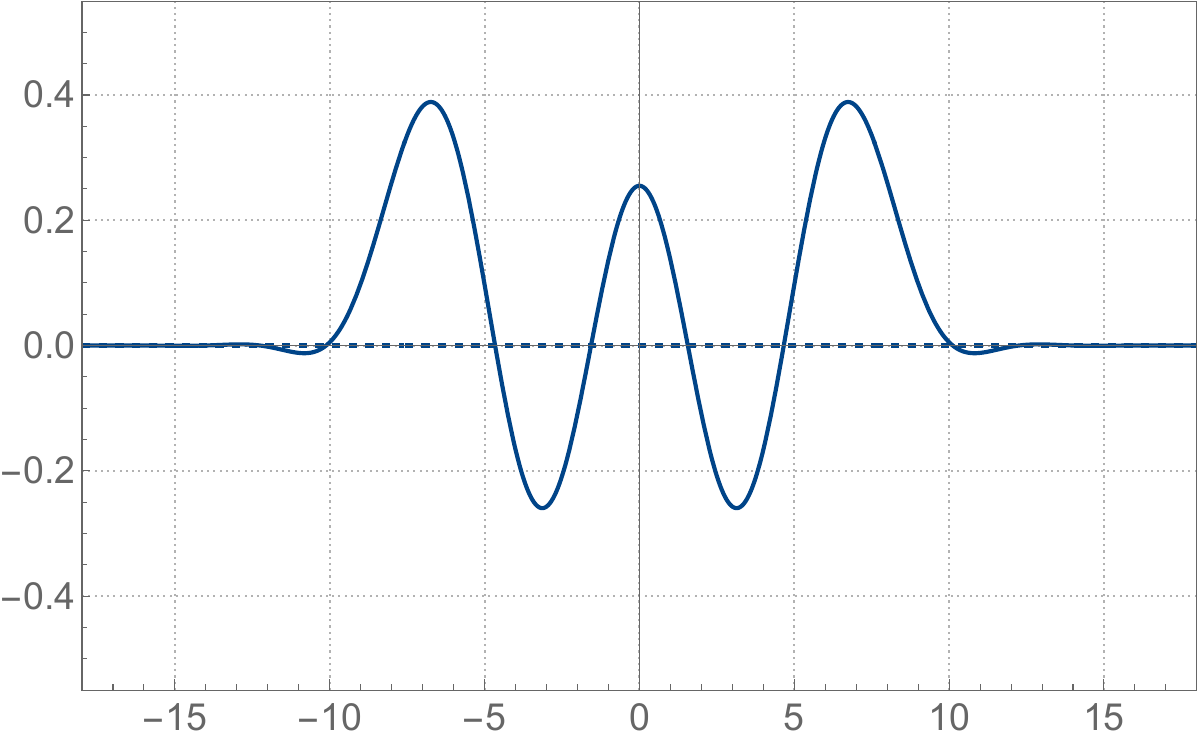}
    \includegraphics[width=0.44\textwidth]{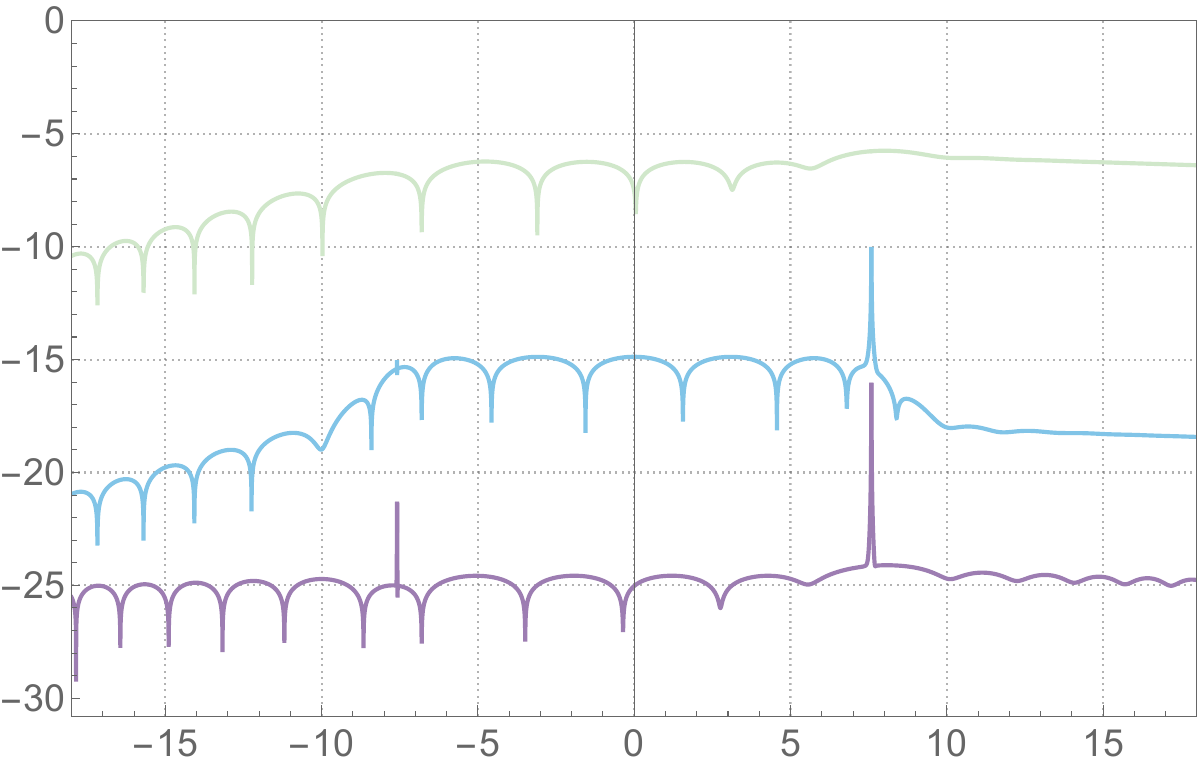}
    \caption{The on-shell eigenfunction from topological strings for $\xi = 1 / 7$ and $\hbar = 5 \pi / 2$ at energy $E = E_4 \approx 7.87$\,. See the explanation at the beginning of \autoref{sec:plots}.}
\end{figure}

\begin{figure}[h!]
    \centering
    \includegraphics[width=0.44\textwidth]{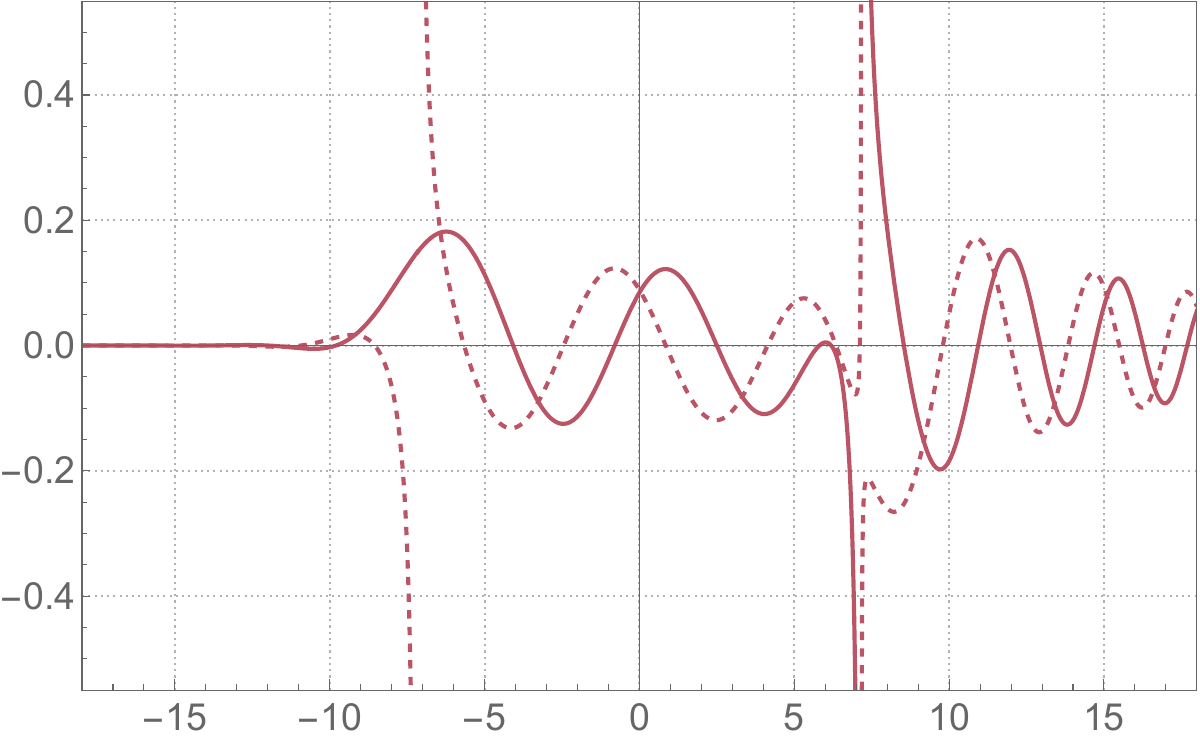}
    \includegraphics[width=0.44\textwidth]{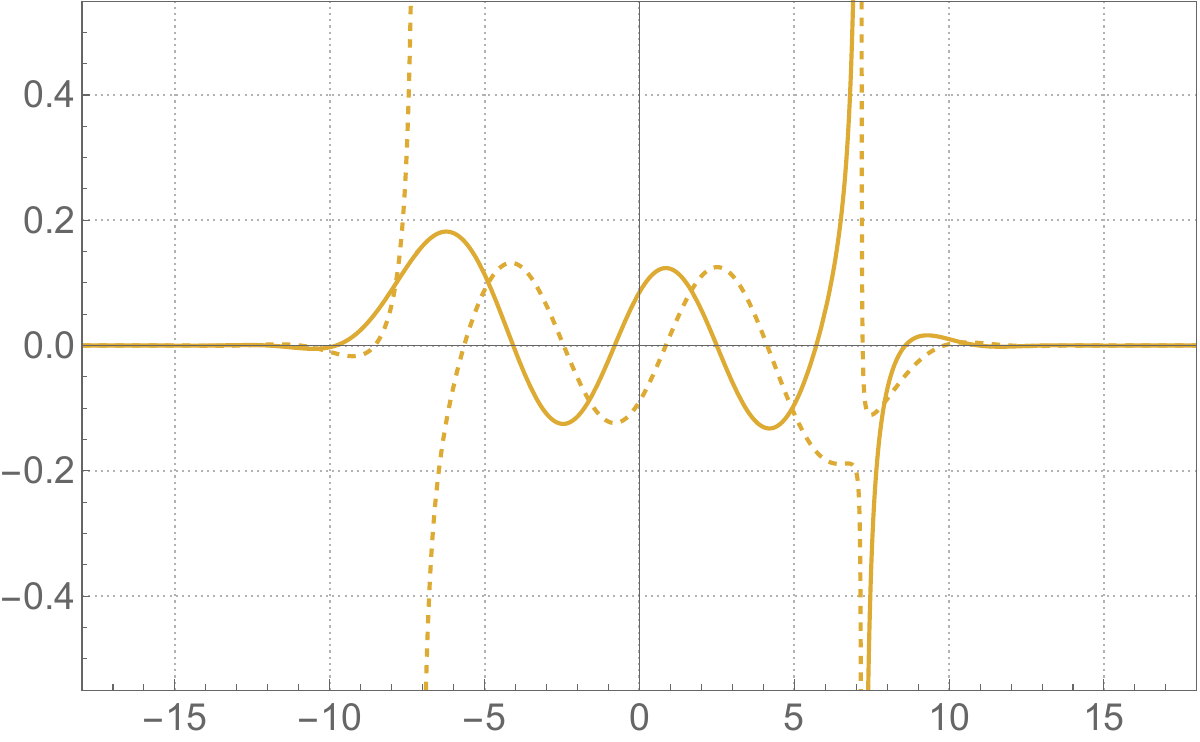}
    \includegraphics[width=0.44\textwidth]{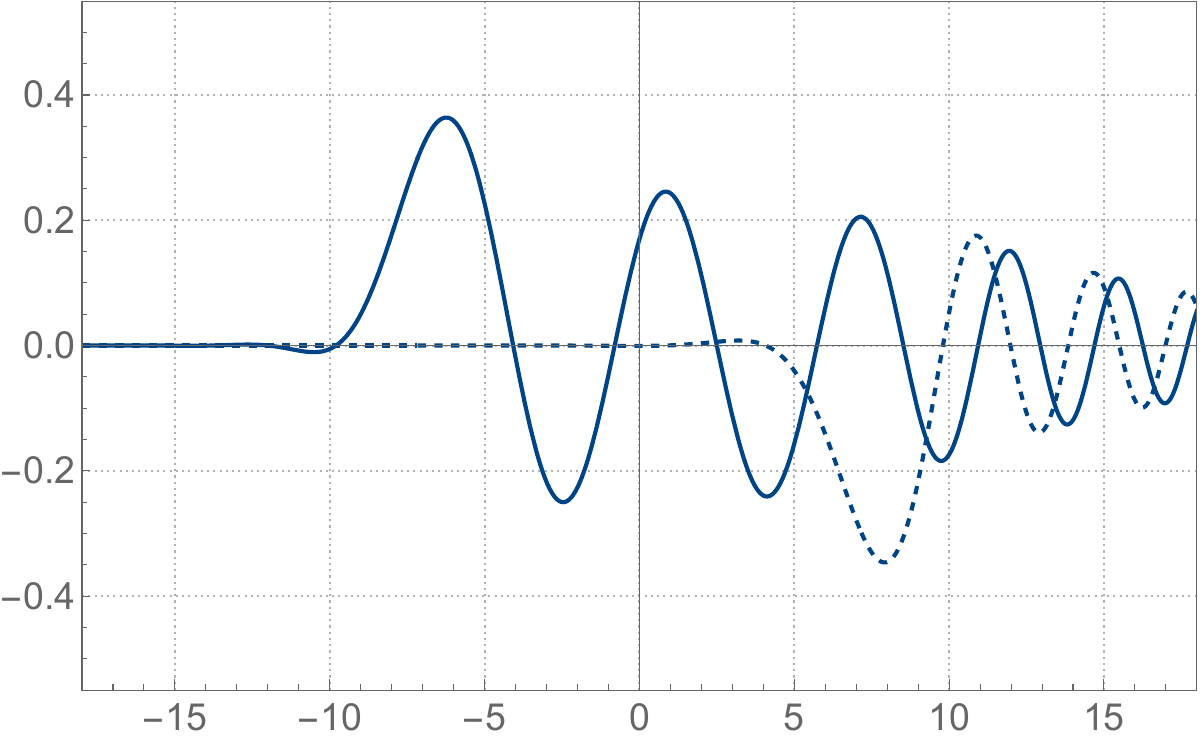}
    \caption{The off-shell eigenfunction from topological strings for $\xi = 1 / 7$ and at energy $E = 149/20$\,, which is between $E_3$ and $E_4$. See the explanation at the beginning of \autoref{sec:plots}.}
\end{figure}

\clearpage

\subsection{\texorpdfstring{The case $\xi = 2 / 3$ and $\hbar = 4 \sqrt{2}$}{The case \pdfxi~=~2/3 and \pdfhbar~=~4\pdfsqrt2}}

\begin{figure}[h!]
    \centering
    \includegraphics[width=0.44\textwidth]{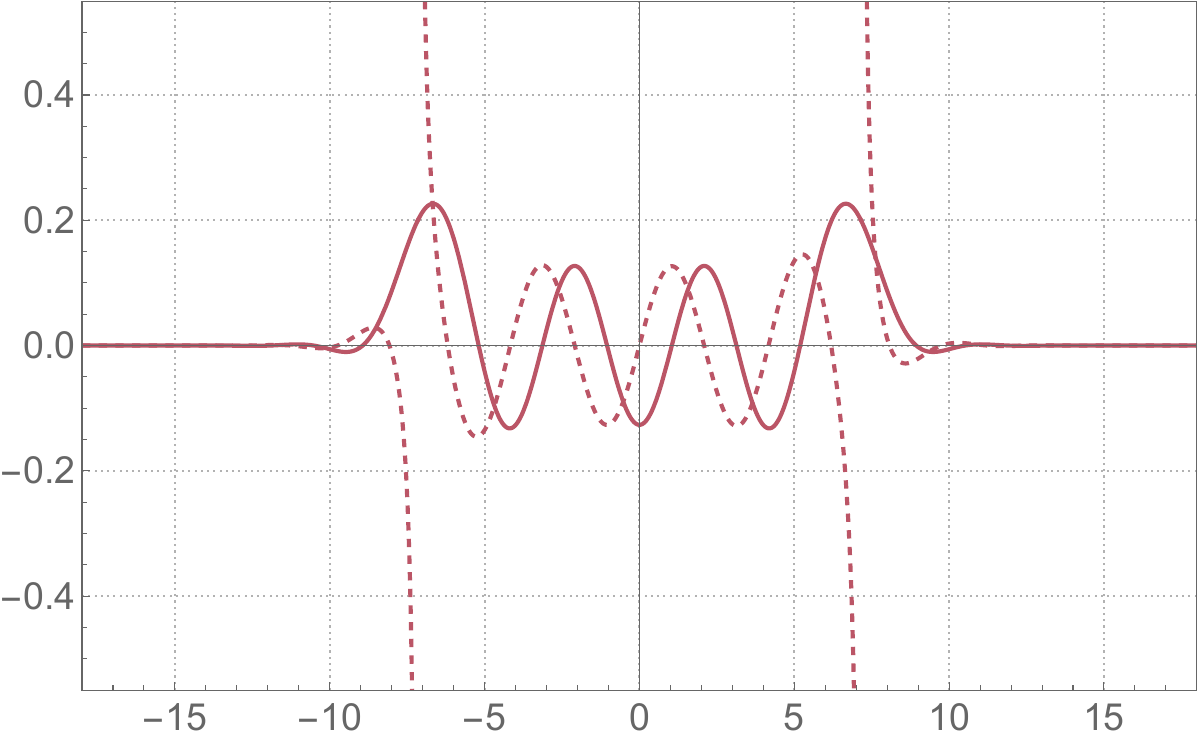}
    \includegraphics[width=0.44\textwidth]{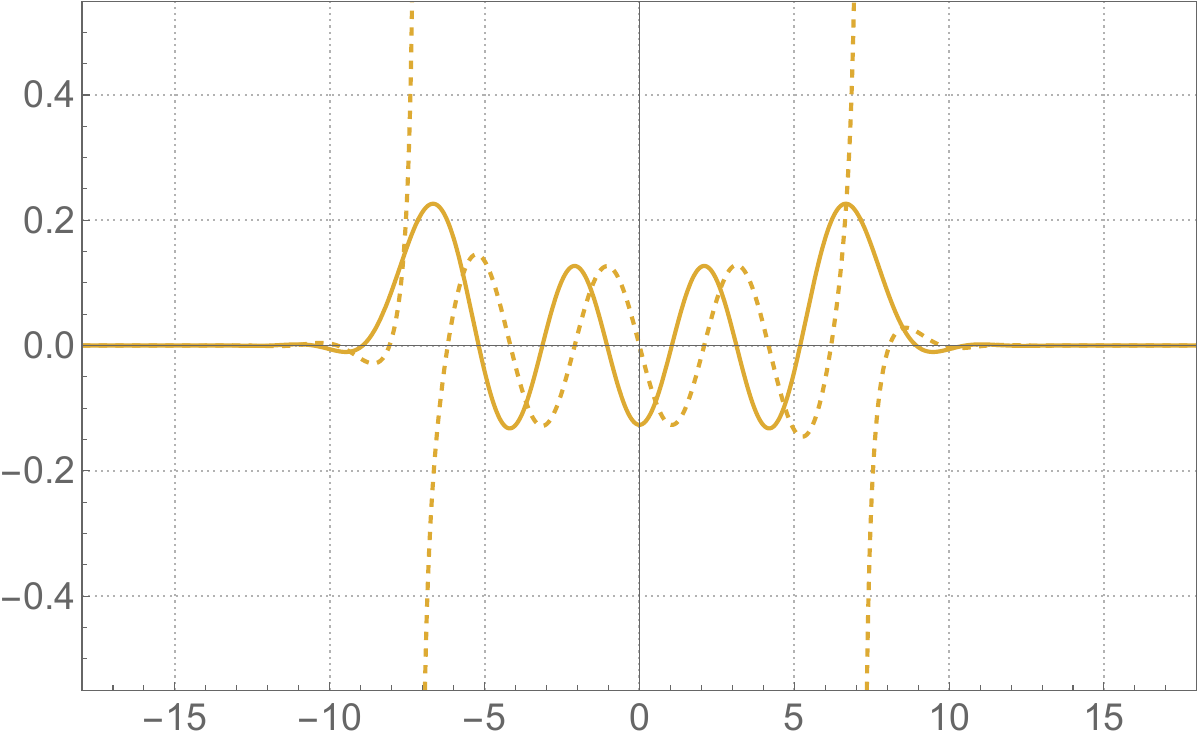}
    \includegraphics[width=0.44\textwidth]{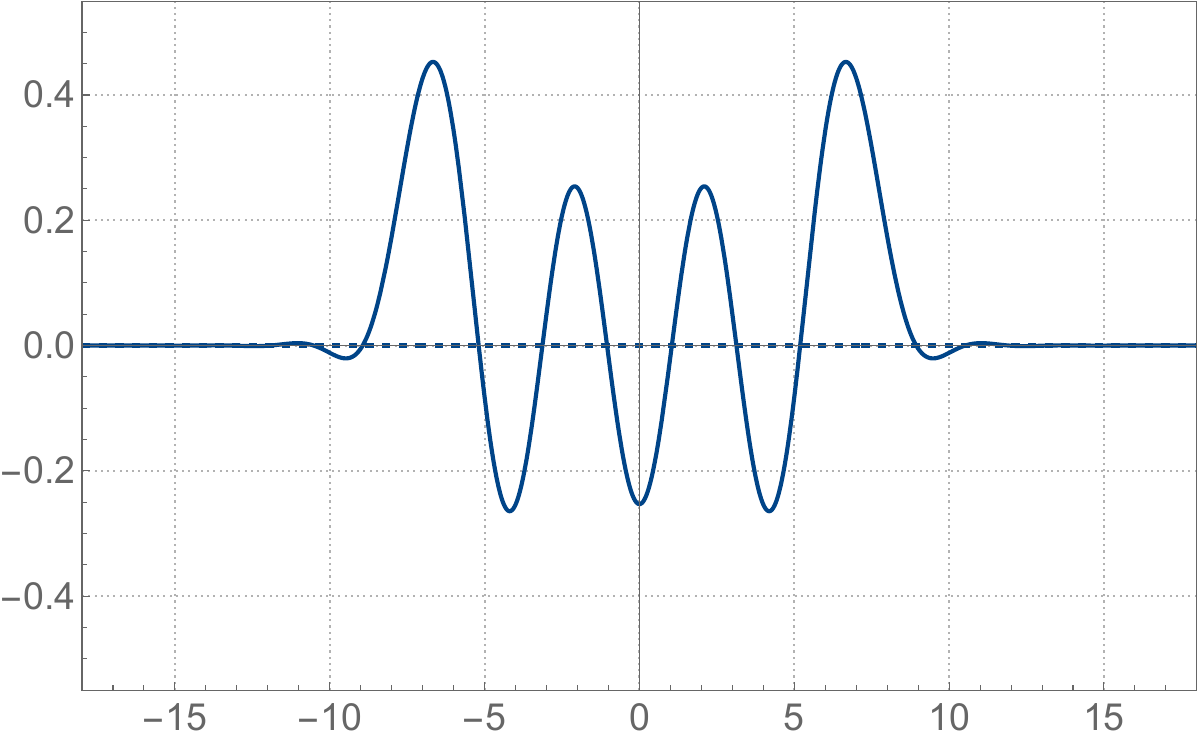}
    \includegraphics[width=0.44\textwidth]{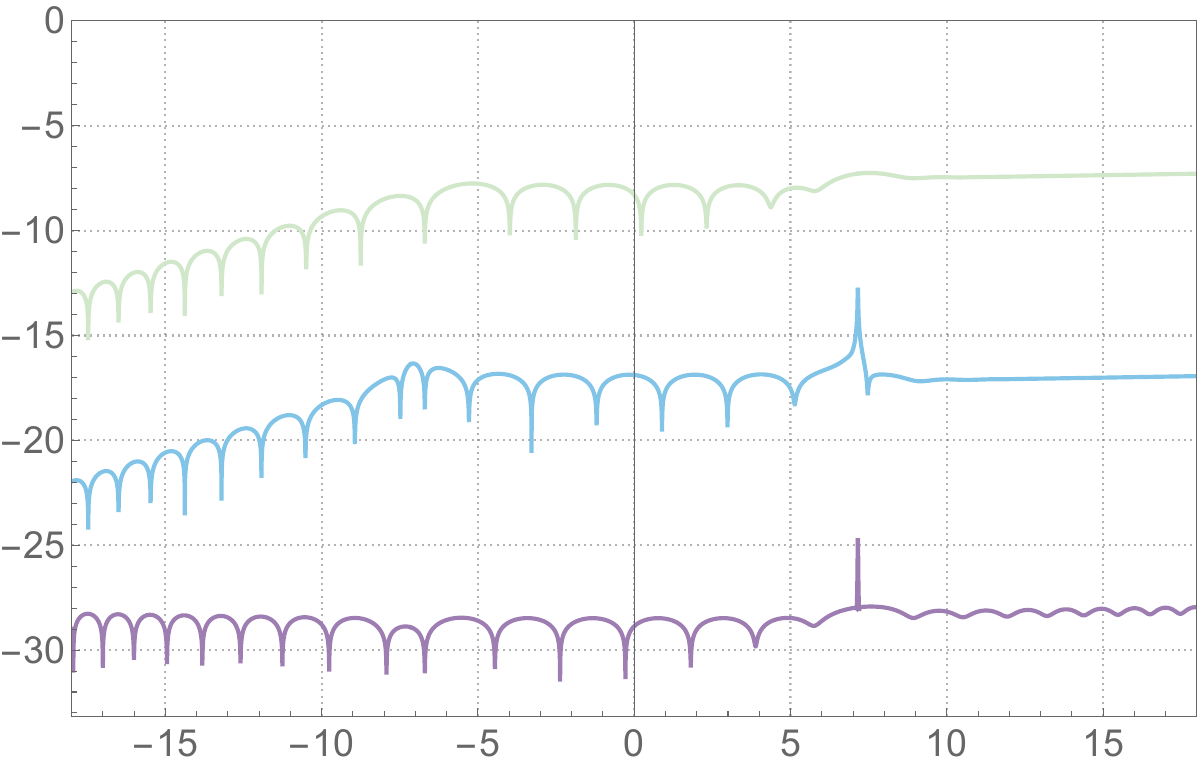}
    \caption{The on-shell eigenfunction from topological strings for $\xi = 2/3$ and $\hbar = 4 \sqrt{2}$ at energy $E = E_6 \approx 8.49$\,. See the explanation at the beginning of \autoref{sec:plots}.}
\end{figure}

\begin{figure}[h!]
    \centering
    \includegraphics[width=0.44\textwidth]{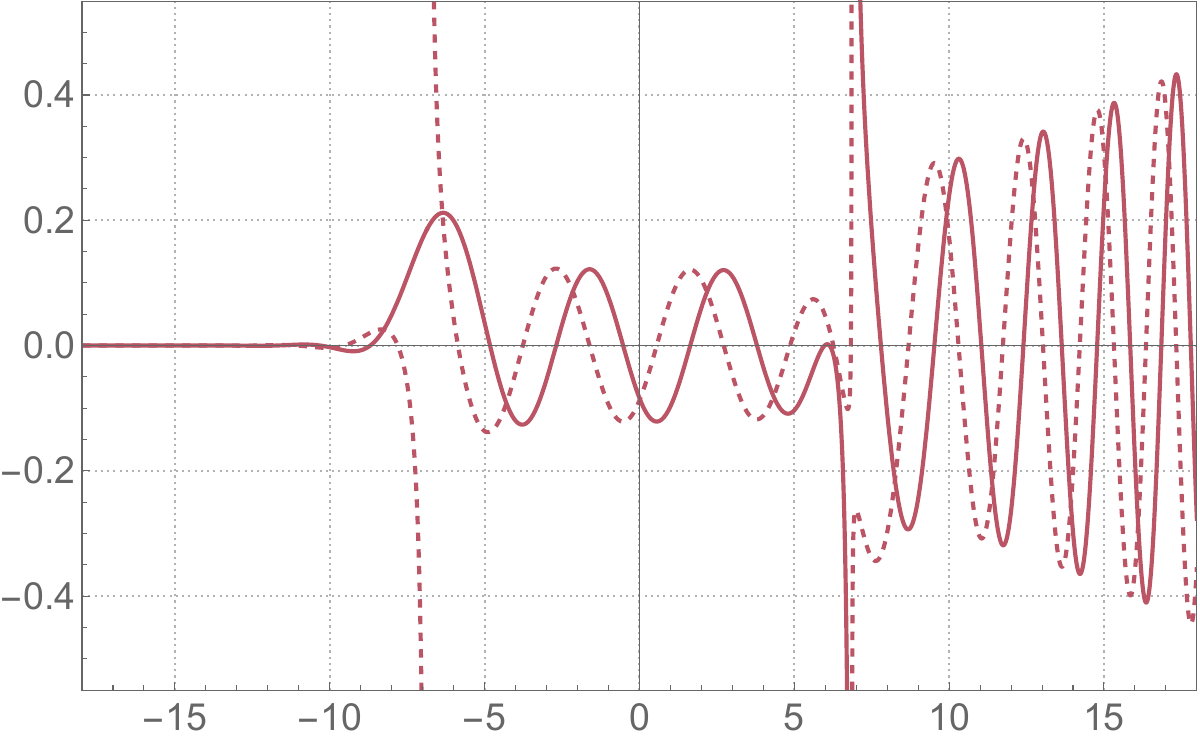}
    \includegraphics[width=0.44\textwidth]{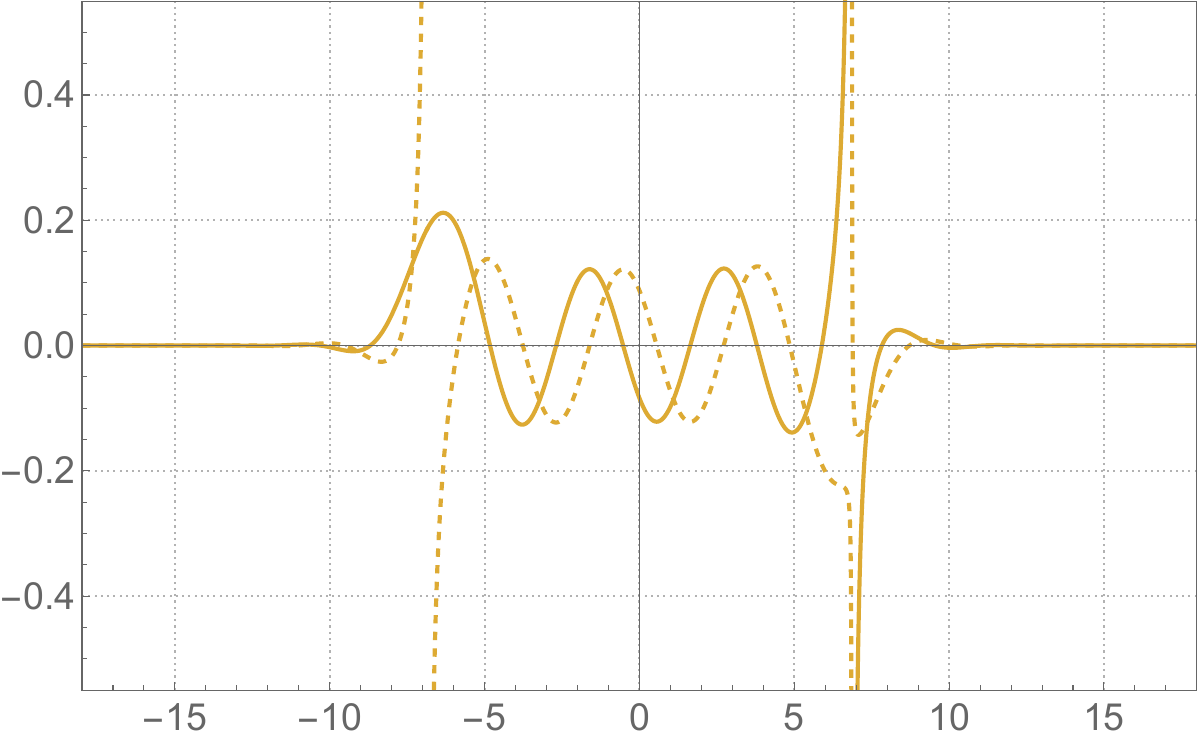}
    \includegraphics[width=0.44\textwidth]{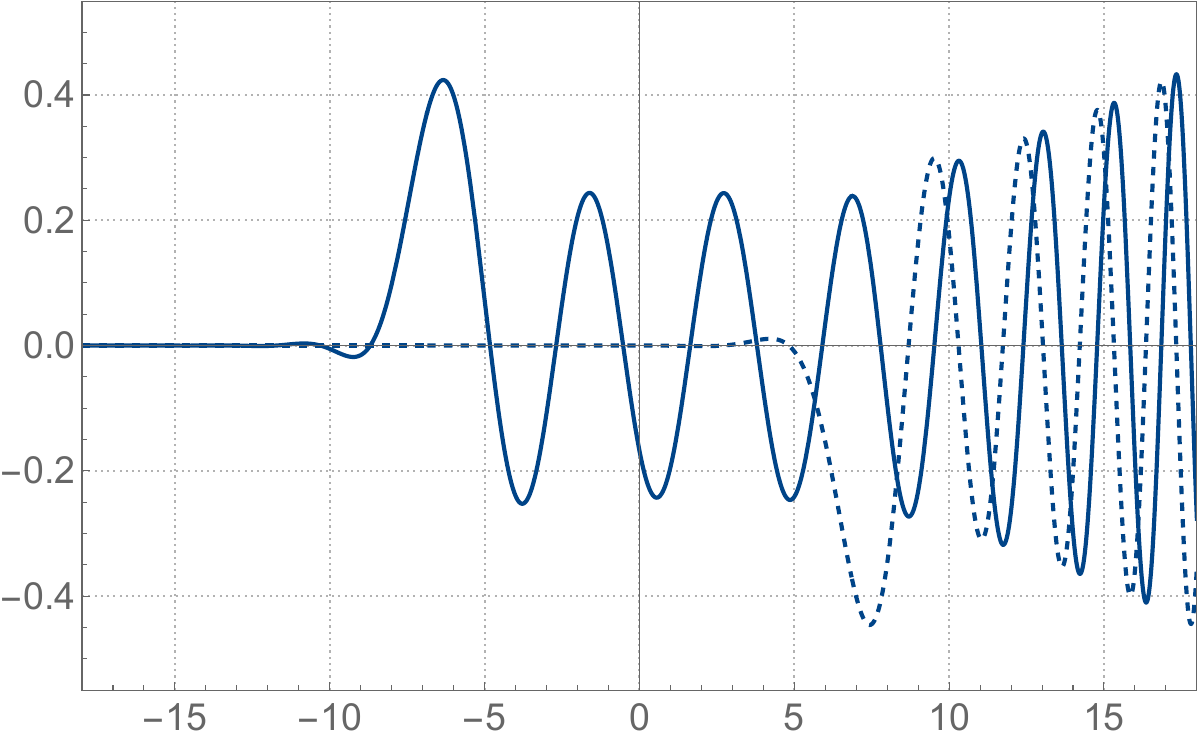}
    \caption{The off-shell eigenfunction from topological strings for $\xi = 2 / 3$ and $\hbar = 4 \sqrt{2}$ at energy $E = 172/21$, between $E_5$ and $E_6$. See the explanation at the beginning of \autoref{sec:plots}.}
\end{figure}

\clearpage

\bibliographystyle{JHEP}

\phantomsection 
\pdfbookmark[1]{References}{References} 

\bibliography{biblio}

\end{document}